# Hypoconstrained Jammed Packings of Nonspherical Hard Particles: Ellipses and Ellipsoids

Aleksandar Donev,[1,2] Robert Connelly,[3] Frank H. Stillinger,[4] and Salvatore Torquato[1,2,4,5,*]

[1]Program in Applied and Computational Mathematics, Princeton University, *Princeton NJ 08544*
[2]PRISM, Princeton University, *Princeton NJ 08544*
[3]Department of Mathematics, Cornell University, *Ithaca NY 14853*
[4]Department of Chemistry, Princeton University, *Princeton NJ 08544*
[5]Princeton Center for Theoretical Physics, Princeton University, *Princeton NJ 08544*

Continuing on recent computational and experimental work on jammed packings of hard ellipsoids [*Donev et al., Science, vol. 303, 990-993*] we consider jamming in packings of smooth strictly convex nonspherical hard particles. We explain why the isocounting conjecture, which states that for large disordered jammed packings the average contact number per particle is twice the number of degrees of freedom per particle ($\bar{Z} = 2d_f$), does not apply to nonspherical particles. We develop first- and second-order conditions for jamming, and demonstrate that packings of nonspherical particles can be jammed even though they are hypoconstrained ($\bar{Z} < 2d_f$). We apply an algorithm using these conditions to computer-generated hypoconstrained ellipsoid and ellipse packings and demonstrate that our algorithm does produce jammed packings, even close to the sphere point. We also consider packings that are nearly jammed and draw connections to packings of deformable (but stiff) particles. Finally, we consider the jamming conditions for nearly spherical particles and explain quantitatively the behavior we observe in the vicinity of the sphere point.

## I. INTRODUCTION

Jamming in disordered hard-sphere packings has been studied intensely in past years [1–3], and recently packings of non-spherical particles have been investigated as well [4, 5]. Computer simulations and experiments performed for packings of hard ellipsoids in Ref. [4] showed that asphericity, as measured by the deviation of the *aspect ratio* $\alpha$ from unity, dramatically affects the properties of jammed packings. In particular, it was observed that for frictionless particles the *packing fraction* (density) at jamming $\phi_J$ and the average *coordination* (*contact*) *number* per particle $\bar{Z}$ increase sharply from the typical sphere values $\phi_J \approx 0.64$ and $\bar{Z} = 6$ when moving away from the sphere point $\alpha = 1$. If one views $\phi_J$ and $\bar{Z}$ as functions of the particle shape, they have a cusp (i.e., they are non-differentiable) minimum at the sphere point.

It has been argued in the granular materials literature that large disordered jammed packings of hard frictionless spheres are *isostatic* [6–8], meaning that the total number of interparticle contacts (constraints) equals the total number of degrees of freedom and that all of the constraints are (linearly) independent. This property implies that the average number of contacts per particle is equal to the number of degrees of freedom per particle, $\bar{Z} = 2d$. This prediction has been verified computationally with very high accuracy [2, 3]. Most of previous discussions of isostaticity have been restricted to systems of spheres [6, 7], frictional systems [9], or systems of deformable particles [10]. For a general hard frictionless particle shape, the obvious generalization of the arguments that have been used for hard frictionless spheres would produce the expectation $\bar{Z} = 2d_f$, where $d_f$ is the number of degrees of freedom per particle ($d_f = 2$ for disks, $d_f = 3$ for ellipses, $d_f = 3$ for spheres, $d_f = 5$ for spheroids, and $d_f = 6$ for general ellipsoids). Although it has already been noted in Ref. [8] that the arguments for isostaticity only rigorously apply to perfectly spherical systems, this does not appear to be widely appreciated, and there appears to be a wide-spread expectation that $\bar{Z} \approx 2d_f$ for large disordered jammed packings of hard frictionless particles. We refer to this expectation as the *isocounting conjecture*, since it is based on the expectation that the total number of (independent) constraints equals the total number of degrees of freedom, that is, that the packings are *isoconstrained*. We have referred to this conjecture in the past as the *isostatic conjecture* [4]; however, here we give a more mathematically precise meaning to the term "isostatic", as explained in Section IV.

Since $d_f$ increases discontinuously with the introduction of rotational degrees of freedom as one makes the particles non-spherical, the isocounting conjecture would predict that $\bar{Z}$ would have a jump at $\alpha = 1$. Such a discontinuity was not observed in Ref. [4], rather, it was observed that ellipsoid packings are *hypoconstrained*, $\bar{Z} < 2d_f$, near the sphere point, and only become close to isoconstrained for large aspect ratios (but still remain hypoconstrained). These findings support the theoretical predictions in Ref. [8] that in general systems of non-spherical particles would be hypoconstrained and that the properties of packings should depend mildly on the exact particle shape.

The isocounting conjecture, as expressed by several of our colleagues, appears to be based on several wrong assumptions arising because of the use of linearization in the treatment of the interparticle impenetrability con-straints. Reference [8] terms this linearization as the "approximation of small displacements" (ASD). Firstly, it has been stated that a hypoconstrained packing *cannot* be rigid (jammed) due to the existence of floppy modes [10], which are unjamming motions (mechanisms) derived within a linear theory of rigidity. Additionally, various force-based arguments have been given [9] without realizing that forces themselves are first-order Lagrange multipliers and do not necessarily exist when one considers perfectly hard particles outside of the linear (first-order) approximation. Reference [8] states that the ASD approximation is "indispensable if one wishes to deal with linear problems...In the case of granular systems, it will also lead to a linearization of the problems, for the curvature of configuration spaces will be ignored." The observation that terms of order higher than first need to be considered is emphasized in Ref. [8]: "When floppy modes exist..., they appear as marginally unstable and one cannot tell whether, to higher orders, they actually destabilize the equilibrium configuration." However, the mathematical analysis extending beyong the ASD is not developed except for spheres [8]. If the curvature of the particles at the point of contact is included in a second-order approximation (still for infinitesimaly small displacements), then it can be seen that hypoconstrained packings of nonspherical hard packings can in fact be rigid, jammed, or stable (these terms are defined in Section IV). One does not need to consider particle deformability, friction, large displacements, or stability under a specific applied load such as gravity, in order to see why packings of nonspherical particles are generally hypoconstrained. Through the second-order mathematical analysis it will become clear that pre-existing (internal) stresses inside the packing are essential, as already realized in Ref. [10]. While this prestress is merely a mathematical tool for static packings of perfectly hard frictionless particles, for realistic systems particle deformability, history of preparation, and applied loads all bear a strong influence on the internal stresses in the packing and thus the mechanical properties of the system.

*Electronic address: torquato@electron.princeton.edu

In this paper, we generalize our previous theoretical and computational investigations of jamming in sphere packings [2, 11] to packings of nonspherical particles, and in particular, packings of hard ellipsoids. We generalize the mathematical theory of rigidity of tensegrity frameworks [12, 13] to packings of nonspherical particles, and demonstrate rigorously that the computer-generated ellipsoid packings we studied in Ref. [4] are jammed even very close to the sphere point. Armed with this theoretical understanding of jamming, we also obtain a quantitative understanding of the cusp-like behavior of $\phi_J$ and $\bar{Z}$ around the sphere point. Specifically, we:

- Explain why the isocounting conjecture does not apply to nonspherical particles.

- Develop first- and second-order conditions for jamming, and demonstrate that packings of nonspherical particles can be jammed even though they are hypoconstrained.

- Design an algorithm that uses the jamming conditions to test whether computer-generated hypoconstrained ellipsoid and ellipse packings are jammed, and demonstrate numerically that our algorithm does produce jammed packings, even close to the sphere point.

- Study the thermodynamics of packings that are nearly jammed and draw connections to packings of deformable (but stiff) particles.

- Develop first-order expansions for nearly spherical particles and explain quantitatively the behavior we observe in the vicinity of the sphere point.

### A. Random Jammed Packings of Hard Ellipsoids

The packing-generation algorithm we employ generalizes the Lubachevsky-Stillinger (LS) sphere-packing algorithm [14] and is described in detail in Ref. [15]. The method is a hard-particle molecular dynamics (MD) algorithm for producing dense disordered packings. Initially, small particles are randomly distributed and randomly oriented in a box with periodic boundary conditions and without any overlap. The particles are given velocities (including angular velocities) and their motion followed as they collide elastically and also expand uniformly. As the density approaches the jamming density, the collision rate diverges. In the jamming limit, the particles touch to form the *contact network* of the packing, exerting compressive forces on each other but not being able to move despite thermal agitation (shaking). If the rate of particle growth, or *expansion rate* $\gamma$, is initially sufficiently large to suppress crystallization, and small enough close to jamming to allow for local relaxation necessary for true jamming, the final packings are disordered and representative of the maximally random jammed[46] (MRJ) state [16] (corresponding to the least ordered among all jammed packings).

Note that the computational methodology presented in Ref. [2] applies to ellipsoids as well and we do not repeat the details presented there. The ellipsoid packings produced by the algorithm do not show signs of local or global crystallization. The exact phase diagram for hard ellipsoids is not known, and in particular, it is not known what the high-density crystal phase is [17]; however, it is expected that nematic ordering is present at high densities. The produced packings do not show (global) nematic order to within statistical accuracy [4, 18]. A more detailed analysis of the local (translational and orientational) correlations in truly jammed ellipsoid packings has not been performed to our knowledge, however, based on our experience with spheres we expect our algorithm to supress crystallization under appropriate conditions [2]. Sphere packings have been observed to have a substantial fraction of rattling particles ($\sim 2.5\%$) [2], and such rattlers are also observed in packings of nearly-spherical ellipsoids. However, the fraction of rattlers appear to rapidly decrease as asphericity is increased, so that the majority of ellipsoid packings we have generated do not have any rattlers at all. For spheres, the packings produced with the MD algorithm can be very close to the jamming point, so that the interparticle gaps are close to numerical precision ($\sim 10^{-15}$) [2]. Similar precision can be achieved for ellipsoids, however, this takes at least an order of magnitude more computational effort (or even two orders of magnitude for very aspherical ellipsoids). Typically we have jammed the packings to a reduced pressure $p \sim 10^6 - 10^9$, which ensures that the distance to jamming is on order of $10^{-9} - 10^{-6}$. To really identify the exact contact network in the jamming limit requires even higher pressures for larger packings due to existence of a multitude (more specifically, a power-law divergence) of near contacts in disordered packings [2]. However, with reasonable effort the average coordination number $\bar{Z}$ can be identified within one percent even for systems of $N = 10^5$ ellipsoids. Those packings for which we perform an exact analysis of the contact network (such as, for example, rigorously testing for jamming) have been prepared carefully and are sufficiently close to the jamming point to exactly identify all of the true contacts.

In Fig. 1 we show newer results than those in Ref. [4] for the jamming density $\phi_J$ and contact number $\bar{Z}$ of jammed monodisperse packings of hard ellipsoids in three dimensions. The ellipsoid semiaxes have ratios $a : b : c = 1 : \alpha^\beta : \alpha$ where $\alpha > 1$ is the *aspect ratio* (for general particle shapes, $\alpha$ is the ratio of the radius of the smallest circumscribed to the largest inscribed sphere), and $0 \leq \beta \leq 1$ is the "oblateness", or *skewness* ($\beta = 0$ corresponds to prolate and $\beta = 1$ to an oblate spheroid). It is seen that the density rises as a linear function of $\alpha - 1$ from its sphere value $\phi_J \approx 0.64$, reaching densities as high as $\phi_J \approx 0.74$ for the self-dual ellipsoids with $\beta = 1/2$. The jamming density eventually decreases again for higher aspect ratios, however, we do not investigate that region in this work. The contact number also shows a rapid rise with $\alpha - 1$, and then plateaus at values somewhat below isoconstrained, $\bar{Z} \approx 10$ for spheroids, and $\bar{Z} \approx 12$ for nonspheroids. In Section IX we will need to revert to two dimensions (ellipses) in order to make some analytical calculations possible. We therefore also generated jammed packings of ellipses, and show the results in Fig. 2. Since monodisperse packings of disks always crystallize and do not form disordered jammed packings, we used a binary packing of particles with one third of the particles being 1.4 times larger than the remaining two thirds. The ellipse packings show exactly the same qualitative behavior as ellipsoids.

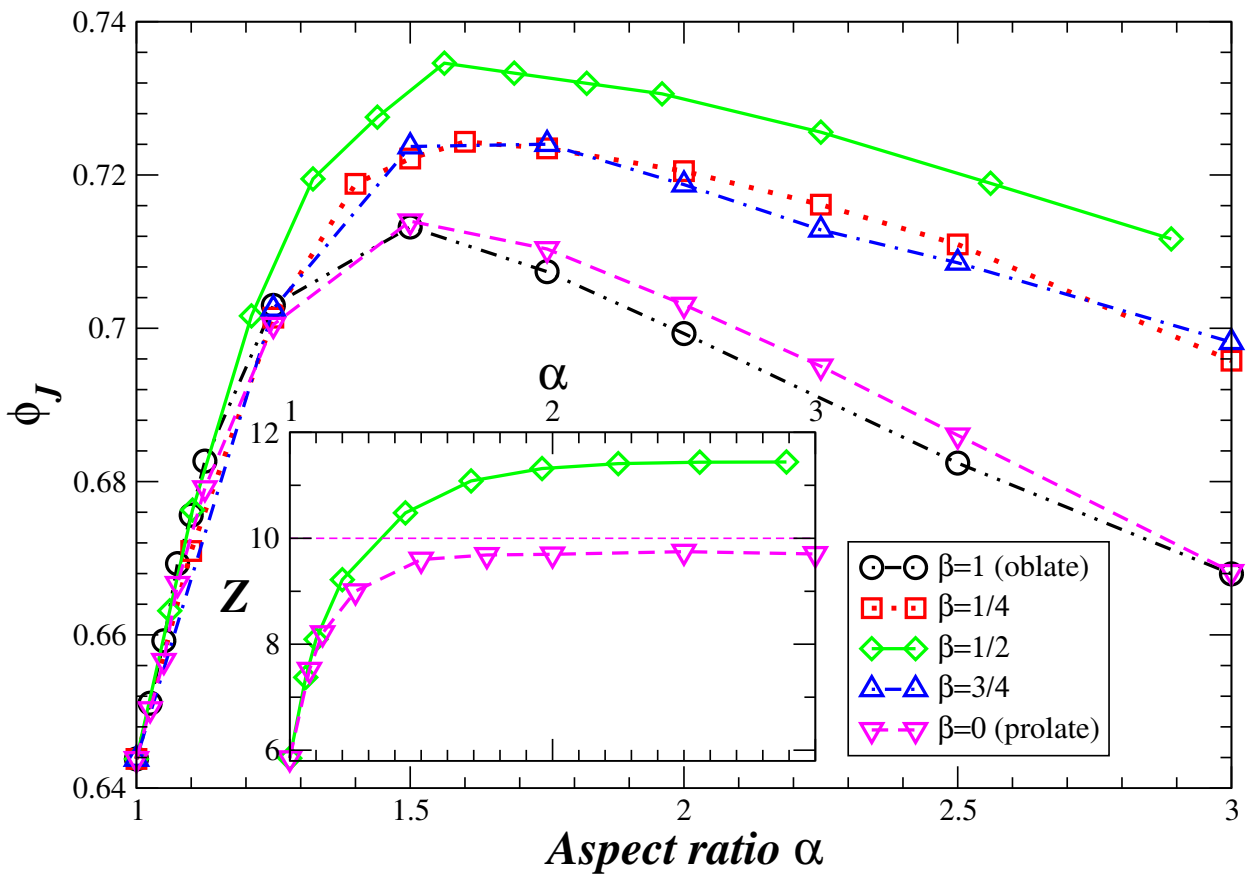

Figure 1: (Color online) Jamming density and average contact number (inset) for packings of $N = 10000$ ellipsoids with ratios between the semiaxes of $1 : \alpha^\beta : \alpha$ [c.f. Fig. (2) in Ref. [4]]. The isoconstrained contact numbers of 10 and 12 are shown as a reference.

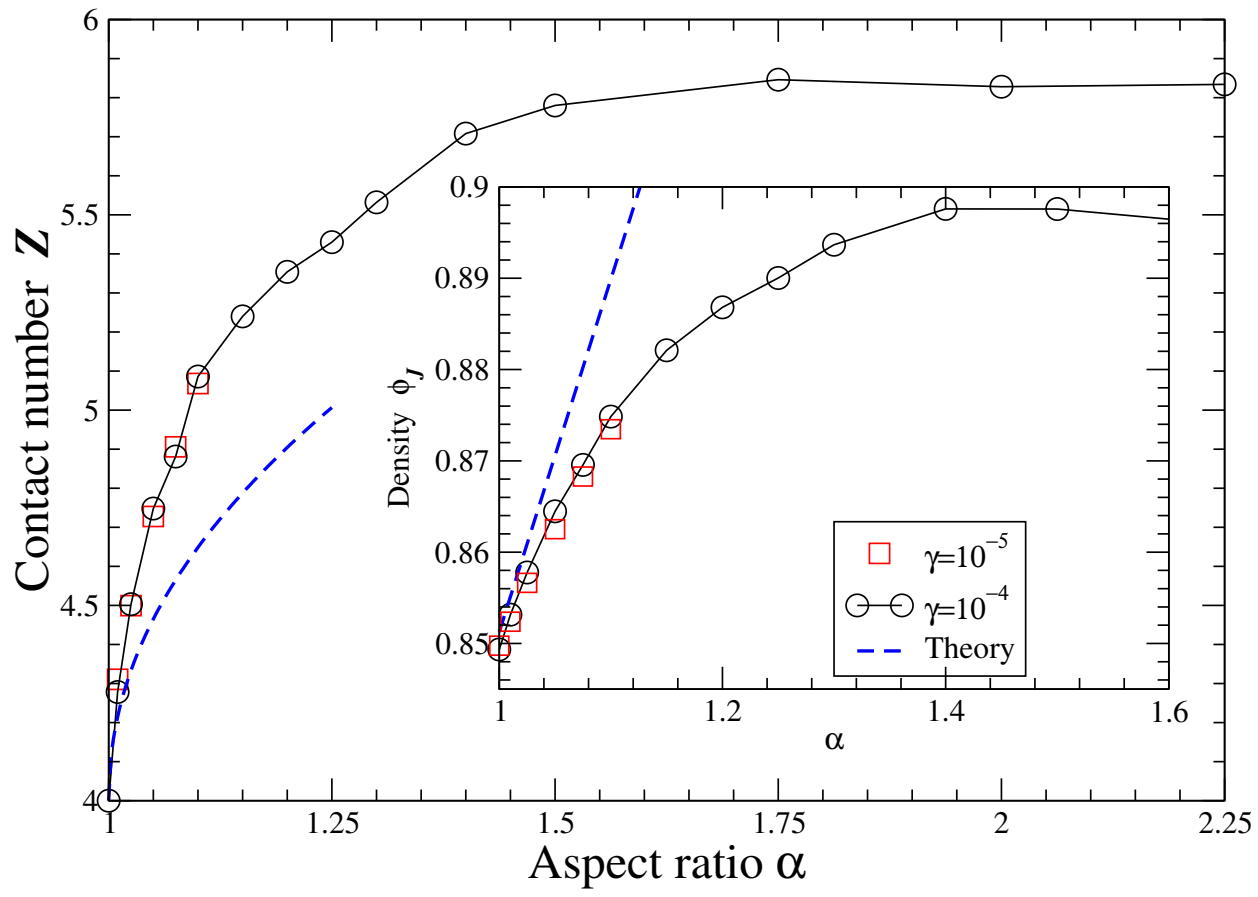

Figure 2: (Color online) Average contact number and jamming density (inset) for bi-disperse packings of $N = 1000$ ellipses with ratios between the semiaxes of $1 : \alpha$, as produced by the MD algorithm using two different expansion rates $\gamma$ (affecting the results only slightly). The isoconstrained contact number is 6. The results of the leading-order (in $\alpha - 1$) theory presented in Section IX are shown for comparison.

### B. Non-Technical Summary of Results

In this Section, we provide a non-technical summary our theoretical results and observations discussed in the main body of the paper. This summary is intended to give readers an intuitive feeling for the mathematical formalism developed in this work and demonstrate the physical meaning and relevance of our results. We will refer the interested reader to appropriate sections to find additional details.

One aim of this paper is to explain the numerical results presented in Section I A. In particular, we will explain why jammed disordered packings of ellipsoids are strongly hypoconstrained near the sphere point, and also why, even far from the sphere point, ellipsoid packings are hypoconstrained rather than isoconstrained as are sphere packings. By a jammed packing we mean a packing in which any motion of the particles, including collective combined translational and rotational displacements, introduces overlap between some particles. Under appropriate qualifications, a jammed packing can also be defined as a rigid packing, that is, a packing that can resolve any externally applied forces through interparticle ones.

#### 1. *Hypostatic Packings of Nonspherical Particles Can be Jammed*

As explained in Section IV, the isoconstrained property is usually justified in two steps. First, non-degeneracy is invoked to demonstrate the inequality $\bar{Z} \leq 2d_f$, then, the converse inequality $\bar{Z} \geq 2d_f$ is invoked to demonstrate the equality $\bar{Z} = 2d_f$. The inequality $\bar{Z} \geq 2d_f$ is usually justified by claiming that a packing cannot be jammed without having more contacts (impenetrability constraints) than degrees of freedom. A hypoconstrained packing necessarily has "floppy" or zero modes, which are collective motions of the particles that preserve the interparticle distances to first order in the magnitudes of the particle displacements. It is claimed that such floppy modes are not blocked by the impenetrability constraints and therefore a hypoconstrained packing cannot be jammed. Alternatively, it is claimed that externally applied forces that are in the direction of such floppy modes cannot be resisted (sufficiently) by the interparticle forces and therefore the packing cannot be rigid. We will now explain, through an example, why these claims are wrong and, in fact, why a hypoconstrained packing can be jammed/rigid if the curvature of the particles at the point of contact is sufficiently flat in order to block the floppy modes.

Consider an isoconstrained jammed packing of hard circular disks, as illustrated in Fig. 3. In reality, the disks would be elastic (soft) but stiff, and let's imagine the system is under a uniform state of compression, so that the particles are exerting compressive forces on each other. If there are no additional external forces, the interparticle forces would be in force equilibrium. The packing is translationally jammed, and the disk centroids are immobile; however, the (frictionless) disks can freely rotate without introducing any additional overlap. That is, if we take into account orientational degrees of freedom, the disk packing would not be jammed. It would possess floppy modes consisting of particles rotating around their own centroids. These floppy modes are however trivial at the circle (sphere) point in that they do not actually change the packing configuration.

Now imagine making the particles non-circular (or nonspherical in three dimensions), and in particular, making them polygons, so that the point contacts between the disks become (extended) contacts between flat sides of the polygons. The floppy modes still remain, in the sense that rotations of the polygons, to first order, simply lead to the two tangent planes at the points of contact sliding along each other without leading to overlap. However, it is clear that this is only a first-order approximation. In reality, the polygons cannot be rotated because such rotation leads to overlap in the *extended* region of contact around the point of contact. To calculate the amount of overlap, one must use second-order terms, that is, consider not only the tangent planes at the point of contact but also the *curvature* of the particles at the point of contact. Low curvature, that is, "flat" contacts, block rotations of the particles. It should be evident that even if the radius of curvature is not infinite, but exceeds a certain threshold[47], the floppy modes would in fact be blocked and the packing would be jammed despite being hypoconstrained. In fact, the packing has exactly as many contacts as the original disk packing.

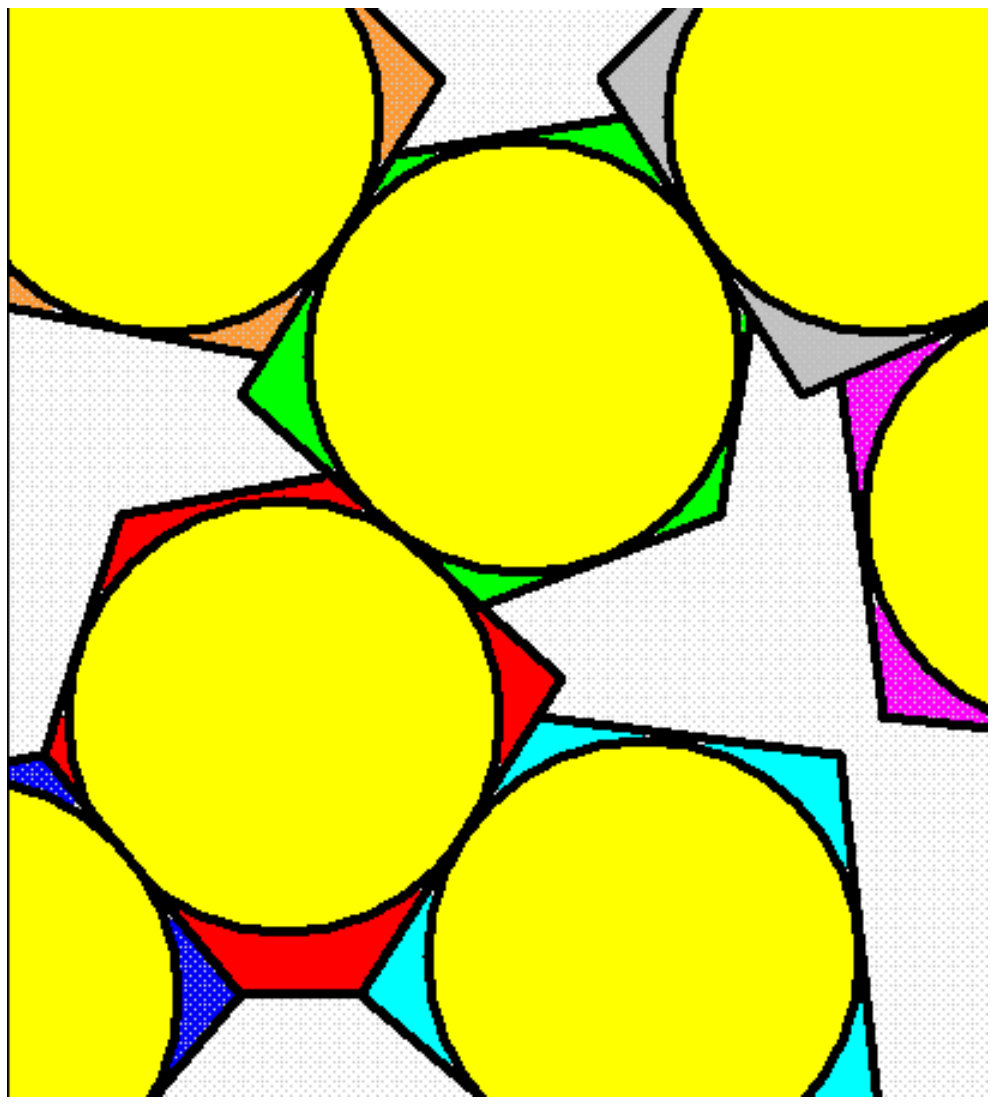

Figure 3: (Color online) A jammed packing of hard disks (colored yellow) is converted into a jammed packing of nonspherical particles by converting the disks to polygons (colored in different colors), without changing the contact network or contact forces. This preserves the jamming property since the floppy modes composed of pure particle rotations are blocked by the flat contacts. Jamming would also be preserved if the disks swell between the original shape and the polygonal shape, so that the curvature of the particle surfaces at the point of contact is sufficiently flat.

It is important to note that contact curvature cannot block purely translational particle displacements unless one of the particles is curved outward, i.e., is concave (e.g., imagine a dent in a table and a sphere resting in it, not being able to slide translationally). If the particles shapes are convex, a packing cannot have fewer contacts than there are translational degrees of freedom, that is, $\bar{Z} \geq 2d$. This explains why hypersphere packings are indeed isoconstrained. It is only when considering rotational degrees of freedom that jammed packings can be hypoconstrained.

Those that prefer to think about rigidity (forces) would consider applying external forces and torques on the particles in the example from Fig. 3. The forces would clearly be resisted just like they were in the jammed disk packing. However, at first sight, it appears that torques would not be resisted. In fact, it would seem that torques *cannot* be resisted by interparticle forces since, for each of the particles, the normal vectors at the points of contact all intersect at a single point (the center of the hard disks) and therefore the net torque is identically zero. This argument, however, neglects an important physical consideration: the deformability of the particles. Namely, no matter how stiff the particles are, they will deform slightly under an applied load. In particular, upon application of torques, the particles will rotate and the normal vectors at the points of contact would change and no longer intersect at a single point, and the packing will be able to resist the applied torques. One may be concerned about the amount of rotation necessary to resist the applied load. If the packing needs to deform significantly to resist applied loads, should it really be called rigid?

To answer such concerns, one must calculate the particle displacements needed to resist the load. Such a calculation, carried out for deformable particles in Section VIII, points to the importance of the pre-existing (i.e., internal) contact forces. This is easy to understand physically. If the packing is under a high state of compression, the interparticle forces would be large and even a small change in the packing geometry (deformation) would resist large torques. If, on the other hand, the internal forces (stresses) are small, the particles would have to deform sufficiently to both induce sufficiently large contact forces *and* to change the normal vectors sufficiently. This kind of stability, requiring sufficiently large internal stresses, is well-known for engineering structures called *tensegrities*. These structures are built from elastic cables and struts, and are stabilized by stretching the cables so as to induce internal stresses. Beautiful and intriguing structures can be built that are rigid even though they appear not to be sufficiently braced (as bridges or other structures would have to be).

While the above discussion focused on packings of macroscopic elastic particles, similar arguments apply also to systems such as glasses. For such systems, floppy modes are manifested as zero-frequency vibrational modes, that is, zero eigenvectors of the dynamical matrix. The calculations in Section VIII show that for nonspherical particles, the dynamical matrix contains a term proportional to the internal forces and involving the contact curvatures. If the system is at a positive pressure, the forces will be nonzero and this term contributes to the overall dynamical matrix. In fact, it is this term that makes the dynamical matrix positive definite, i.e., that eliminates zero-frequency modes despite the existence of floppy modes.

### *2. Translational Versus Rotational Degrees of Freedom*

Having explained that hypoconstrained packings of nonspherical particles can be jammed if the interparticle contacts are sufficiently flat, we now try to understand why packings of nearly spherical particles are hypoconstrained. The analysis will also demonstrate why packings of hard ellipsoids are necessarily denser than the corresponding sphere packings.

The first point to note is that disordered isoconstrained packings of nearly spherical ellipsoids are hard to construct. In particular, achieving isocounting near the sphere point requires such high contact numbers [specifically, $\bar{Z} = d(d+1)$] that translational ordering will be necessary. Translationally maximally random jammed (MRJ) sphere packings have $\bar{Z} = 2d$, and even if one considers the observed multitude of near contacts [2], they fall rather short of $\bar{Z} = d(d+1)$. It seems intuitive that translational crystallization would be necessary in order to raise the contact number that much. In other words, in order to gain sufficiently many constraints, one would have to sacrifice translational disorder. Furthermore, there is little reason to expect packings of nearly spherical particles to be rotationally jammed. Near the jamming point, it is expected that particles can rotate significantly even though they will be translationally trapped and rattle inside small cages, until of course the actual jamming point is reached, at which point rotational jamming will also come into play. Therefore, it is not surprising that near the sphere point, the translational structure of the packings changes little.

Mathematically, jamming is analyzed by using a Taylor expansion of the interparticle distances in the particle displacements. At the first-order level, this expansion contains first-order terms coming from translations and from rotations and involving the contact points and contact normals. The expansion also contains second-order terms from translations, rotations and combined motions, involving additionally the contact curvatures. And of course, there are even more complicated higher-order terms. One should be careful of such a Taylor expansion for two reasons. First, the expansion assumes that terms coming from translations and rotations are of the *same* order. This is clearly not true for neither the case of perfectly spherical particles, when rotational terms are identically zero, nor for the case of rods or plates, where even a small rotation can cause very large overlap. Second, the expansion assumes that various quantities related to the particle and contact geometry (for example, the contact curvature radii) are of similar order. This fails, for example, for the case of planar (flat) contacts, where even a small rotation of the particles leads to significant overlap far from the point of contact. These subtle points arise only when considering aspherical particles and should caution one from blindly generalizing the mathematical formalism of jamming developed and tested only within the context of sphere packings.

In Section IX, we will consider packings of nearly spherical ellipsoids as a perturbation of jammed sphere packings in which the particles, following a slight change of the particle shape away from perfect spherical symmetry, translate and rotate in order to re-establish contacts and jamming. While the necessary particles' translations are small, the particle rotations are large. In fact, rotational symmetry is broken, and particles must orient themselves correctly, so that contacts can be re-established, and also so that forces *and* torques become balanced. This symmetry breaking is the cause of the *cusp-like non-analyticity* of the density as a function of particle shape [4]. We will see that the particle orientations in the final jammed packing of nearly spherical ellipsoids are not random, but rather, they are determined by the structure of the initial sphere packing. Of course, as aspect ratio increases, rotations become more and more on equal footing with translations, and the packings become both truly translationally and orientationally disordered.

This picture of jamming in the vicinity of the sphere point also explains why the density rises sharply near the sphere point for ellipsoids. Start with a jammed sphere packing and apply an affine transformation to obtain an aligned (nematic) packing with exactly the same density. This packing will not be rotationally jammed, and by displacing the particles one will be able to open up free volume between them and therefore increase the density. We will show that in fact the maximal increase in the density is obtained for the choice of particle orientations that balances the torques on the particles in addition to the forces. Therefore, the jammed disordered ellipsoid packings we obtain near the sphere point are the densest perturbation of the corresponding sphere packings. The added rotational degrees of freedom allow one to increase the density beyond that of the aligned (nematic) packing, which for ellipsoids has exactly the same density as the sphere point.

In conclusion, near the sphere point, there is a competition between translational and rotational jamming and also between translational and rotational disorder. At the sphere point $\alpha = 1$, and in this neighborhood, translational degrees of freedom win. As one moves away from the sphere point, however, translational and rotational degrees of freedom start to play an equal role. For very large aspect ratios, $\alpha \gg 1$, it is expected that rotational degrees of freedom will dominate, although we do not investigate that region here.

### C. Contents

Before proceeding, we give an overview of our notation in Section II. In Section III, we discuss the non-overlap conditions between convex hard particles. In Section IV, we define jamming and investigate the reasons for the failure of the isocounting conjecture for nonspherical particles. In Section V, we develop the first- and second-order conditions for jamming in a system of nonspherical particles, and then design and use a practical algorithm to test these conditions for ellipsoid packings in Section VI. In Section VII, we consider the thermodynamical behavior of hypoconstrained packings that are close to, but not quite at, the jamming point. In Section VIII, we discuss the connections between jammed packings of hard particles and strict energy minima for systems of deformable particles. In Section IX, we focus on packings of nearly spherical ellipsoids, and finally, offer conclusions in Section X.

It is important to note that Sections III, V, and VI are highly technical, and may be either skipped or skimmed by readers not interested in the mathematical formalism of jamming. Readers interested in specific examples of hypoconstrained packings are referred to Section IV B 2 and Appendix A.

## II. NOTATION

We have tried to develop a clear and consistent notation, however, in order to avoid excessive indexing and notation complexity we will often rely on the context for clarity. The notation is similar to that used in Ref. [15] and attempts to unify two and three dimensions whenever possible. We refer to reader to Ref. [15] or Ref. [19] for details on representing particle orientations and rotations in both two and three dimensions.

We will use matrix notation extensively, and denote vectors and matrices with bolded letters, and capitalize matrices in most cases. Infinite-dimensional or discrete quantities such as sets or graphs will typically be denoted with script letters. We will often capitalize the letter denoting a vector to denote a matrix obtained from that vector. Matrix multiplication is assumed whenever products of matrices or a matrix and a vector appear. We prefer to use matrix notation whenever possible and do not carefully try to distinguish between scalars and matrices of one element. We denote the dot product $\mathbf{a} \cdot \mathbf{b}$ with $\mathbf{a}^T\mathbf{b}$, and the outer product $\mathbf{a} \otimes \mathbf{b}$ with $\mathbf{a}\mathbf{b}^T$. We denote a vector with all entries unity by $\mathbf{e} = 1$, so that $\sum_i a_i = \mathbf{e}^T\mathbf{a}$. We consider matrices here in a more general linear operator sense, and they can be of order higher than two (i.e., they do not necessarily have to be a rectangular two-dimensional array). We refer to differentials as gradients even if they are not necessarily differentials of scalar functions. Gradients of scalars are considered to be column vectors and gradients of vectors or matrices are matrices or matrices (linear operators) of higher rank.

### A. Particle Packings

A jammed particle packing has a *contact network* indicating the touching pairs of particles $\{i, j\}$. We will sometimes talk about a particular particle $i$ or a particu-lar contact $\{i,j\} \equiv ij$ and we will usually let the context determine what specific particle or contact is being referred to, or, if deemed necessary, put subscripts such as $i$ or $ij$ to make it specific what particle or contact is being referred to. The contact $ji$ is physically the same *undirected* contact as $ij$, but the two *directed contacts* are considered distinct.

There are two primary kinds of vectors $\mathbf{x}$, *particle vectors* $\mathbf{X} = (\mathbf{x}_i) = (\mathbf{x}_1, \ldots, \mathbf{x}_N)$, which are obtained by concatenating together the vectors $\mathbf{x}_i$ (typically of size of the order of the space dimensionality $d$) corresponding to each of the $N$ particles, and *contact vectors* $\mathbf{y} = (y_{ij}) = (y_1, \ldots, y_M)$, obtained by concatenating together the (typically scalar) values $y_{ij}$ corresponding to each of the $M$ contacts (numbered in arbitrary order from 1 to $M$). Note the capitalization of particle vectors, which we will often do implicitly, to indicate that one can view $\mathbf{X}$ as a matrix where each row corresponds to a given particle. If a contact vector agglomerates a vector quantity attached to each contact, for example, the common normal vector $\mathbf{n}$ at the point of contact of two particles, it too would be capitalized, e.g., $\mathbf{N} = (\mathbf{n}_{ij})$.

### 1. Packing Configuration

A packing is a collection of $N$ hard particles in $\mathcal{R}^d$ such that no two particles overlap. Each particle $i$ has $d_f$ configurational *degrees of freedom*, for a total of $N_f = N d_f$ degrees of freedom. A packing $\mathcal{Q} = (\mathbf{Q}, \phi)$ is characterized by the *configuration* $\mathbf{Q} = (\mathbf{q}_1, \ldots, \mathbf{q}_N) \in \mathcal{R}^{N_f}$, determining the positions of the centroid and the orientations of each particle, and the *packing fraction* (*density*), $\phi$, determining the size of the particles. For spheres $\mathbf{Q} \equiv \mathbf{R}$ corresponds to only the positions of the centroids, and $d_f = d$. For nonspherical particles without any axes of symmetry there are an additional $d(d-1)/2$ rotational degrees of freedom, for a total of $d_f = d(d+1)/2$ degrees of freedom. In actual numerical codes particle orientation is represented using unit quaternions, which are redundant representations in the sense that they use $d(d-1)/2+1$ coordinates to describe orientation. Here we will be focusing on displacements of the particles $\Delta\mathbf{Q}$ from a reference jammed configuration $\mathbf{Q}_J$, and therefore we will represent particle orientations as a rotational displacement from a reference orientation $\Delta\boldsymbol{\varphi}$. In two dimensions $\Delta\boldsymbol{\varphi} = \Delta\varphi$ simply denotes the angle of rotation in the plane, and in three dimensions the direction of $\Delta\boldsymbol{\varphi}$ gives the axis of rotation and its magnitude determines the angle of rotation. For simplicity, we will sometimes be sloppy and not specifically separate centroid positions from orientations, and refer to $\mathbf{q}_i$ as (a generalized) *position*; similarly, we will sometimes refer to both forces and torques as (generalized) *forces.*

### 2. Rigidity Matrix

For the benefit of readers not interested in the mathematical formalism, we briefly introduce the concepts and notation developed in more detail in Section III.

We denote the distance, or gap, between a pair of hard particles $i$ and $j$ with $\zeta_{ij}$. When considering all of the $M$ contacts together, the gradient of the distance function $\boldsymbol{\zeta} = (\zeta_{ij})$ with respect to the positions (i.e., displacements) of the particles is the rigidity matrix $\mathbf{A} = \boldsymbol{\nabla}_{\mathbf{Q}}\boldsymbol{\zeta}$. This linear operator connects, to first order, the change in the interparticle gaps to the particle displacements, $\Delta\boldsymbol{\zeta} = \mathbf{A}^T\Delta\mathbf{Q}$. We denote the magnitudes of the compressive (positive) interparticle forces carried by the particle contacts with $\mathbf{f} = (f_{ij})$, $f_{ij} \geq 0$, where it is assumed that the force vectors are directed along the normal vectors at the point of contact (since the particles are frictionless). The total forces and torques exerted on the particles $\mathbf{B}$ (alternatively denoted by $\Delta\mathbf{B}$ if thought of as force imbalance) are connected to the interparticle forces via a linear operator that can be shown to be the conjugate (transpose) of the rigidity matrix, $\mathbf{B} = \mathbf{A}\mathbf{f}$.

A subtle point that we will return to later is the role of the density $\phi$. Since we are interested in (locally) maximally dense disordered packings, we will sometimes consider the density as an additional kinematic degree of freedom. That is, we will sometimes include the change in density $\Delta\phi$ in the deformation $\Delta\mathbf{Q}$. This effectively adds an additional row to the rigidity matrix. One may similarly include additional global degrees of freedom, such as boundary conditions, and add further rows to the rigidity matrix. This also adds generalized forces (stresses) as the conjugate variables to those additional kinematic degrees of freedom [19].

## B. Cross Products

In three dimensions, the cross product of two vectors is a linear combination of them that can be thought of as matrix-vector multiplication

$$\mathbf{a} \times \mathbf{b} = \mathbf{A}\mathbf{b} = -\mathbf{b} \times \mathbf{a} = -\mathbf{B}\mathbf{a} \tag{1}$$

where

$$\mathbf{A} = \lfloor \mathbf{a} \rfloor_{\times} = \begin{bmatrix} 0 & -a_z & a_y \\ a_z & 0 & -a_x \\ -a_y & a_x & 0 \end{bmatrix} = -\mathbf{A}^T$$

is a skew-symmetric matrix which is characteristic of the cross product and is derived from a vector. We will simply *capitalize* the letter of a vector to denote the corresponding *cross product matrix* (like $\mathbf{A}$ above corresponding to $\mathbf{a}$), or use $\lfloor \mathbf{a} \rfloor_{\times}$ when capitalization is not possible. In two dimensions, there are two "cross products". The first one gives the velocity of a point $\mathbf{r}$ in a system which rotates around the origin with an angular frequency $\boldsymbol{\omega}$

(which can also be considered a scalar $\omega$),

$$\mathbf{v} = \boldsymbol{\omega} \boxtimes \mathbf{r} = \begin{bmatrix} -\omega \mathbf{r}_y \\ \omega \mathbf{r}_x \end{bmatrix} = \boldsymbol{\Omega}\mathbf{r}, \tag{2}$$

where

$$\boldsymbol{\Omega} = \begin{bmatrix} 0 & -\omega \\ \omega & 0 \end{bmatrix} = -\boldsymbol{\Omega}^T$$

is a cross product matrix derived from $\boldsymbol{\omega}$. The second kind of "cross product" gives the torque around the origin of a force $\mathbf{f}$ acting at a point (arm) $\mathbf{r}$,

$$\boldsymbol{\tau} = \mathbf{f} \times \mathbf{r} = -\mathbf{r} \times \mathbf{f} = [f_x r_y - f_y r_x] = \mathbf{F}^L \mathbf{r}, \tag{3}$$

where

$$\mathbf{F}^L = \begin{bmatrix} -f_y & f_x \end{bmatrix} = -\left(\mathbf{F}^R\right)^T$$

is another cross product matrix derived from a vector (the $L$ and $R$ stand for left and right multiplication, respectively). Note that in three dimensions all of these coincide, $\mathbf{F}^L = \mathbf{F}^R = \mathbf{F}$, and also $\boxtimes \equiv \times$, while in two dimensions they are related via $\mathbf{a} \boxtimes \mathbf{b} = \mathbf{A}\mathbf{b} = -\mathbf{B}^R a$.

## III. NONOVERLAP CONSTRAINTS AND INTERPARTICLE FORCES

In this section we will discuss hard-particle overlap potentials used to measure the distance between a pair of hard particles. These potentials will be used to develop analytic expansions of the non-overlap conditions in the displacements of the particles. This section is technical and may be skipped or skimmed by readers not interested in the mathematical formalism of jamming. Interested readers can find additional technical details on the material summarized in this Section in Chapter 2 of Ref. [19].

### A. Overlap Potentials

The nonoverlap condition between a pair of particles $A$ and $B$ can be thought of as an inequality between the positions and orientations of the particles. For this purpose, we measure the distance between the two ellipsoids using the *overlap potential* $\zeta(A, B) = \zeta(\mathbf{q}_A, \mathbf{q}_B)$, whose sign not only gives us an overlap criterion,

$$\begin{cases} \zeta(A,B) > 0 \text{ if } A \text{ and } B \text{ are disjoint} \\ \zeta(A,B) = 0 \text{ if } A \text{ and } B \text{ are externally tangent} \\ \zeta(A,B) < 0 \text{ if } A \text{ and } B \text{ are overlapping,} \end{cases}$$

but which is also at least twice continuously differentiable in the positions and orientations of $A$ and $B$. An additional requirement is that $\zeta(A, B)$ be defined and easy to compute for *all* positions and orientations of the particles.

We define and compute the overlap conditions using a procedure originally developed for ellipsoids by Perram and Wertheim [20]. This procedure is easily generalized to any convex particle shape given by the inequality $\zeta(\mathbf{r}) \leq 1$, where the *shape function* $\zeta$ is strictly convex and defined through

$$\zeta(\mathbf{r}) = [\mu(\mathbf{r})]^2 - 1,$$

where $\mu$ is the *scaling factor* by which the particle needs to be resized in order for the point $\mathbf{r}$ to lie on its surface. The unnormalized normal vector to the surface at a given point $\mathbf{r}$, if the particle is rescaled so that it passes through it, is $\mathbf{n}(\mathbf{r}) = \boldsymbol{\nabla}\zeta(\mathbf{r})$. Define also the displacement between the particle centroids $\mathbf{r}_{AB} = \mathbf{r}_A - \mathbf{r}_B$, and the unit vector joining the two particle centroids with $\mathbf{u}_{AB} = \mathbf{r}_{AB} / \|\mathbf{r}_{AB}\|$.

The Perram and Wertheim (PW) overlap potential is defined through

$$\zeta = \mu^2 - 1 = \max_{0 \leq \lambda \leq 1} \min_{\mathbf{r}_C} \left[\lambda \zeta_A(\mathbf{r}_C) + (1-\lambda)\zeta_B(\mathbf{r}_C)\right].$$

For every multiplier $\lambda$, the solution of the inner optimization over $\mathbf{r}_C$ is unique due to the strict convexity of $\mathbf{r}_C$, and satisfies the gradient condition

$$\lambda \mathbf{n}_A(\mathbf{r}_C) = -(1-\lambda)\mathbf{n}_B(\mathbf{r}_C),$$

which shows that the normal vectors are parallel (with opposite directions). The solution of the outer optimization problem over $\lambda$ is given through the condition

$$\zeta = \zeta_A(\mathbf{r}_C) = \zeta_B(\mathbf{r}_C),$$

which means that when the particles are rescaled by a common scaling factor $\mu = 1 + \Delta\mu = \sqrt{1+\zeta}$ they are in external tangency, sharing a common normal direction $\mathbf{n} = \mathbf{n}_A / \|\mathbf{n}_A\|$ (i.e., normalized to unit length and directed from $A$ to $B$), and sharing a *contact point* $\mathbf{r}_C$. When focusing on one particle we can measure $\mathbf{r}_C$ with respect to the centroid of the particle, or otherwise specifically denote $\mathbf{r}_{AC} = \mathbf{r}_C - \mathbf{r}_A$ and $\mathbf{r}_{BC} = \mathbf{r}_C - \mathbf{r}_B$. This is illustrated for ellipses in Fig. 4. If the particles are touching then $\mu = 1$ and the procedure described above gives us the geometric contact point and therefore the common normal vector. In the case of spheres of radius $O$ the PW overlap potential simply becomes

$$\zeta_{AB} = \frac{(\mathbf{r}_A - \mathbf{r}_B)^T (\mathbf{r}_A - \mathbf{r}_B)}{(O_A + O_B)^2} - 1 = \frac{l_{AB}^2}{(O_A + O_B)^2} - 1, \tag{4}$$

which avoids the use of square roots in calculating the distance between the centers of $A$ and $B$, $l_{AB}$, and is easily manipulated analytically.

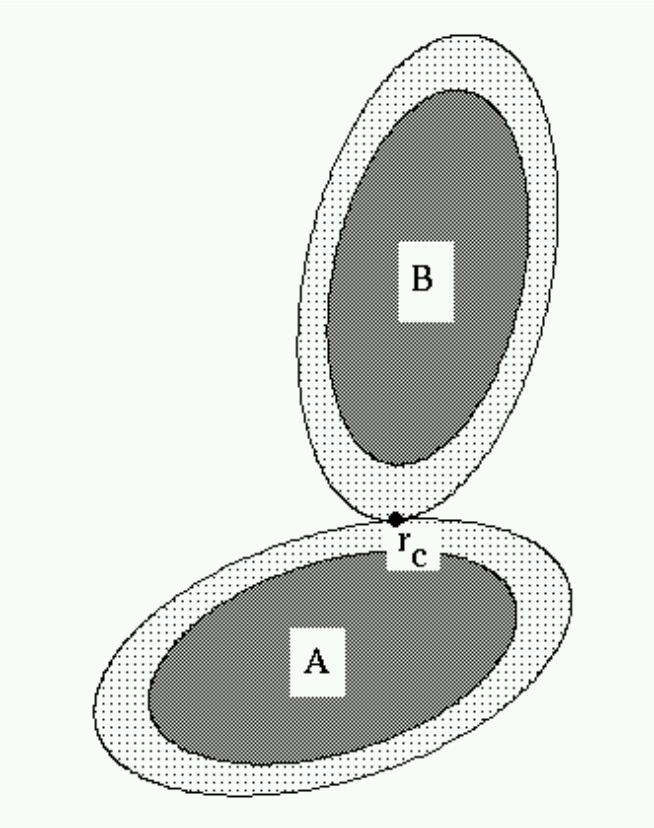

Figure 4: Illustration of the common scaling $\mu$ that brings two ellipses (dark gray) into external tangency at the contact point $\mathbf{r}_C$.

### 1. Derivatives of the Overlap Potentials

We will frequently need to consider derivatives of the overlap function with respect to the (generalized) positions of the particles, either first order,

$$\boldsymbol{\nabla}_{\mathbf{q}_i}\zeta = \boldsymbol{\nabla}_i\zeta = \left(\frac{\partial\zeta}{\partial q_i}\right),$$

or second order

$$\boldsymbol{\nabla}^2_{\mathbf{q}_i\mathbf{q}_j}\zeta = \boldsymbol{\nabla}^2_{ij}\zeta = \left[\frac{\partial^2\zeta}{\partial q_i\partial q_j}\right].$$

To first order, the particles can be replaced by their (parallel) tangent planes at the point of contact and the first order derivatives can be expressed in terms of quantities relating to the two tangent planes. To second order, the particles can be replaced by paraboloids that have the same tangent plane, as well as the same principal curvature axes and the same radii of curvatures as the two particles at the point of contact. It is therefore possible to derive general expressions for the derivatives in terms of quantities relating to the normal vectors and surface curvatures of the particles at the point of contact.

The first order derivatives can easily be expressed in terms of the position of the contact point $\mathbf{r}_C$ and the (normalized and outwardly-directed) contact normal vector $\mathbf{n}$. For this purpose, it is easier to measure the distance between two particles in near contact via the Euclidian *interparticle gap h* giving the (minimal) surface-to-surface distance between the particles along the normal vector. Moving one of the particles by $\Delta\mathbf{q} = (\Delta\mathbf{r}, \Delta\boldsymbol{\varphi})$ displaces the contact point by $\Delta\mathbf{r}_C = \Delta\mathbf{r} + \Delta\boldsymbol{\varphi} \boxtimes \mathbf{r}_C$ and therefore changes the gap by $\Delta h = -\mathbf{n}^T\Delta\mathbf{r}_C = -\mathbf{n}^T\Delta\mathbf{r} - (\mathbf{r}_C \times \mathbf{n})^T\Delta\boldsymbol{\varphi}$, giving the gradient

$$\boldsymbol{\nabla}_{\mathbf{q}}h = -\begin{bmatrix}\mathbf{n}\\ \mathbf{r}_C \times \mathbf{n}\end{bmatrix}.$$

The relation between the (small) Euclidian gap $h$ and the (small) gap as measured by the PW overlap potential $\zeta$ can be seen by observing that scaling an ellipsoid by a factor $\mu$ displaces the contact point by $\Delta\mathbf{r}_C = \Delta\mu\mathbf{r}_C$. Therefore, the scaling factor needed to close the interparticle gap is

$$\mu \approx \frac{\zeta}{2} \approx \frac{h}{(\mathbf{r}_{BC} - \mathbf{r}_{AC})^T\mathbf{n}} = \frac{h}{\mathbf{r}_{AB}^T\mathbf{n}},$$

giving the gradient of the overlap potential $\boldsymbol{\nabla}_{\mathbf{q}}\zeta = 2\left(\boldsymbol{\nabla}_{\mathbf{q}}h\right)/\left(\mathbf{r}_{AB}^T\mathbf{n}\right)$,

$$\boldsymbol{\nabla}_{A/B}\zeta = \mp\frac{2}{\mathbf{r}_{AB}^T\mathbf{n}}\begin{bmatrix}\mathbf{n}\\ \mathbf{r}_{(A/B)C} \times \mathbf{n}\end{bmatrix}.$$

For spheres the cross product is identically zero and rotations can be eliminated from consideration.

The second-order derivatives are not as easily evaluated for a general particle shape. In two dimensions, or in three dimensions when the principal radii of curvatures at the point of contact are equal, one can replace the particle around the point of contact with a sphere of the appropriate position and radius. However, when the radii of curvatures are different this is not as easy to do. We will give explicit expressions for the second-order derivatives of $\zeta$ for ellipsoids in Section VI A 2. Related first- and second-order geometric derivatives have been derived for general particle shapes (i.e., using the normal vectors and curvature tensors of the particles at the point of contact) in the granular materials literature in more general contexts [21, 22]; here we specialize to the case of hard frictionless ellipsoids.

## B. The Rigidity Matrix

When considering all of the $M$ contacts together, the gradient of the overlap potential $\boldsymbol{\zeta} = (\zeta_{ij})$ is the important *rigidity matrix*

$$\mathbf{A} = \boldsymbol{\nabla}_{\mathbf{Q}}\boldsymbol{\zeta}.$$

This $[N_f \times M]$ matrix connects, to first order, the change in the interparticle gaps to the particle displacements, $\Delta\boldsymbol{\zeta} = \mathbf{A}^T\Delta\mathbf{Q}$. It may sometimes be more convenient to work with surface-to-surface interparticle gaps, $\Delta\mathbf{h} = \mathbf{A}_E^T\Delta\mathbf{Q}$ (the subscript $E$ stands for Euclidian), especially if second-order terms are not considered [11]. The rigidity matrix is sparse and has two blocks of $d_f$ non-zero entries in the column corresponding to the particle contact $\{i, j\}$, namely, $\boldsymbol{\nabla}_i\zeta_{ij}$ in the block row corresponding to particle $i$ and $\boldsymbol{\nabla}_j\zeta_{ij}$ in the block row corresponding to particle $j$ (unless one of these particles is frozen). Represented schematically:

$$\mathbf{A} = \begin{array}{c} \\ i \to \\ \\ j \to \\ \\ \end{array} \overset{\{i,j\}}{\overset{\downarrow}{\begin{bmatrix} \vdots \\ \nabla_i \zeta_{ij} \\ \vdots \\ \nabla_j \zeta_{ij} \\ \vdots \end{bmatrix}}}.$$

### C. Interparticle Forces

Hard particles in contact can exert a compressive (positive) *contact force* $\mathbf{f} = f\mathbf{n}$, $f \geq 0$, directed along the normal vector (for frictionless particles). The total excess force and torque exerted on a given particle $i$ by the contacts with its neighbors $\mathcal{N}(i)$ is

$$\Delta\mathbf{b}_i = -\sum_{j\in\mathcal{N}(i)} f_{ij} \begin{bmatrix} \mathbf{n}_{ij} \\ \left(\mathbf{r}_{iC}^{ij} \times \mathbf{n}_{ij}\right) \end{bmatrix} = \sum f_{ij} \left(\nabla_i h_{ij}\right),$$

or, considering all particles together

$$\Delta\mathbf{B} = \mathbf{A}_E \mathbf{f}.$$

The fact that the matrix (linear operator) connecting force imbalances to contact forces is the transpose of the rigidity matrix is well-known and can also be derived by considering the work done by the contact forces to displace the particles

$$\begin{aligned} W &= \Delta\mathbf{b}^T \Delta\mathbf{Q} = \left(\tilde{\mathbf{A}}\mathbf{f}\right)^T \Delta\mathbf{Q} = \mathbf{f}^T \left(\tilde{\mathbf{A}}^T \Delta\mathbf{Q}\right) = \\ &= \mathbf{f}^T \Delta\mathbf{h} = \mathbf{f}^T \left(\mathbf{A}_E^T \Delta\mathbf{Q}\right), \end{aligned}$$

showing that $\tilde{\mathbf{A}} = \mathbf{A}_E^T$. In this work we will use forces $\mathbf{f}$ that are a rescaled version of the physical forces $\mathbf{f}_E$, $f_{ij} = \left(\mathbf{r}_{ij}^T \mathbf{n}_{ij}\right) f_{ij}^E/2$, so that $\mathbf{A}\mathbf{f} = \mathbf{A}_E \mathbf{f}_E$. This scaling is more natural for our choice of overlap potential, and does not affect any of the results.

In static packings that are under an applied load $\mathbf{B}$, the force/torque equilibrium condition

$$\mathbf{A}\mathbf{f} = -\mathbf{B} \text{ and } \mathbf{f} \geq \mathbf{0}$$

must be satisfied. The actual magnitudes of the forces are determined by external loads (for example the applied pressure for a system of deformable particles), history of the packing preparation, etc. However, the relation between the forces at different contacts is determined by the packing geometry, or more specifically, by $\mathbf{A}$. Typically forces are rescaled to a mean value of unity, $\mathbf{e}^T\mathbf{f} = M$, and it has been observed that the distribution of rescaled contact forces has some universal features, for example, there is an exponential tail of contacts carrying a large force, and also a large number of contacts supporting nearly zero force [2, 23]. We will see later that these *force chains*, or *internal stresses*, are an essential ingredient of jamming for hard particles.

## IV. THE ISOCOUNTING CONJECTURE

In the granular materials literature special attention is often paid to so-called isostatic packings. There are several different definitions of isostaticity, and most of the discussions in the literature are specifically applied to mechanical structures composed of elastic bars, to packings of hard spheres, or to packings of frictional particles. In this section we summarize several relevant definitions of and arguments for isostaticity and generalize them to nonspherical particles.

We define a packing to be *isoconstrained* if the number of constraints (contacts) is equal to the total number of degrees of freedom

$$N_c = N_f + 1,$$

where for jammed packings one should count the density $\phi$ as a single degree of freedom, in addition to the degrees of freedom due to the particles and boundary $N_f$, as discussed further in Sections IV A 1 and IV A 2. Packings with fewer contacts than isoconstrained are called *hypoconstrained*, and packings with more contacts than isoconstrained are *hyperconstrained*. The *isocounting conjecture states that large jammed disordered packings of hard particles are isoconstrained.* Defining what precisely is meant by a disordered packing is difficult in itself [16, 24]. Intuitively, in a disordered packing there is only the minimal degree of correlations between particles, as necessitated by the constraints of impenetrability and jamming. Therefore, it is expected that in a certain sense disordered packings are "generic" [48], and that "special" configurations with geometric degeneracies will not appear. Note that for large systems the majority of the degrees of freedom come from the particles themselves, $N_f \approx Nd_f$, and the majority of constraints come from contacts shared between two particles, $N_c \approx M = N\bar{Z}/2$, giving the isocounting property

$$\bar{Z} = 2d_f. \tag{5}$$

Equation (5) has been verified to very high accuracy for jammed hard-sphere packings [2]. However, disordered packings of hard ellipsoids are always hypoconstrained and thus contradict the isocounting conjecture [4].

The notion of an isoconstrained packing is very closely related to the concept of an isostatic packing, which considers the (linear) independence of the constraints in addition to their number. An *isostatic packing is defined as a packing that has an invertible (and thus square) rigidity matrix.* This definition has not been formally stated in the literature, and it is the obvious generalization of the definition commonly used for systems of spheres. One can include the density as an additional degree of freedom when forming the rigidity matrix, or exclude it, depending on the definition of jamming that is adopted, as discussed in the next section. This choice changes the counting arguments by one. This definition of the term isostatic implicitly relies on the linearization of the impenetrability constraints. We try to make our definitions independent of the order of *approximation* used in some particular expansion. Therefore, we use the simple definition of "isoconstrained" based on counting, and qualify it with "jammed" or "rigid".

In this Section we attempt to deconstruct previous discussions of isostaticity and jamming in hard-particle packings, and we hope that through our discussions it will become clear why previous "proofs" of the isocounting conjecture do not apply to nonspherical particles, or to put it the other way around, what makes disordered sphere packings isoconstrained.

### A. Jamming, Rigidity and Stability

An essential initial step is defining more precisely what is meant by a stable, rigid, or jammed packing. All of these terms have been used in the literature, and in fact we equate each of them with a particular perspective on jamming:

**Kinematic** A packing is *jammed* if none of the particles can be displaced in a non-trivial way without introducing overlap between some particles.

**Static** A packing is *rigid* if it can resolve any externally applied forces through interparticle ones, without changing the packing configuration.

**Perturbation** A packing is *stable* if the structure of the packing changes smoothly for small perturbations of the packing.

We will consider each of these approaches separately. It will shortly become clear that all of them are closely related, and under certain mild conditions they are actually equivalent. We will use the term jamming as an umbrella term, and later give our preferred definition of jamming, which is based on the kinematic perspective. We note that it is important to precisely specify the boundary conditions applied regardless of the view used in considering jamming; different boundary conditions lead to different jamming categories, specifically *local*, *collective* or *strict* jamming [11, 25]. Here, we will sometimes use local jamming in simple examples but mostly focus on collective jamming; all collective particle motions are blocked by the impenetrability constraints subject to periodic boundary conditions with fixed lattice vectors. In order to eliminate trivial uniform translations of the systems, we can freeze the centroid of one of the particles, to obtain a total of

$$N_f = Nd_f - d$$

internal degrees of freedom. The exact boundary conditions affect the counting of constraints and degrees of freedom, however, the correction is not extensive in $N$ and therefore is negligible for large system when considering per-particle quantities such as $\bar{Z}$.

Readers should observe that the terms "stable", "rigid" and "jammed" are defined differently by different authors. These different definitions are, however, mathematically closely related. For example, Ref. [8] defines a rigid packing as a packing which has no floppy modes, thus relying on linearization of the impenetrability constraints. We prefer to use the term "jammed" for kinematic considerations, and not involve linear approximations so that all definitions apply to systems of nonspherical particles. Reference [8] defines a stable packing as one which is a strict local potential energy minimum (where the potential energy is, for example, gravity). A precise definition of jamming based on stability is developed mathematically in Reference [21]. Since a packing can be at a stable energy minimum without being jammed (see for example, Fig. 13 in Ref. [8]), we use a more stringent definition. We have chosen the more stringent definition because our focus is on locally maximally dense packings, that is, packings where the density cannot be increased by continuously displacing the particles. Such packings are relevant to understanding granular materials that have been vibrated/shaken for long periods of time [26, 27], and also to understanding the inherent structures of glassy materials [3, 28]. They can be produced computationally with our molecular dynamics algorithm and experimentally by shaking the packing container while adding more particles [4, 18]. It is important to point out that we do not wish to promote our definition of jamming as the "correct" one. It is equally "correct" to define a jammed packing as one stable under a particular applied load, and then study the particle rearrangements that result when the direction of the external applied forces change. In fact, realistic granular assemblies are not jammed according to our strict definitions, and particles typically rearrange in response to external perturbations. In the limit of infinite compaction, however, the rearrangements will cease and the assembly will become jammed. We focus here on understanding this terminal idealized jammed state as an important first step in understanding more realistic systems. Additionally, we are interested in the mathematics and physics of maximally dense disordered packings in their own right [4].

An important point to note is that the above definitions of jamming treat all degrees of freedom identically, in particular, translational motion (forces) is treated on the same footing as rotational motion (torques). This is not necessarily the most appropriate definition, as is easily seen by considering the case of spheres, which can rotate in place freely even though they are (translationally) jammed. This distinction between translations and rotations will become important in Section VII when considering packings that are nearly, but not quite jammed. It should also be mentioned that jammed random particle packings produced experimentally or in simulations typically contain a small population of *rattlers*, i.e., particles trapped in a cage of jammed neighbors but free to move within the cage. For present purposes we shall assume that these have been removed before considering the (possibly) jammed remainder. This idea of excluding rattlers can be further extended to rattling clusters of particles, i.e., groups of particles that can be displaced collectively even though the remainder of the packing is jammed. In fact, we will consider any packing which has a jammed subpacking (called *backbone*) to be jammed.

#### 1. Kinematic View

The kinematic perspective considers a packing jammed if it is not possible to continuously displace the particles in a non-trivial way without introducing overlap. We have focused on this perspective in our work, see Refs. [11, 25]. That is, the impenetrability conditions preclude any motion of the particles. Here trivial motions are those that do not change the distances between any two particles, such as global translations when periodic boundary conditions are used. We can assume that such trivial motions have been eliminated via some artificial constraint, such as fixing the centroid of one particle externally when using periodic boundary conditions.

Mathematically, for any continuous motion $\Delta\mathbf{Q}(t)$ there exists a $T > 0$ such that at least one of the impenetrability constraints between a touching pairs of particles

$$\boldsymbol{\zeta}\left[\mathbf{Q}_J + \Delta\mathbf{Q}(t)\right] \geq \mathbf{0} \tag{6}$$

is violated for all $0 < t < T$. A motion $\Delta\mathbf{Q}(t)$ such that for all $0 < t < T$ none of the constraints are violated is an *unjamming motion*. One can in fact restrict attention to analytic paths $\Delta\mathbf{Q}(t)$, and also show that a jammed packing is in a sense isolated in configuration space, since the only way to get to a different packing is via a discontinuous displacement $\|\Delta\mathbf{Q}\| > 0$ [12].

A similar definition of jamming was used by Alexander in Ref. [10]. He considers a packing to be *geometrically rigid* if it cannot be "deformed continuously by rotating and translating the constituent grains without deforming any of them and without *breaking* the contacts between any two grains". This definition implies that a packing in which particles can be moved so as to break contacts (for example, imagine a pebble resting on other pebbles in gravity, and moving it upward away from the floor) is jammed. Later in the manuscript Alexander talks about adding constraints to block motions that break contacts. We in fact have in a certain sense a choice in the matter, determining whether we work with inequality or equality constraints. We choose to work with inequality constraints, since this is the natural choice for frictionless hard particles; there is no cohesion between the particles maintaining contacts. In effect, when counting degrees of freedom for packings, we count the density $\phi$ (i.e., the possible collective rescaling of the particle shapes necessary to maintain contacts) as a single degree of freedom, as discussed further in Section IV A 2.

#### 2. Static View

The static perspective considers a packing *rigid* if it can resolve any applied forces through interparticle ones. This is sometimes referred to as *static rigidity*, to be contrasted with *kinematic rigidity* as defined in the previous section. For hard particles, there is no scale for the forces, and so the actual magnitude of the forces does not matter, only the relative magnitudes and the directions. The particles do not deform, but can exert an arbitrary positive contact force.

Mathematically, we consider the existence of a solution to the force-equilibrium equations

$$\mathbf{Af} = -\mathbf{B}, \text{ where } \mathbf{f} \geq 0, \tag{7}$$

for all resolvable external loads $\mathbf{B}$. The space of resolvable loads is determined by the boundary conditions: certain forces such as pulling on the walls of a container cannot be resolved by any packing and need to be excluded. This is similar to the definition used in Ref. [7]: A packing is *mechanically stable* "if there is a nonzero measure set of external forces which can be balanced by interbead ones." The problem with this definition of rigidity and in particular Eq. (7) is that it does not take into account the fact that the geometry of the packing, i.e., the rigidity matrix $\mathbf{A}$, changes when an external load is applied on the packing. Physically, forces arise only through deformation, and this deformation, however small, *together* with the pre-existing forces in the packing, may need to be taken into account. Forces are in essence Lagrange multipliers associated with the impenetrability constraints in Eq. (6); the very existence of such Lagrange multipliers may require a change in the packing configuration.

The above formulation also neglects the existence of small interparticle gaps, which cannot be neglected when analyzing the response of packings to applied loads, especially for granular materials [8, 11]. While mathematically we talk about *ideal jammed packings*, where geometric contacts are perfect, in reality one should really analyze packings that are almost jammed, i.e., where the contacts are almost closed. This is more appropriate for granular materials, where there is typically some room for the particles to move freely. Alternatively, one should analyze packings where all the contacts are indeed closed, however, the system is under some form of global compression. This is appropriate for glassy systems under a uniform external pressure. When interparticle gaps are present, particles must displace slightly to close the gaps so that they can exert positive contact forces on one another and resist the applied load. The set of contacts $\{i, j\}$ that are closed (i.e., have a positive force $f_{ij}$) is the set of *active contacts*. Different applied loads will be supported by different active contact networks, and for sufficiently small interparticle gaps finding the active set requires solving a linear program, as discussed in Section V D 1. When there is a global external compression (pressure) in the system that keeps all contacts closed, one has one more additional force-equilibrium equation in Eq. (7) that has the pressure $p$ on the right hand side. Mathematically, the pressure is the conjugate (dual) variable of the density (viewed as a degree of freedom) [19].

Various counting arguments related to force equilibrium constraints, starting with the seminal work of Maxwell, have appeared in the engineering literature on mechanical structures [29]. There are, however, some important differences between elastic structures and packings of hard particles. Most significantly, the non-negativity of the contact forces is an added condition, and it effectively adds +1 to the number of contacts needed to ensure static rigidity, i.e., adds a single degree of freedom in various counting arguments. For classical structures of elastic bars, an isostatic framework is such that it has exactly as many bars, i.e., unknown internal bar forces, as there are force-equilibrium equations, $M = N_f$. That is, the rigidity matrix is square and the solution to the force-equilibrium equations is $\mathbf{f} = -\mathbf{A}^{-1}\mathbf{B}$. Finding the internal forces therefore does not require knowing anything about the specific elastic properties of the bars: the structure is *statically determinate* [49]. Reference [8] defines "isostatic structures" as "such that all problems are isostatic, whatever the choice of the load. More precisely, one requires all loads orthogonal to the overall rigid body degrees of freedom to be supportable with a unique determination of internal forces."

On the other hand, a jammed isoconstrained packing, as we have defined it, has $M = N_f + 1$ contacts, and the additional one contact is needed in order to ensure that any applied load can be resolved by *non-negative* interparticle forces in the active contact network. Assuming the existence of (infinitesimally) small but positive interparticle gaps, under certain mild non-degeneracy conditions, it can be demonstrated that if one applies a specific load only $N_f$ of the contacts will actually be active, and one contact will be broken and will carry no force. Different contacts will be broken for different loads, however, once it is known which contact is broken (see Section V D 1) the active contact network is isostatic in the classical structural mechanics sense and the forces can be determined, $\mathbf{f} = -\mathbf{A}^{-1}_{\text{active}}\mathbf{B}$, without resorting to constitutive elastic equations for the contacts. The additional contact appears because of our choice of definition of jamming; if one considers stability under a single external load, then all the contacts will be active, $M = N_f$. This difference has some subtle effects that may lead to confusion when comparing to previous results in the literature. For example, as we will see in Section V, ideal jammed packings posseses a non-trivial internal stress, or *self-stress* $\mathbf{f} > \mathbf{0}$, $\mathbf{A}\mathbf{f} = \mathbf{0}$. In elastic structures such an internal stress is associated with overconstrained (sub)structures, and such stresses do not appear in unloaded granular piles. The fact that we observe only a single self-stress for packings means that upon removal of any contact from the packing there will no longer be self-stresses left, i.e., the system will no longer be overconstrained.

Note that a positive internal (self) stress does appear in glasses under a uniform external pressure [23], and in those systems indeed all $N_f + 1$ contacts are active and participate in the resolution of applied loads. The magnitude of the internal stresses is determined by the external pressure. For high pressures, depending on the stiffness of the packing elements, additional active contacts may form as particles deform and one would have to consider the constitutive elastic equations for the contacts in order to determine the interparticle forces.

### 3. Perturbation View

The perturbation perspective considers a packing to be stable if the structure of the packing changes smoothly for small perturbations of the packing. In particular, the structure of the packing includes the positions of the particles *and* the contact force network. Perturbations to be considered should include changes in the grain internal geometry (deformation), strain, and stress (external forces due to shaking, vibration, or a macroscopic load). In great generality we can restrict our perturbations to small perturbations of the distances between contacting particles *combined* with small perturbations of the applied forces. Such a perspective on jamming was recently presented in Ref. [21]. In this work, however, only perturbations of the applied forces were considered. However, it is realized in Ref. [21] that deformations of the boundary conditions can easily be incorporated without changing the stability conditions. In fact, arbitrary external perturbations of the geometry of the contacts can be considered in addition to the applied load perturbations without any significant complication.

Mathematically, we consider the sensitivity of the configuration and force chains to *all* perturbations of the interparticle gaps $\Delta\boldsymbol{\zeta}$ and applied forces $\Delta\mathbf{B}$ away from zero, i.e., we look for solutions of the *coupled* system of equations of preserving contacts and maintaining force equilibrium:

$$\begin{aligned} \left[\mathbf{A}\left(\mathbf{Q}+\Delta\mathbf{Q}\right)\right]\left(\mathbf{f}+\Delta\mathbf{f}\right) &= -\varepsilon\Delta\mathbf{B} \\ \boldsymbol{\zeta}\left(\mathbf{Q}+\Delta\mathbf{Q}\right)-\Delta\zeta_{\mu} &= -\varepsilon\Delta\boldsymbol{\zeta} \\ \mathbf{e}^{T}\Delta\mathbf{f} &= \mathbf{0}, \end{aligned} \tag{8}$$

where $\varepsilon > 0$ is a small number and we have assumed $\mathbf{f} > \mathbf{0}$. Note that in Ref. [21], $\Delta\mathbf{f}$ are called the "basic statical unknowns" and $\Delta\mathbf{Q}$ are called the "basic kinematical unknowns."

Similarly to the external forces, the space of resolvable gap perturbations is determined by the boundary conditions: global expansions will lead to gaps that cannot all be closed unless the particles grow by a certain scaling factor $\mu = 1 + \Delta\mu$. It is therefore convenient to include $\Delta\zeta_\mu \approx 2\Delta\mu$ as an additional variable. An added constraint is that the normalization $\mathbf{e}^T\mathbf{f} = M$ be maintained. It is important to note that we explicitly account for the dependence of the rigidity matrix on the configuration in the force-balance equation. Notice that when we combine perturbations of the geometry and forces together, the total number of variables is $M + N_f$, and the total number of constraints is also $M + N_f$ (here we include the global particle rescaling $\Delta\zeta_\mu$ as a degree of freedom). Therefore there are no underdetermined (linear) systems as found in counting arguments that consider geometry and forces separately, as is typically done in the literature.

## B. Isocounting

In this section we will attempt to deconstruct previous arguments in justification of an isocounting conjecture, mostly in the context of sphere packings, and try to identify the problems when the same arguments are applied to nonspherical particles.

The isocounting conjecture (property) is usually justified in two steps. First, an inequality $\bar{Z} \leq 2d_f$ is demonstrated, then, the converse inequality $\bar{Z} \geq 2d_f$ is invoked to demonstrate the equality $\bar{Z} = 2d_f$. We will demonstrate that it is the second of these steps that fails for non-spherical particles, however, first we recall some typical justifications for the inequality $\bar{Z} \leq 2d_f$. The observation that the inequality $\bar{Z} \geq 2d_f$ does not generally apply to nonspherical particles is already made by Roux in Ref. [8], as we point out below. Roux also discusses the applicability of the converse inequality $\bar{Z} \leq 2d_f$ in significant detail; here we present our own summary for completeness.

### 1. *Why $\bar{Z} \leq 2d_f$ applies*

A packing with $\bar{Z} > 2d_f$ is overconstrained, and in a certain sense geometrically degenerate and thus not "random". It can be argued that such a packing is not stable against small perturbations of the packing geometry, since all contacts cannot be maintained closed without deforming some of the particles. For example, Tkachenko and Witten [7] consider hard-sphere packings with a small polydispersity, so that particles have slightly different sizes, to conclude that "the creation of a contact network with coordination number higher than $2d$ occurs with probability zero in an ensemble of spheres with a continuous distribution of diameters." Moukarzel [6, 30] considers how the actual stiffness modulus of deformable particles affects the interparticle forces and concludes that making the particles very stiff will eventually lead to negative forces and thus breaking of contacts, until the remaining contact network has $\bar{Z} \leq 2d_f$ [50]: "The contact network of a granular packing becomes isostatic when the stiffness is so large that the typical self-stress...would be much larger than the typical load-induced stress...granular packings will only fail to be isostatic if the applied compressive forces are strong enough to close interparticle gaps establishing redundant contacts." A similar argument is made by Sir Edwards in Ref. [9] for frictional grains: "if $z > 4$ then there is a solution with no force on $z - 4$ contacts, and there is no reason why other solutions would have validity."

These arguments apply also to nonspherical particles, however, it is important to point out that they specifically only apply to truly hard-particle packings or to packings of deformable particles in the limit of zero applied pressure ($\mathbf{f} \to \mathbf{0}$). In real physical systems particles will have a finite stiffness and the applied forces will be non-negligible, and such packings will have more contacts than the idealized hard-particle construction.

### 2. *Why $\bar{Z} \geq 2d_f$ does not apply*

The converse inequality, stating that a minimum of $M = N_f + 1$ contacts is necessary for jamming (rigidity), does not apply to nonspherical particles. We can demonstrate this vividly with a simple example of an ellipse jammed between three other stationary (fixed) ellipses, as shown in Fig. 5. This example was also given in Ref. [31], however, a detailed explanation was not provided.

Jamming a disk requires at least three touching disks; the additional rotational degree of freedom of the ellipse would seem to indicate that four touching ellipses would be needed in order to jam an ellipse. However, this is not true: if the normal contact vectors intersect at a single point, three ellipses can trap another ellipse, as shown in Fig. 5. We will shortly develop tools that can be used to demonstrate rigorously that this example is indeed jammed. Another simple example demonstrating that $\bar{Z} \geq 2d_f$ does not apply is the rectangular lattice of ellipses, which is collectively jammed even though $\bar{Z} = 4$, the minimum necessary even for discs. This example is discussed in Appendix A, where we also demonstrate that, in fact, any isostatic packing of spheres can be converted into a jammed packing of nonspherical particles.

The above example shows that the claim of Ref. [10] that "One requires $4(= 3 + 1)$ contacts to fix the DOF [degrees of freedom]...of an ellipse in the plane" is wrong. Similarly, it shows that the argument in Ref. [6], namely, that the minimum number of contacts needed for a packing of $N$ spheres in $d$ dimensions to be rigid is $dN$, cannot be generalized to nonspherical particles by simply replacing $d$ with $d_f$. Claims that the number of constraints must be larger than the number of degrees of freedom have been made numerous times within the kinematic perspective on jamming, for example, in Ref. [9]. Our careful analysis of the conditions for jamming in the next section will elucidate why this is correct for spheres but not necessarily correct for nonspherical particles, and under what conditions a hypoconstrained packing can be jammed.

The example in Fig. 5 is a geometrically-degenerate configuration which would usually be dismissed as a probability-zero configuration. However, we will explain in later sections why such apparently non-generic (degen-erate) configurations must appear for sufficiently small aspect ratios for a variety of realistic packing protocols. In Ref. [29] geometrically-peculiar examples such as this one are presented, however, they are considered to be in unstable equilibrium, i.e., stable only under special types of loading. This type of argument, made within the static perspective on jamming [c.f. Eq. (7)], is given in the context of granular materials in Ref. [7]: "The number of equilibrium equations $Nd$ should not exceed the number of force variables $N_c$; otherwise these forces would be overdetermined." The example in Fig. 5 demonstrates why this argument cannot be applied to nonspherical grains. Since the normal vectors at the points of contact intersect at a point, a torque around that point cannot be resolved by any set of normal forces between the particles. Yet the packing is jammed, and if built in the laboratory it will resist the torque by slight deformations of the particles, so that the normal vectors no longer intersect in one point and the contact forces can resist the applied torque. The connection between the geometry of the contact network, i.e., $\mathbf{A}$, and the packing configuration $\mathbf{Q}$, *as well* as the pre-existing stresses (forces) in the packing, must be taken into account when considering the response of hypoconstrained packings to external perturbations. This important observation was also recently pointed out independently in Ref. [21], and we elaborate on it in the next section.

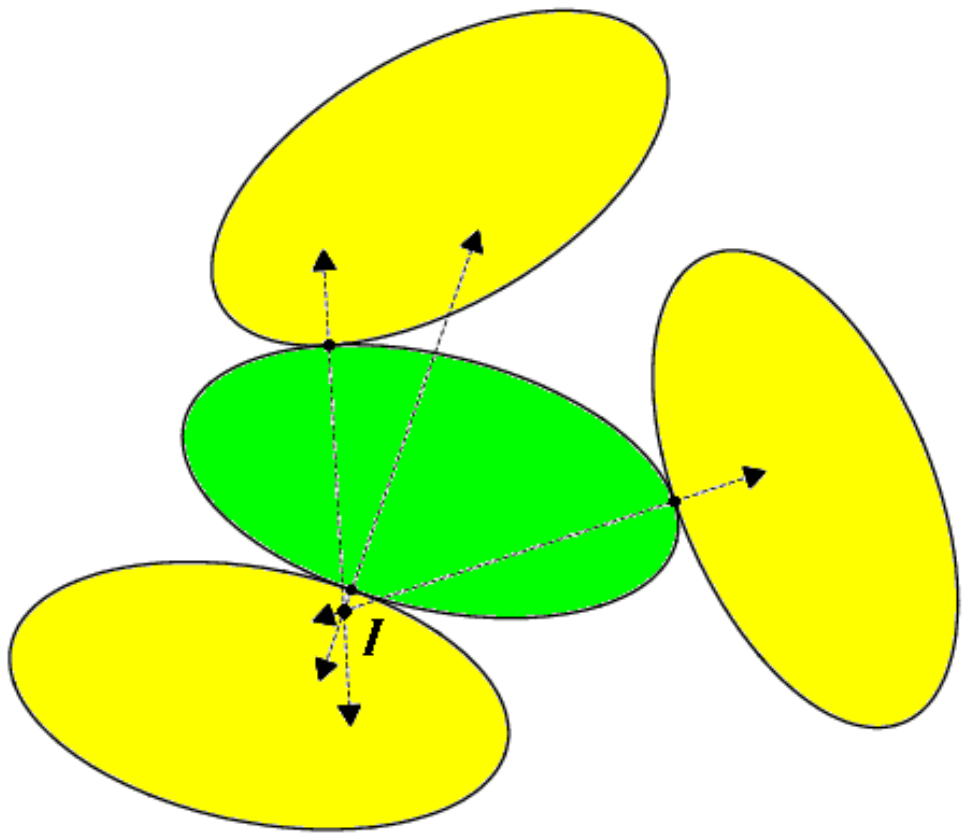

Figure 5: (Color online) A mobile ellipse (green) jammed between three fixed ellipses (yellow). All ellipses are of the same size and have an aspect ratio $\alpha = 2$. This packing was produced by a Lubachevsky-Stillinger type algorithm, where the three particles were kept fixed by giving them infinite mass and no initial velocities. The normal vectors at the points of contact intersect at a common point $I$, as is necessary to achieve torque balance. For the corresponding disk example, shown in Fig. 7, the number of force balance constraints is two, while the number of unknown forces is three. For the ellipse case the number of unknown forces is the same, while the number of force balance constraints is two, and the number of torque constraints is one, giving a total of three equilibrium constraints. However, due to the geometric degeneracy there are only two independent equations of mechanical equilibrium; the torques are always balanced. In the notation described in Section V A 2, for the ellipse example above $N_{\text{stresses}} = N_{\text{floppy}} = 1$, while for the corresponding disk case, $N_{\text{stresses}} = 1$ but $N_{\text{floppy}} = 0$.

## V. CONDITIONS FOR JAMMING

In this Section we develop first and second order conditions for jamming, using a kinematic approach. Statics (forces) will emerge through the use of duality theory. The discussion here is an adaptation of the theory of first-order, pre-stress, and second-order rigidity developed for tensegrities in Ref. [12]. This section is technical and may be skipped or skimmed by readers not interested in the mathematical formalism of jamming. In Section VIII the rigorous hard-particle results are explained more simply by considering the conditions for local (stable) energy minima in soft-particle systems.

We consider an analytic motion of the particles

$$\Delta\mathbf{Q}\,(t) = \dot{\mathbf{Q}}t + \ddot{\mathbf{Q}}\frac{t^2}{2} + O(t^3),$$

where $\dot{\mathbf{Q}}$ are the velocities, and $\ddot{\mathbf{Q}}$ are the accelerations. Expanding the distances between touching particles to second-order, and taking into account that $\boldsymbol{\zeta}\,(\mathbf{Q}_J) = \mathbf{0}$, gives

$$\boldsymbol{\zeta}(t) \approx \mathbf{A}^T\dot{\mathbf{Q}}t + \left[\dot{\mathbf{Q}}^T\mathcal{H}\dot{\mathbf{Q}} + \mathbf{A}^T\ddot{\mathbf{Q}}\right]\frac{t^2}{2} = \dot{\boldsymbol{\zeta}}t + \ddot{\boldsymbol{\zeta}}\frac{t^2}{2}, \quad (9)$$

where the Hessian $\mathcal{H} = \boldsymbol{\nabla}^2_{\mathbf{Q}}\boldsymbol{\zeta} = \boldsymbol{\nabla}_{\mathbf{Q}}\mathbf{A}$ can be thought of as a higher-rank symmetric matrix.

### A. First-Order Terms

Velocities $\dot{\mathbf{Q}} \neq \mathbf{0}$ for which $\dot{\boldsymbol{\zeta}} = \mathbf{A}^T\dot{\mathbf{Q}} \geq \mathbf{0}$ represent a *first-order flex* (using the terminology of Ref. [12]). If we can find an unjamming motion $\dot{\mathbf{Q}}$ such that $\dot{\boldsymbol{\zeta}} > \mathbf{0}$ (note the strict inequality), then the packing is *first-order flexible*, and there exists a $T > 0$ such that none of the impenetrability conditions [c.f. Eq. (6)] are violated for $0 \leq t < T$. We call such a $\dot{\mathbf{Q}}$ a *strict first-order flex.* If on the other hand for at least one constraint $\dot{\zeta} < 0$ for every $\dot{\mathbf{Q}}$, then the packing is jammed, since every non-trivial movement of the particles violates some impenetrability condition for all $0 < t < T$ for some $T > 0$. We call such a packing *first-order jammed.* Finally, a $\dot{\mathbf{Q}}$ such that $\dot{\boldsymbol{\zeta}} = \mathbf{0}$ is a *null first-order flex*, often referred to as *zero* or *floppy mode* in the physics literature.

A packing is first-order jammed if and only if there are no (non-trivial) first order flexes. A packing is first-order flexible if there exists a strict first-order flex. Some packings are neither first-order jammed nor first-order flexible; One must consider higher-order terms to access whether such packings are jammed, and if they are not, to identify an unjamming motion. We will consider the second-order terms later; in this section we develop conditions and algorithms to verify first-order jamming and identify first-order flexes if they exist. The algorithms are closely based on work in Ref. [11].

### 1. Strict Self-Stresses

Let us first focus on a single contact $\{i,j\}$, and ask whether one can find a first order flex that is strict on that contact, i.e.,

$$\dot{\zeta}_{ij} = \left(\mathbf{A}^T\dot{\mathbf{Q}}\right)_{ij} = \left(\mathbf{A}^T\dot{\mathbf{Q}}\right)^T \mathbf{e}_{ij} = \left(\mathbf{A}\mathbf{e}_{ij}\right)^T \dot{\mathbf{Q}} > 0,$$

where $\mathbf{e}_{ij}$ denotes a vector that has all zero entries other than the unit entry corresponding to contact $\{i,j\}$. If it exists, such a flex can be found by solving the linear program (LP)

$$\max_{\dot{\mathbf{Q}}} \; \left(\mathbf{A}\mathbf{e}_{ij}\right)^T \dot{\mathbf{Q}}$$
$$\mathbf{A}^T\dot{\mathbf{Q}} \geq \mathbf{0}. \tag{10}$$

If this LP has optimal objective value of zero, then there is no first-order flex that is strict on the contact in question. Otherwise, the LP is unbounded, with an infinite optimal objective value. The dual LP of (10) is a feasibility problem

$$\mathbf{A}\left(\tilde{\mathbf{f}} + \mathbf{e}_{ij}\right) = \mathbf{0}$$
$$\tilde{\mathbf{f}} \geq \mathbf{0}, \tag{11}$$

where the contact forces $\tilde{\mathbf{f}}$ are the Lagrange multipliers corresponding to the impenetrability constraints $\mathbf{A}^T\dot{\mathbf{Q}} \geq \mathbf{0}$. If the dual LP (11) is feasible, then the primal LP (10) is bounded. If we identify $\mathbf{f} = \tilde{\mathbf{f}} + \mathbf{e}_{ij} \geq \mathbf{0}$, $\tilde{f}_{ij} \geq 1$, we are naturally led to consider the existence of non-trivial solutions to the force-equilibrium equations

$$\mathbf{A}\mathbf{f} = \mathbf{0} \text{ and } \mathbf{f} \geq \mathbf{0}. \tag{12}$$

A set of non-negative contact forces $\mathbf{f} \neq \mathbf{0}$ that are in equilibrium as given by Eq. (12) is called a *self-stress* [51]. In Ref. [12] these are called *proper self-stresses*, as opposed to self-stresses which are not required to be non-negative. Self-stresses can be scaled by an arbitrary positive factor, so we will often add a normalization constraint that the average force be unity, $\mathbf{e}^T\mathbf{f} = M$. A self-stress that is strictly positive on a given contact is *strict* on that contact. A self-stress $\mathbf{f} > 0$ is a *strict-self stress*. The existence of a (strict) self-stress can be tested by solving the linear program

$$\max_{\mathbf{f},\varepsilon} \; \varepsilon$$
$$\mathbf{A}\mathbf{f} = \mathbf{0}$$
$$\mathbf{f} \geq \varepsilon\mathbf{e}$$
$$\mathbf{e}^T\mathbf{f} = M \tag{13}$$

and seeing whether the optimal value is negative (no self-stress exists), positive (a strict self-stress exists), or zero (a self-stress exists). What we showed above using linear duality is that if there is a self stress that is strict on a given contact, there is no flex strict on that contact. In particular, this means that *packings that have a strict self-stress can only have null first-order flexes.*

We can also show that there is a first-order flex that is strict on all contacts that do not carry a force in any self-stress (i.e., no self-stress is strict on them). To this end, we look for a first-order flex that is strict on a given subset of the contacts, as denoted by the positions of the unit entries in the vector $\tilde{\mathbf{e}}$

$$\max_{\dot{\mathbf{Q}},\varepsilon} \; \varepsilon$$
$$\mathbf{A}^T\dot{\mathbf{Q}} \geq \varepsilon\tilde{\mathbf{e}}. \tag{14}$$

The dual program is the feasibility problem

$$\mathbf{A}\mathbf{f} = \mathbf{0}$$
$$\tilde{\mathbf{e}}^T\mathbf{f} = 1$$
$$\mathbf{f} \geq \mathbf{0}, \tag{15}$$

which is infeasible if there is no self-stress that is positive on at least on the contacts under consideration, since $\tilde{\mathbf{e}}^T\mathbf{f} \equiv 0$. Therefore the primal problem (13) is unbounded, that is, one can find a self-stress that is strict (since $\varepsilon \to \infty$) on the given set of contacts. This shows that *packings that do not have a self-stress are first-order flexible.* In other words, the existence of force chains in a packing is a *necessary* criterion for jamming.

In summary, if a packing has no self-stress, it is not jammed, and one can easily find a strict first-order flex by solving a linear program [11]. The analysis is simplified if the packing has a strict self-stress, since in that case all first-order flexes are null, i.e., they are solutions of a linear system of equalities $\mathbf{A}^T\dot{\mathbf{Q}} = \mathbf{0}$. This is the case of practical importance to jammed packings, so we will focus on it henceforth.

### 2. Floppy Modes

The linear system $\mathbf{A}^T\dot{\mathbf{Q}} = \mathbf{0}$ has $N_{\text{floppy}} = N_f - r$ solutions, where $r = M - N_{\text{stresses}}$ is the rank of the rigidity matrix, and $N_{\text{stresses}}$ is the number of (not necessarily proper) self-stresses (more precisely, the dimensionality of the solution space of $\mathbf{A}\mathbf{f} = 0$). We know that $N_{\text{stresses}} \geq 1$ for a jammed packing. If the packing is not hypoconstrained, or more precisely, if the number of contacts is sufficiently large

$$M = N_f + N_{\text{stresses}} \geq N_f + 1,$$

then there are no non-trivial null first-order flexes (floppy modes), $N_{\text{floppy}} = 0$. Therefore, *a packing that has a strict self-stress and a rigidity matrix of full-rank is (first-order) jammed.* We will later show that this sufficient condition for jamming is also necessary for sphere packings, that is, *jammed sphere packings are never hypoconstrained.*

However, we will see that jammed ellipsoid packings may be hypoconstrained, $M < N_f + 1$. Such a hypoconstrained packing always has floppy modes,

$$N_{\text{floppy}} = N_f + N_{\text{stresses}} - M \geq N_f + 1 - M.$$

Every floppy mode can be expressed as a linear combination of a set of $N_{\text{floppy}}$ basis vectors, i.e.,

$$\dot{\mathbf{Q}} = \mathbf{V}\mathbf{x} \text{ for some } \mathbf{x}, \tag{16}$$

where the matrix $\mathbf{V}$ is a basis for the null-space of $\mathbf{A}^T$. To determine whether any of the null first-order flexes can be extended into a true unjamming motion, we need to consider second-order terms, which we do next.

### B. Second-Order Terms

Consider a given null first-order flex $\mathbf{A}^T\dot{\mathbf{Q}} = \mathbf{0}$. We want to look for accelerations $\ddot{\mathbf{Q}}$ that make the second-order term in the expansion (9) non-negative, i.e.,

$$\mathbf{A}^T\ddot{\mathbf{Q}} \geq -\dot{\mathbf{Q}}^T\mathcal{H}\dot{\mathbf{Q}}. \tag{17}$$

If we cannot find such a $\ddot{\mathbf{Q}}$ for any first-order flex, then the packing is *second-order jammed*. If we find a $\ddot{\mathbf{Q}}$ such that all inequalities in (17) are strict, than we call the unjamming motion $\left(\dot{\mathbf{Q}}, \ddot{\mathbf{Q}}\right)$ a *strict second-order flex*, and the packing is *second-order flexible*, since there exists a $T > 0$ such that none of the impenetrability conditions [c.f. Eq. (6)] are violated for $0 \leq t < T$. If for all first-order flexes $\dot{\mathbf{Q}}$ at least one of the inequalities in (17) has to be an equality, then we need to consider even third- or higher-order terms, however, we will see that for sphere and ellipsoid packings this is not necessary.

#### 1. The Stress Matrix

In order to find a strict second-order flex, we need to solve the LP

$$\begin{aligned} \max_{\ddot{\mathbf{Q}},\varepsilon} \quad & \varepsilon \\ \mathbf{A}^T\ddot{\mathbf{Q}} \;\geq\; & \varepsilon\mathbf{e} - \dot{\mathbf{Q}}^T\mathcal{H}\dot{\mathbf{Q}}, \end{aligned} \tag{18}$$

the dual of which is

$$\begin{aligned} \min_{\mathbf{f}} \quad & \left(\dot{\mathbf{Q}}^T\mathcal{H}\dot{\mathbf{Q}}\right)^T \mathbf{f} \\ \mathbf{A}\mathbf{f} \quad & = \quad \mathbf{0} \\ \mathbf{e}^T\mathbf{f} \quad & = \quad 1 \\ \mathbf{f} \quad & \geq \quad \mathbf{0}, \end{aligned} \tag{19}$$

where the common optimal objective function is

$$\varepsilon^* = \left(\dot{\mathbf{Q}}^T\mathcal{H}\dot{\mathbf{Q}}\right)^T \mathbf{f} = \dot{\mathbf{Q}}^T\left(\mathcal{H}\mathbf{f}\right)\dot{\mathbf{Q}} = \dot{\mathbf{Q}}^T\mathbf{H}\dot{\mathbf{Q}},$$

where $\mathbf{H} = \mathcal{H}\mathbf{f}$ is a form of reduced Hessian that incorporates information about the contact force and the curvature of the touching particles. The $[N_f \times N_f]$ matrix $\mathbf{H}$ plays an essential role in the theory of jamming for hypoconstrained ellipsoid packing and we will refer to it as the *stress matrix* following Ref. [12].

The stress-matrix has a special block structure, where all of the blocks are of size $[d_f \times d_f]$, and both the block-rows and the block-columns correspond to particles. The block entry corresponding to the pair of particles $(i, j)$ is nonzero if and only if there is a contact between them. Written explicitly, the stress matrix is a force-weighted sum of contributions from all the contacts

$$\mathbf{H} = \sum_{\{i,j\}} f_{ij}\mathbf{H}_{ij},$$

where the contribution from a given contact $\{i, j\}$ is

$$\mathbf{H}_{ij} = \begin{array}{c} \\ \\ i \rightarrow \\ \vdots \\ j \rightarrow \end{array} \begin{array}{c} \begin{array}{ccc} i & \cdots & j \\ \downarrow & \cdots & \downarrow \end{array} \\ \begin{bmatrix} \nabla^2_{ii}\zeta_{ij} & \cdots & \nabla^2_{ji}\zeta_{ij} \\ \vdots & \ddots & \vdots \\ \nabla^2_{ij}\zeta_{ij} & \cdots & \nabla^2_{jj}\zeta_{ij} \end{bmatrix} \end{array}. \tag{20}$$

If $\dot{\mathbf{Q}}^T\mathbf{H}\dot{\mathbf{Q}} < 0$ then $\varepsilon^* < 0$ and therefore the first-order flex $\dot{\mathbf{Q}}$ cannot be extended into a second-order flex. We say that the stress matrix *blocks* the flex $\dot{\mathbf{Q}}$. If on the other hand $\dot{\mathbf{Q}}^T\mathbf{H}\dot{\mathbf{Q}} > 0$, then $\varepsilon^* > 0$ and by solving the LP (18) we can find an unjamming motion, i.e., the packing is second-order flexible. Therefore, finding an unjamming motion at the second-order level essentially consists of looking for a null first-order flex (floppy mode) $\dot{\mathbf{Q}}$, $\mathbf{A}^T\dot{\mathbf{Q}} = \mathbf{0}$, that is also a positive curvature vector for the stress matrix.

Recalling that every floppy mode can be expressed as $\dot{\mathbf{Q}} = \mathbf{V}\mathbf{x}$ [c.f. Eq. (16)], we see that

$$\dot{\mathbf{Q}}^T\mathbf{H}\dot{\mathbf{Q}} = \mathbf{x}^T\left(\mathbf{V}^T\mathbf{H}\mathbf{V}\right)\mathbf{x} = \mathbf{x}^T\mathbf{H}_V\mathbf{x}.$$

If the matrix $\mathbf{H}_V$ is negative-definite, than the packing is second-order jammed. In Ref. [12] such packings are called *pre-stress stable*, since the self-stress $\mathbf{f}$ rigidifies the packing (i.e., blocks all of the floppy modes). If $\mathbf{H}_V$ is indefinite, than the packing is second-order flexible since any of the positive-curvature directions can be converted into a strict self-stress by solving the LP (18).

If a packing has more than one (proper) self-stress, than it is not clear which one to use in the stress-matrix. One can try to find a self-stress that provides for jamming (pre-stress stability) by looking for a solution to Eq. (13) such that $\mathbf{H}_V \preceq \mathbf{0}$ (i.e., $\mathbf{H}_V$ is negative-semidefinite). This is known as semidefinite programming (SDP), and is a powerful generalization of linear programming that has received lots of attention recently [32]. It is however possible that different self-stresses are needed to block different portions of the space of floppy modes, and this general case of a second-order jammed packing is difficult to test for algorithmically. In our study of disordered sphere and ellipsoid packings, we will see that in practice the jammed packings only have one strict self-stress. In this case, testing for jamming reduces to calculating the smallest eigenvalue of $\mathbf{H}_V$. We will discuss actual numerical algorithms designed for ellipsoid packings in subsequent sections, but first we explain what makes sphere packings special.

#### 2. *The Stress Matrix for Hard Spheres*

For hard spheres it is easy to write down the explicit form for $\mathbf{H}_{ij}$ since the overlap function is given explicitly by Eq. (4) and its second-order derivatives are trivial,

$$\nabla^2_{ii} F_{ij} = \nabla^2_{jj} F_{ij} = -\nabla^2_{ij} F_{ij} = -\nabla^2_{ji} F_{ij} = \frac{2\mathbf{I}_d}{(O_i + O_j)^2},$$

where $\mathbf{I}_d$ is the $[d \times d]$ identity matrix. This implies that $\mathbf{H}_{ij}$ is a positive-definite matrix, since

$$\dot{\mathbf{R}}^T \mathbf{H}_{ij} \dot{\mathbf{R}} = (\dot{\mathbf{r}}_i - \dot{\mathbf{r}}_j)^T (\dot{\mathbf{r}}_i - \dot{\mathbf{r}}_j) \geq 0.$$

Therefore, any first-order flex in fact represents a true unjamming motion, since $\dot{\mathbf{Q}}^T \mathcal{H} \dot{\mathbf{Q}} \geq \mathbf{0}$ and we can trivially use $\ddot{\mathbf{Q}} = \mathbf{0}$ in Eqs. (18). In other words, a sphere packing is jammed *if and only if it is first-order jammed*, and therefore it cannot be hypoconstrained. To test for jamming in hard-sphere packings we need only focus on the velocities of the sphere centroids and associated linear programs in Section V A. This important conclusion was demonstrated using a simple calculation in Ref. [11].

For general particle shapes, however, $\mathbf{H}_{ij}$ may be indefinite for some contacts, and testing for jamming may require considering second-order terms. If one considers general convex particle shapes but freezes the orientations of the particles, the packing will behave like a hard-sphere packing. In particular, a jammed packing of nonspherical particles must have at least as many contacts as the corresponding isoconstrained packing of spheres would, that is,

$$\bar{Z} \geq 2d$$

for any large jammed packing of convex hard particles.

### C. Testing for Jamming

We now summarize the theoretical conditions for jamming developed in this section in the form of a procedure for testing whether a given packing of non-spherical particles is jammed. We assume that the contact network of the packing is known and available as input. For spherical particles, as already discussed, second-order terms never need to be considered, and testing for jamming can be done by solving one or two linear programs, as discussed in detail in Ref. [11]. In the formulation below, we avoid solving linear programs unless necessary, but rather use basic linear algebra tools whenever possible.

1. Find a basis $\mathbf{F}$ for the null-space of the rigidity matrix $\mathbf{A}$, i.e., find $N_{\text{stresses}}$ linearly independent solutions to the linear system of equations $\mathbf{Af} = \mathbf{0}$, normalized to mean of unity. This can be done, for example, by looking for zero eigenvalues and the associated eigenvectors of the matrix $\mathbf{A}^T\mathbf{A}$. If

   (a) $N_{\text{stresses}} = 0$,

   (b) $N_{\text{stresses}} = 1$ but the unique self-stress is not non-negative, or

   (c) $N_{\text{stresses}} > 1$ but the linear feasibility program (13) is infeasible,

   then declare the packing not jammed (first-order flexible), optionally identify an unjamming motion by solving the linear feasibility program $\mathbf{A}^T\dot{\mathbf{Q}} \geq \mathbf{e}$, and terminate the procedure. Otherwise, if the identified self-stress $\mathbf{f}$ is not strict, declare the test inconclusive and terminate.

2. If $N_{\text{floppy}} = N_f + N_{\text{stresses}} - M = 0$, then declare the packing (first-order) jammed and terminate the procedure. Otherwise, find a basis $\mathbf{V}$ for the null-space of $\mathbf{A}^T$, i.e. $N_{\text{floppy}}$ linearly-independent solutions to the linear system of equations $\mathbf{A}^T\Delta\mathbf{Q} = \mathbf{0}$. Compute the stress matrix $\mathbf{H}$ using the previously-identified strict self-stress $\mathbf{f}$, and compute its projection $\mathbf{H}_V$ on the space of null first-order flexes.

3. Compute the smallest eigenvalue $\lambda_{\min}$ and associated eigenvector $\mathbf{x}_{\min}$ of the matrix $\mathbf{H}_V$. If $\lambda_{\min} < 0$, declare the packing (second-order) jammed and terminate the procedure. If $\lambda_{\min} > 0$ and $N_{\text{stresses}} = 1$ declare the packing not jammed (second-order flexible), optionally compute an unjamming motion by solving the LP (18) with $\dot{\mathbf{Q}} = \mathbf{V}\mathbf{x}_{\min}$, and terminate the procedure. Otherwise, declare the test inconclusive and terminate.

We will discuss the actual numerical implementation of this algorithm later, and see that in practice we do not need to solve linear programs to test for jamming in hypoconstrained ellipsoid packings. Essentially, the packings we encounter in our work with disordered packings of hard ellipsoids always have a single strict self-stress and a negative-definite $\mathbf{H}_V$. The rectangular lattice of ellipses offers a different kind of example, namely, one with simple regular geometry but multiple self-stresses, and we analyze this example theoretically in Appendix A.

### D. Outside the Kinematic Perspective

It is worthwhile to briefly consider the connections between the jamming criteria developed above using the kinematic approach to jamming, and the static and perturbation approaches.

#### 1. Static View

We have already seen that forces appear naturally as Lagrange multipliers corresponding to impenetrability constraints, in the form of a strict self-stress $\mathbf{f} > 0$. In the static view, we ask whether a packing can support a given applied external force $\mathbf{B}$ by a set of non-negative interparticle forces. The key observation is that we can add an arbitrary positive multiple of a self-stress to any set of interparticle forces that support $\mathbf{B}$ in order to make them non-negative, without affecting force balance. Therefore, if the rigidity matrix $\mathbf{A}$ is of full-rank, as it has to be for jammed sphere packings, any (supportable) load $\mathbf{B}$ can be balanced with non-negative interparticle forces, and kinematic and static rigidity become equivalent [33].

The addition of arbitrary multiples of the self-stress to $\mathbf{f}$ is, however, a product of the mathematical idealization of the packing. In fact, each specific applied load in an isoconstrained packing with $M = N_f + 1$ contacts will be supported by a well-defined $\mathbf{f}$. The self-stress is only physical if all $N_f + 1$ contacts are active, which requires that the packing already be compressed by some pre-existing applied load. Otherwise, the density will be slightly smaller than the jamming density and upon application of an external load one of the contacts will break and only $N_f$ of the contacts will be active. In general, finding the active set of contacts requires solving the linear program [11]

$$\begin{aligned} \min_{\mathbf{f}} \mathbf{e}^T\mathbf{f} & \quad \text{for virtual work} \\ \text{such that} \quad \mathbf{A}\mathbf{f} = -\mathbf{B} & \quad \text{for equilibrium} \\ \mathbf{f} \geq \mathbf{0} & \quad \text{for repulsion only.} \end{aligned} \tag{21}$$

At the solution, modulo degenerate situations, only $N_f$ of the forces will be positive, the remaining ones will be zero.

For jammed hypoconstrained ellipsoid packings, such as the one in Fig. 5, supporting some loads may require a small deformation of the packing, such as a slight rotation of the mobile ellipse in the example in Fig. 5. After this small deformation, the normal vectors at the points of contact will change slightly and the interparticle forces $\mathbf{f}$ can support the applied force $\mathbf{B}$. The larger the magnitude of the forces is, the smaller the deformation needed to support the load is. Therefore every jammed packing can support any applied force in a certain generalized sense. Another way to look at this is to observe that, if the interparticle forces are much larger than the applied ones, the applied load will act as a small perturbation to the packing and the static view becomes equivalent to the perturbation view (with $\Delta\boldsymbol{\zeta} = \mathbf{0}$). We consider the perturbation view next and show how the stress matrix appears in the response of the packing to perturbations.

#### 2. Perturbation View

In the perturbation view we consider how the configuration *and* the contact forces respond to perturbations consisting of small changes of the contact geometry and small applied forces. Counting geometric and force constraints separately, as done in the literature, is incorrect when $\mathbf{f} > 0$: There is coupling between the particle positions and the interparticle forces as represented by the Hessian $\mathbf{H} = \mathcal{H}\mathbf{f}$.

With this in mind, we can expand Eq. (8) to first order in $\{\|\Delta\mathbf{Q}\|, \|\Delta\mathbf{f}\|\}$, to get the linear system of equations

$$\begin{bmatrix} \mathbf{A} & -\mathbf{H} & 0 \\ \mathbf{0} & \mathbf{A}^T & -2\mathbf{e} \\ \mathbf{e} & \mathbf{0} & 0 \end{bmatrix} \begin{bmatrix} \Delta\mathbf{f} \\ \Delta\mathbf{Q} \\ \Delta\mu \end{bmatrix} = -\varepsilon \begin{bmatrix} \Delta\mathbf{B} \\ \Delta\boldsymbol{\zeta} \\ 0 \end{bmatrix}. \tag{22}$$

It can be demonstrated that if the reduced Hessian $\mathbf{H}_V$ is definite, this system will have a solution for any $\Delta\mathbf{B}$ and $\Delta\boldsymbol{\zeta}$. Furthermore, if $\mathbf{H}_V$ is negative-definite the response to perturbations will be stable, in the sense that applied forces will do a positive work in order to perturb the packing. This is explained in greater detail in Ref. [21], where the conditions $\|\Delta\mathbf{Q}\| = O(\|\Delta\mathbf{B}\|)$ and $\Delta\mathbf{B}^T\Delta\mathbf{Q} < 0$ are stated in a more general setting, and then a linearization of the response of the packing to perturbations is considered (recall that in Ref. [21] $\Delta\boldsymbol{\zeta} \equiv \mathbf{0}$).

Equation (22) can be used to find the jamming point starting with a packing that is nearly jammed, i.e., a packing that has nonzero interparticle gaps $\varepsilon\Delta\boldsymbol{\zeta}$ and a self-stress that has a small imbalance $\varepsilon\Delta\mathbf{B} = \mathbf{A}\mathbf{f}$. This works well for small packings, however, for large disordered packings, the force chains are very sensitive to small changes in the geometry and the linearization of the perturbation response is not a good approximation even for packings very close to the jamming point. Additionally, we note that to first order in $\varepsilon$, the solution to Eq. (22) has $\Delta\mu/\varepsilon = \mathbf{f}^T\Delta\boldsymbol{\zeta}/2M = \mathbf{f}_E^T\mathbf{h}/2M$, which can be used to quickly estimate the jamming gap of a nearly-jammed packing from just the interparticle gaps $\Delta\boldsymbol{\zeta} = \boldsymbol{\zeta}$ and the interparticle forces, without knowing the actual jamming point [2].

## VI. NUMERICALLY TESTING FOR JAMMING IN HYPOCONSTRAINED ELLIPSOID PACKINGS

In this section we will apply the criteria for jamming and the algorithm to test for jamming from Section V C to our computationally-generated hypoconstrained packings of ellipsoids. This section is technical and may be skipped or skimmed by readers not interested in the mathematical formalism of jamming. The numerical results show that the packings are indeed second-order jammed, even very close to the sphere point. Before discussing the numerical details of the algorithm, we need to calculate the first and second-order derivatives of the overlap potential for ellipsoids.

### A. Overlap Potentials for Ellipsoids

Numerical algorithms for calculating the PW overlap potential $\zeta = \mu^2 - 1$ for ellipsoids are presented in the second part of Ref. [15]. Here we review the essential notation and give the first and second-order derivatives of the overlap potential, necessary to build the rigidity and stress matrices for a given packing.

An ellipsoid is a smooth convex body consisting of all points $\mathbf{r}$ that satisfy the quadratic inequality

$$(\mathbf{r}-\mathbf{r}_0)^T \mathbf{X} (\mathbf{r}-\mathbf{r}_0) \leq 1, \tag{23}$$

where $\mathbf{r}_0$ is the position of the center (centroid), and $\mathbf{X}$ is a characteristic *ellipsoid matrix* describing the shape and orientation of the ellipsoid,

$$\mathbf{X} = \mathbf{Q}^T \mathbf{O}^{-2} \mathbf{Q}, \tag{24}$$

where $\mathbf{Q}$ is the rotational matrix describing the orientation of the ellipsoid, and $\mathbf{O}$ is a diagonal matrix containing the major semi-axes of the ellipsoid along the diagonal. Consider two ellipsoids $A$ and $B$ and denote

$$\mathbf{Y} = \lambda \mathbf{X}_B^{-1} + (1-\lambda) \mathbf{X}_A^{-1}, \tag{25}$$

where $\lambda$ is defined in Section III. The *contact point* $\mathbf{r}_C$ of the two ellipsoids is

$$\mathbf{r}_C = \mathbf{r}_A + (1-\lambda) \mathbf{X}_A^{-1} \mathbf{n} = \mathbf{r}_B - \lambda \mathbf{X}_B^{-1} \mathbf{n}, \tag{26}$$

where

$$\mathbf{n} = \mathbf{Y}^{-1} \mathbf{r}_{AB} \tag{27}$$

is the *unnormalized* common *normal vector* at the point of contact.

In principle the overlap potential is a function of the normalized quaternions describing the particle orientations, and derivatives of $\zeta$ need to be projected onto the unit quaternion sphere. This projection can be avoided if we do not do a traditional Taylor series in the quaternions, namely an additive perturbation $\Delta\mathbf{q}$, but rather consider a multiplicative perturbation to the quaternions in the form of a small *rotation* from the current configuration $\Delta\boldsymbol{\varphi}$.

#### 1. First-Order Derivatives

The gradient of the overlap potential, which enters in the columns of the rigidity matrix, can be shown to be

$$\nabla_B \zeta = -\nabla_A \zeta = \begin{bmatrix} \nabla_{\mathbf{r}_B} \zeta \\ \nabla_{\boldsymbol{\varphi}_B} \zeta \end{bmatrix} = 2\lambda(1-\lambda) \begin{bmatrix} \mathbf{n} \\ \mathbf{r}_{BC} \times \mathbf{n} \end{bmatrix},$$

as we derived in Section III A 1 for a general convex particle shape by using the normalized normal vector $\hat{\mathbf{n}} = \mathbf{n}/\|\mathbf{n}\|$ [note that $\zeta = \lambda(1-\lambda)\mathbf{r}_{AB}^T \mathbf{n} - 1 = 0$]. Additionally, it is useful to know the derivatives of $\lambda$,

$$\nabla_{\mathbf{r}_B} \lambda = -\frac{2}{f_{\lambda\lambda}} \tilde{\mathbf{n}},$$

where

$$f_{\lambda\lambda} = 2 \frac{\mathbf{r}_{BC}^T \mathbf{Y}^{-1} \mathbf{r}_{AC}}{\lambda(1-\lambda)} < 0,$$

$$\tilde{\mathbf{n}} = \lambda \mathbf{n}_B + (1-\lambda)\mathbf{n}_A = \lambda \mathbf{Y}^{-1} \mathbf{r}_{AC} + (1-\lambda) \mathbf{Y}^{-1} \mathbf{r}_{BC},$$

and

$$\nabla_{\boldsymbol{\varphi}_B} \lambda = -\frac{2}{f_{\lambda\lambda}} \left[ \mathbf{M}_B \mathbf{n}_A - \lambda (\mathbf{r}_{BC} \times \mathbf{n}) \right],$$

where

$$\mathbf{M}_B = \lambda \mathbf{N}^L \mathbf{X}_B^{-1} + \mathbf{R}_{CB}^L.$$

#### 2. Second-Order Derivatives

The explicit expressions for the Hessian of the overlap potential are

$$\nabla^2_{\mathbf{r}_B} \zeta = 2\lambda(1-\lambda)\mathbf{Y}^{-1} - \frac{4}{f_{\lambda\lambda}} \left( \tilde{\mathbf{n}} \tilde{\mathbf{n}}^T \right) \succ \mathbf{0}$$

$$\nabla^2_{\boldsymbol{\varphi}_B \mathbf{r}_B} \zeta = 2\lambda(1-\lambda) \mathbf{M}_B \mathbf{Y}^{-1} + 2 \left[ \left( \nabla_{\boldsymbol{\varphi}_B} \lambda \right) \tilde{\mathbf{n}}^T \right]$$

and finally

$$\begin{aligned} \nabla^2_{\boldsymbol{\varphi}_B} \zeta \;=\; & -f_{\lambda\lambda} \left[ \left( \nabla_{\boldsymbol{\varphi}_B} \lambda \right) \left( \nabla_{\boldsymbol{\varphi}_B} \lambda \right)^T \right] + 2\lambda(1-\lambda) \cdot \\ & \left\{ \left[ \frac{1}{2} \left( \mathbf{r}_{BC} \mathbf{n}^T + \mathbf{n} \mathbf{r}_{BC}^T \right) - \left( \mathbf{r}_{BC}^T \mathbf{n} \right) \mathbf{I} \right] + \right. \\ & \left. \lambda \mathbf{N}^L \mathbf{X}_B^{-1} \mathbf{N}^R + \mathbf{M}_B \mathbf{Y}^{-1} \mathbf{M}_B^T \right\}. \end{aligned}$$

The derivatives with respect to the position and orientation of particle $A$ can be obtained by simply exchanging the roles of particles $A$ and $B$, however, there are also mixed derivatives involving motion of both particles

$$\begin{aligned} \nabla^2_{\boldsymbol{\varphi}_B \mathbf{r}_A} \zeta &= -\nabla^2_{\boldsymbol{\varphi}_B \mathbf{r}_B} \zeta \\ \nabla^2_{\boldsymbol{\varphi}_A \mathbf{r}_B} \zeta &= -\nabla^2_{\boldsymbol{\varphi}_A \mathbf{r}_A} \zeta \\ \nabla^2_{\boldsymbol{\varphi}_B \boldsymbol{\varphi}_A} \zeta &= -\nabla^2_{\boldsymbol{\varphi}_B} \zeta + \left( \nabla^2_{\boldsymbol{\varphi}_B \mathbf{r}_B} \zeta \right) \mathbf{R}_{AB}^R - \frac{1}{2} \left| \nabla_{\boldsymbol{\varphi}_\mathbf{B}} \zeta \right|_\times . \end{aligned}$$

The stress-matrix is built from these blocks as given in Eq. (20), where each of the four blocks $\nabla^2_{\alpha\beta}\zeta$ ( $\alpha$ and $\beta$ denote either $A$ or $B$) involves both translations and rotations,

$$\nabla^2_{\alpha\beta} \zeta = \begin{bmatrix} \nabla^2_{\mathbf{r}_\alpha} \zeta & \nabla^2_{\boldsymbol{\varphi}_\alpha \mathbf{r}_\beta} \zeta \\ \nabla^2_{\mathbf{r}_\alpha \boldsymbol{\varphi}_\beta} \zeta & \nabla^2_{\mathbf{r}_\beta} \zeta \end{bmatrix}.$$

### B. Numerically Testing for Jamming

The numerical implementation of the algorithm given in Section V C poses several challenges. The most important issue is that that algorithm was designed for ideal packings, that is, it was assumed that the true contact network of the packing is known. Packings produced by the MD algorithm, although very close to jamming (i.e., very high pressures), are not ideal. In particular, it is not trivial to identify which pairs of particles truly touch at the jamming point. Disordered packings have a multitude of near contacts that play an important role in the rigidity of the packing away from the jamming point [34], and these near contacts can participate in the backbone (force-carrying network) even very close to the jamming point. Additionally, not including a contact in the contact network can lead to the identification of spurious unjamming motions, which are actually blocked by the contact that was omitted in error.

For hard spheres, the algorithms can use linear programming to handle the inclusion of false contacts [11]. For ellipsoids, we look at the smallest eigenvalues of $\mathbf{A}^T\mathbf{A}$, i.e., the least-square solution to $\mathbf{A}\mathbf{f} = \mathbf{0}$. The solution will be positive if we have identified the true contact network, $\mathbf{f} > 0$, but the inclusion of false contacts will lead to small negative forces on those false contacts. The problem comes about because the calculation of the self-stress by just looking at the rigidity matrix does not take into account the actual proximity to contact between the particles. One way to identify the true contact network of the packing is to perform a long molecular dynamics run at a fixed density at the highest pressure reached, and record the list of particle neighbors participating in collisions as well as average the total transfer of collisional momentum between them in order to obtain the (positive) contact forces [2].

Once the contact network is identified, we want to look for null-vectors of the rigidity matrix. This can be done using specialized algorithms that ensure accurate answers [35], however, we have found it sufficient in practice to simply calculate the few smallest eigenvalues of the semi-definite matrix $\mathbf{A}^T\mathbf{A}$. We used MATLAB's sparse linear algebra tools to perform the eigenvalue calculation (internally MATLAB uses the ARPACK library, which implements the Implicitly Restarted Arnoldi Method). We consistently found that the smallest eigenvalue is about $3-6$ orders of magnitude smaller than the second-smallest eigenvalue, indicating that there is a near linear-dependency among the columns of $\mathbf{A}$ in the form of a self-stress. The self-stress, which is simply the eigenvector corresponding to the near-zero eigenvalue, was always strictly positive; in our experience, disordered packings of ellipsoids have a *unique strict self-stress* $\mathbf{f}$. This means that there are $N_{\text{floppy}} = N_f + 1 - M$ solutions to $\mathbf{A}^T\Delta\mathbf{Q} = \mathbf{0}$, $N_f - M$ of which are exact, and one which is approximate (corresponding to the approximate self-stress). This can be seen, for example, by calculating the eigenvalues of $\mathbf{A}\mathbf{A}^T$, since $N_f - M$ will be zero to numerical precision, one will be very small, and the remaining ones will be orders of magnitude larger.

#### 1. Verification of Second-Order Jamming

Once a strict self-stress is known, second-order jamming or flexibility can be determined by examining the smallest eigenvalue of $\mathbf{H}_V$, which requires finding a basis for the linear space of floppy modes. However, it is computationally demanding to find a basis for the null-space of $\mathbf{A}^T$ due to the large number of floppy modes, and since sparsity is difficult to incorporate in null-space codes. There are algorithms to find sparse basis for this null-space [35], however, we have chosen a different approach.

Namely, we calculate the smallest eigenvalues of

$$\mathbf{H}_k = k\mathbf{A}\mathbf{A}^T - \mathbf{H},$$

which as we saw in Section VIII B is the Hessian of the potential energy for a system of deformable ellipsoids where the stiffness coefficients are all $k$. For very large $k$ (we use $k = 10^6$), any positive eigenvalue of $\mathbf{A}\mathbf{A}^T$ is strongly amplified and not affected by $\mathbf{H}$, and therefore only the floppy modes can lead to small eigenvalues of $\mathbf{H}_k$, depending on how they are affected by $\mathbf{H}$. We have found that MATLAB's eigs function is not able to converge the smallest eigenvalues of $\mathbf{H}_k$ for large stiffnesses $k$, however, the convergence is quick if one asks for the eigenvalues closest to zero or even closest to $-1$. This typically reveals any negative eigenvalues of $\mathbf{H}_k$ and the corresponding floppy modes.

It is also possible to perform a rigorous numerical test for positive-definiteness of $\mathbf{H}_k$ using properly rounded IEEE machine arithmetic and MATLAB's (sparse) Cholesky decomposition of a numerically reconditioned $\mathbf{H}_k$ [36]. We have used the code described in Ref. [36] to show that indeed for our packings $\mathbf{H}_k \succ \mathbf{0}$ and therefore the packings are second-order jammed. For spheroids, that is, ellipsoids that have an axes of symmetry, there will be trivial floppy modes corresponding to rotations of the particles around their own centroid. These can be removed most easily by penalizing any component of the particle rotations $\Delta\boldsymbol{\varphi}$ that is parallel to the axis of symmetry. For example, one can add to every diagonal block of $\mathbf{H}_k$ corresponding to the rotation of an ellipsoid with axes of symmetry $\mathbf{u}$ a penalization term of the form $k\mathbf{u}\mathbf{u}^T$.

We have not performed a detailed investigation of a very wide range of samples since our goal here was to simply demonstrate that under appropriate conditions the packings we generate using the modified Lubachevsky-Stillinger algorithm are indeed jammed, even though they are very hypoconstrained near the sphere point. In this work we have given the fundamentals of the mathematics of jamming in these packings. A deeper understanding of the mechanical and dynamical properties of nearly-jammed hypoconstrained ellipsoid packings is a subject for future work.

## VII. NEARLY JAMMED PACKINGS

So far we have considered ideal jammed packings, where particles are exactly in contact. Computer-generated packings however always have a packing fraction $\phi$ slightly lower than the jamming packing fraction $\phi_J$, and the particles can rattle (move continuously) to a certain degree if agitated thermally or by shaking [2]. We can imagine that we started with the ideal jammed packing and scaled the particle sizes by a factor $\mu = 1-\delta < 1$, so that the packing fraction is lowered to $\phi = \phi_J (1-\delta)^d$. We call $\delta$ the *jamming gap* or distance to jamming.

It can be shown that if $\delta$ is sufficiently small the rattling of the particles does not destroy the jamming property, in the sense that the configuration point $\mathbf{Q} = \mathbf{Q}_J + \Delta\mathbf{Q}$ remains trapped in a small *jamming neighborhood* or *jamming basin* $\mathcal{J}_{\Delta\mathbf{Q}} \subset \mathcal{R}^{N_f}$ around $\mathbf{Q}_J$, which can be shown rather generally using arguments similar to those in Ref. [13] for tensegrities. In the limit $\delta \to 0$ the set of accessible configurations $\mathcal{J}_{\Delta\mathbf{Q}} \to \{\mathbf{Q}_J\}$, and in fact this is the definition of jamming used by Salsburg and Wood in Ref. [37]. Rewritten to use our terminology, this definition is: "A configuration is *stable* if for some range of densities slightly smaller than $\phi_J$, the configuration states *accessible from* $\mathbf{Q}_J$ lie in the neighborhood of $\mathbf{Q}_J$. More formally, if for any small $\epsilon > 0$ there exists a $\delta > 0$ such that all points $\mathbf{Q}$ accessible from $\mathbf{Q}_J$ satisfy $\|\mathbf{Q} - \mathbf{Q}_J\| < \epsilon$ provided $\phi \geq \phi_J (1-\delta)^d$." We call this the *trapping* view of jamming, most natural one when considering the thermodynamics of nearly jammed hard-particle systems [38]. Note that the trapping definition of jamming is in fact equivalent to our kinematic definition of jamming [13].

To illustrate the influence of the constraint curvature on jamming, we show in Fig. 6 four different cases with two constraints in two dimensions. In all cases a self-stress exists since the normals of the two constraints are both horizontal. If both constraint surfaces are concave (have negative or outward curvature), as constraints always are for hard-spheres, two constraints cannot close a bounded region $\mathcal{J}_{\Delta\mathbf{Q}}$ around the jamming point. One needs at least three constraints and in that case $\mathcal{J}_{\Delta\mathbf{Q}}$ will be a curved triangle. If however at least one of the constraints is convex (has positive curvature), two constraints can bound a closed jamming basin. Specifically, if the sum of the radii of curvatures of the two constraints at the jamming point $R_1 + R_2$ is positive, there is no unjamming motion. On the other hand, if it is negative then there is an unjamming motion in the vertical (floppy) direction. This is equivalent to looking at the smallest eigenvalue of the stress matrix in higher dimensions.

The jamming basin $\mathcal{J}_{\Delta\mathbf{Q}}(\delta)$ for a given jamming gap $\delta$ is the local solution to the relaxed impenetrability equa-

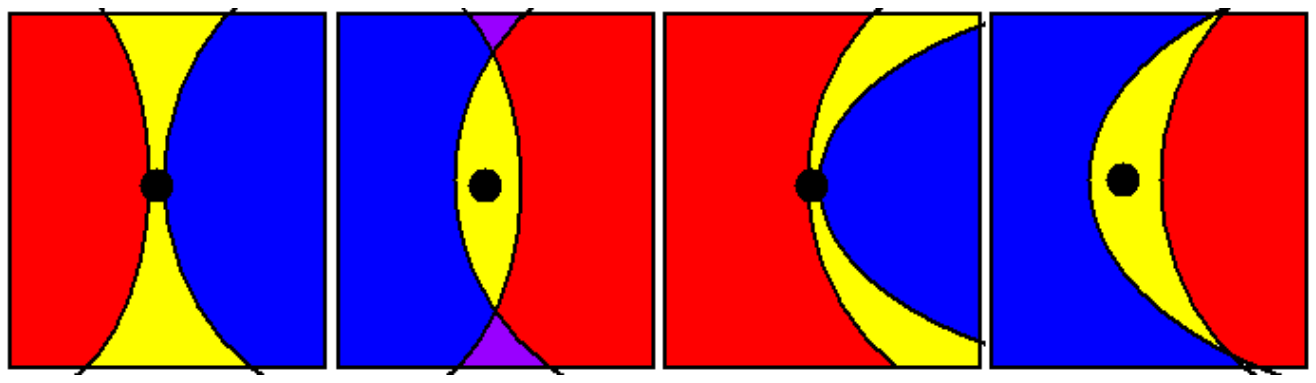

Figure 6: (Color online) The feasible region (yellow) around a jamming point (black circle) for two curved constraints in two dimensions (black circles). The region of the plane forbidden by one of the constraints is colored red, and colored blue for the other constraint. The region forbidden by both constraints is purple. The distance from the jamming point to the constraints is approximately $\delta$ and chosen small. Four cases are shown, going from left to right: (a) Both constraints are concave, and the yellow region is not bounded. Moving along the vertical direction unjams the system (this is typical of hard spheres). (b) Both constraints are convex, and the yellow region is closed, even though it is very elongated along the vertical direction (of order $\sqrt{\delta}$). This is an example of pre-stress stability (second-order jamming). (c) One of the constraints is convex, but not enough to block the unjamming motion in the vertical direction. The motion has to curve to avoid the convex constraint, i.e., a nonzero acceleration is needed to unjam the system (second-order flexible). (d) Only one of the constraints is convex, but enough to close the yellow region (second-order jammed). If the radii of curvatures are very close in magnitude, this region can become a very elongated banana-like shape.

tions

$$\boldsymbol{\zeta}(\Delta\mathbf{Q}) \geq -\zeta_\delta = 1 - \left(\frac{1}{1-\delta}\right)^2.$$

One way to determine $\mathcal{J}_{\Delta\mathbf{Q}}(\delta)$ for a wide range of $\delta$'s is to consider the function of the particle displacements

$$\tilde{\delta}(\Delta\mathbf{Q}) = \sqrt{1 + \min[\boldsymbol{\zeta}(\Delta\mathbf{Q})]} - 1, \tag{28}$$

that is, to calculate by how much the particles need to be shrunk to make a given particle displacement $\Delta\mathbf{Q}$ feasible (preserving non-overlapping). The contours (level-sets) of the function $\tilde{\delta}(\Delta\mathbf{Q})$ denote the boundaries of $\mathcal{J}_{\Delta\mathbf{Q}}(\delta)$, that is, $\mathcal{J}_{\Delta\mathbf{Q}}(\delta) = \left\{\Delta\mathbf{Q} \mid \tilde{\delta}(\Delta\mathbf{Q}) \leq \delta\right\}$.

### A. First-Order Jammed Packings

As a simple but illustrative example, we will consider a single mobile disk jammed between three other stationary disks, as shown in Fig. 7, an analog of the ellipse example from Fig. 5. This packing is first-order jammed, and the figure also shows a color plot of the function $\tilde{\delta}(\Delta\mathbf{Q})$ along with its contours. It is seen that for small $\delta$ the jamming basin $\mathcal{J}_{\Delta\mathbf{Q}}$ is a closed curved triangle.

These observations are readily generalized to higher dimensions. For sufficiently small $\delta$, the jamming basin approaches a convex *jamming polytope* (a closed polyhedron in arbitrary dimension) $\mathcal{P}_{\Delta\mathbf{Q}}$. For spheres all constraint surfaces are concave and therefore $\mathcal{P}_{\Delta\mathbf{Q}} \subseteq \mathcal{J}_{\Delta\mathbf{Q}}$ [37, 39]. The jamming polytope is determined from the linearized impenetrability equations

$$\mathbf{A}^T \Delta\mathbf{Q} \geq -\zeta_\delta \approx -2\delta, \tag{29}$$

and we can see that its volume, which determines the (non-equilibrium) free-energy, scales like $\delta^{N_f}$. This leads to the free-volume divergence of the pressure in the jamming limit

$$p = \frac{PV}{NkT} \approx \frac{d_f}{1-\phi/\phi_J}, \tag{30}$$

which has been verified numerically for disordered isoconstrained hard sphere packings [2].

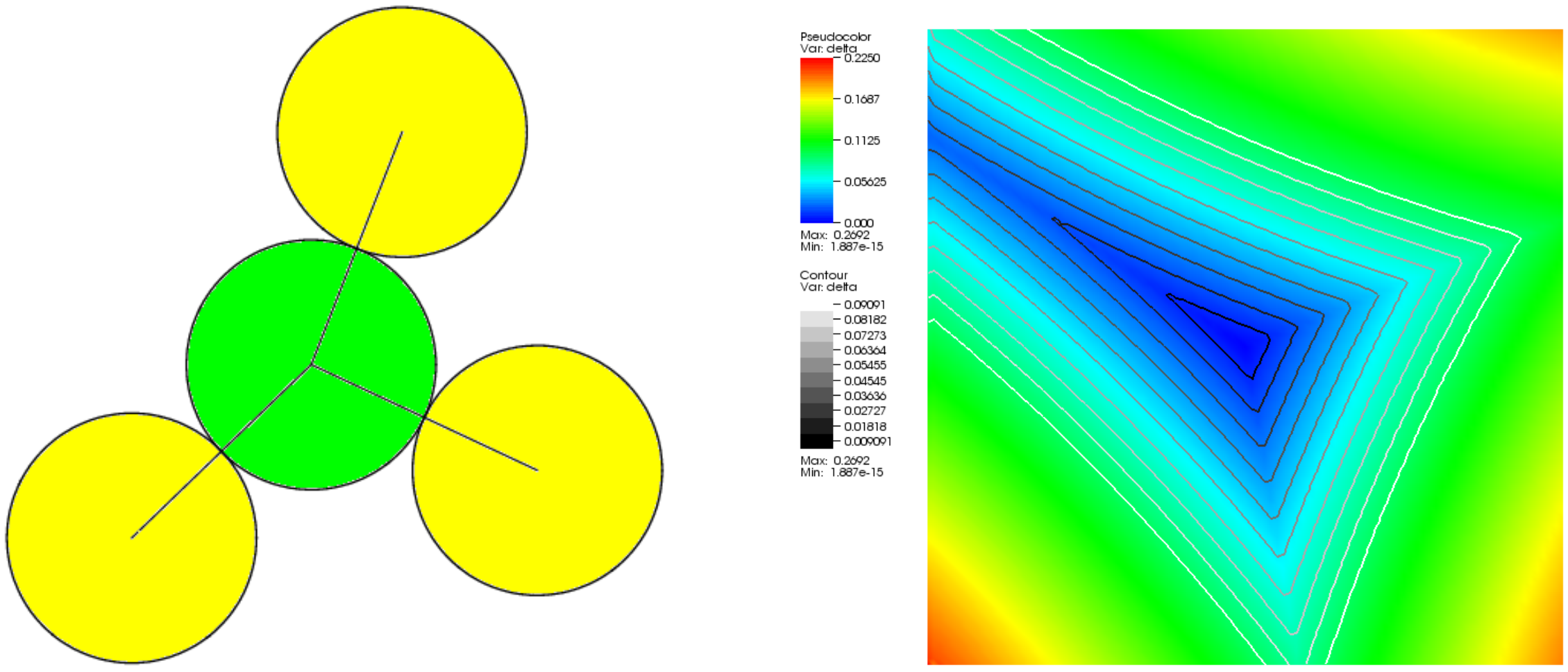

Figure 7: (Color online) (*Left*) An example of a mobile disk (green) jammed between three fixed disks (yellow). This is analogous to the ellipse packing shown in Fig. 5. (*Right*) A color plot of the function $\tilde{\delta}(\Delta\mathbf{Q})$ for this disk packing along with its contours (level sets).

### B. Second-Order Jammed Packings

The ellipse analog from Fig. 5 has three degrees of freedom, two translational and one orientational. If we fix the orientation of the (mobile) ellipse, that is, we take a planar cut through $\tilde{\delta}(\Delta\mathbf{Q})$, the situation is identical to that for the disk example above: For small $\delta$ the jamming basins $\mathcal{J}_{\Delta\mathbf{Q}}$ are closed curved triangles. However, the range of accessible orientations is rather large, on the order of $\sqrt{\delta}$, since even for a small $\delta$ the ellipse can rotate significantly. This is a consequence of the rotation of the ellipse being a floppy mode, and only being blocked by second-order effects as given by the curvature of the impenetrability constraints. In a certain sense, the packing is trapped to a greater extent in the subspace of configuration space perpendicular to the space of floppy modes than it is in the space of floppy modes. This is illustrated in Fig. 8.

### C. Pressure Scaling for Hypostatic Jammed Ellipsoid Packings

The observations in Fig. 8 are readily generalized to higher dimensions, however, it is no longer easy to determine the volume of $\mathcal{J}_{\Delta Q}$ (and thus the free energy) in the jamming limit. If we consider the simple two-constraint example in Fig. 6, we find that the area $A$ of the feasible (yellow) region scales like $\delta^{3/2}$ instead of $\delta^2$,

$$A = \frac{16}{3}\sqrt{\frac{R_1 R_2}{R_1+R_2}}\delta^{3/2}.$$

An obvious generalization of this result to higher dimensions can be obtained by assuming that the jamming basin $\mathcal{J}_{\Delta\mathbf{Q}}$ has extent $\sqrt{\delta}$ along all $N_{\text{floppy}} \approx N_f - M$ directions corresponding to floppy modes, where as it has extent $\delta$ along all other perpendicular directions. The volume would then scale as

$$|\mathcal{J}_{\Delta\mathbf{Q}}| \sim \delta^M \delta^{(N_f-M)/2} = \delta^{N(d_f/2+\bar{Z}/4)} = \delta^{Nd_f(1+s)/2},$$

where we quantify the *hypostaticity* of the packing by $s = \bar{Z}/2d_f$. The corresponding scaling of the pressure in

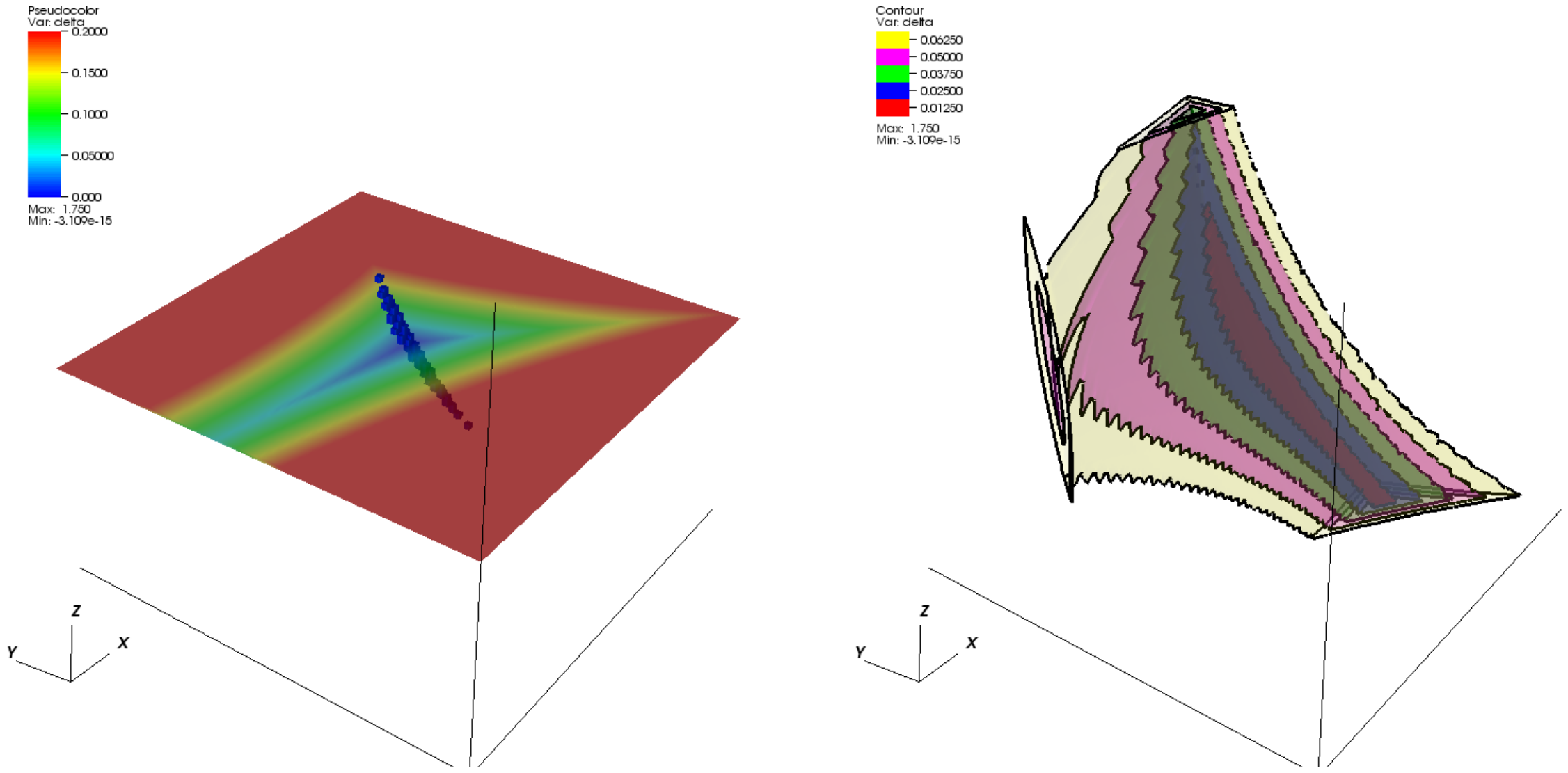

Figure 8: (Color online) (*Left*) A plot of the function $\tilde{\delta}(\Delta\mathbf{Q})$ for the packing from Fig. 5. The horizontal axes correspond to the translational degrees of freedom, and the vertical to the rotational degree of freedom (the rotation angle of the major axes). The $\Delta\mathbf{Q} = \mathbf{0}$ cut is also shown (horizontal colored plane), to be compared to the right part of Fig. 7. We also show the jamming basin $\mathcal{J}_{\Delta\mathbf{Q}}(\delta = 0.0035)$ (dark blue region), illustrating that this region is shaped like a banana, elongated along the direction of the floppy mode. (*Right*) Several contours (iso-surfaces) of $\tilde{\delta}(\Delta\mathbf{Q})$, bounding the banana-shaped regions $\mathcal{J}_{\Delta\mathbf{Q}}(\delta)$.

the jamming limit is

$$p = \frac{PV}{NkT} \approx \frac{d_f(1+s)/2}{1-\phi/\phi_J}.$$

However, as $\delta$ becomes very small, the jamming region becomes so elongated along the space of floppy modes that the time-scales for rattling along the elongated directions becomes much larger than the time for rattling in the perpendicular directions. This manifests itself as a remarkably large and regular oscillation of the "instantaneous" pressure (as measured over time intervals of tens of collisions per particle) during molecular-dynamics runs at a fixed $\delta$, as illustrated in Fig. 9. The oscillations are more dramatic the smaller $\delta$ is, and can span six or more orders of magnitudes of changes in the instantaneous pressure. The period of oscillation, as measured in numbers of collisions per particle, is dramatically affected by the moment of inertia of the ellipsoids $I$, most naturally measured in units of $mO^2$, where $m$ is the particle mass and $O$ is the (say smallest) ellipsoid semiaxis.

We do not understand the full details of these pressure oscillations, however, it is clear that dynamics near the jamming point for the hypoconstrained ellipsoid packings is not ergodic on small time-scales. In particular, as a packing is compressed during the course of the packing algorithm, the time-scale of the compression may be shorter than the time-scale of exploring the full jamming basin. Over shorter time scales the packing can only explore the directions perpendicular to the floppy modes, and in this case we expect that the pressure would scale

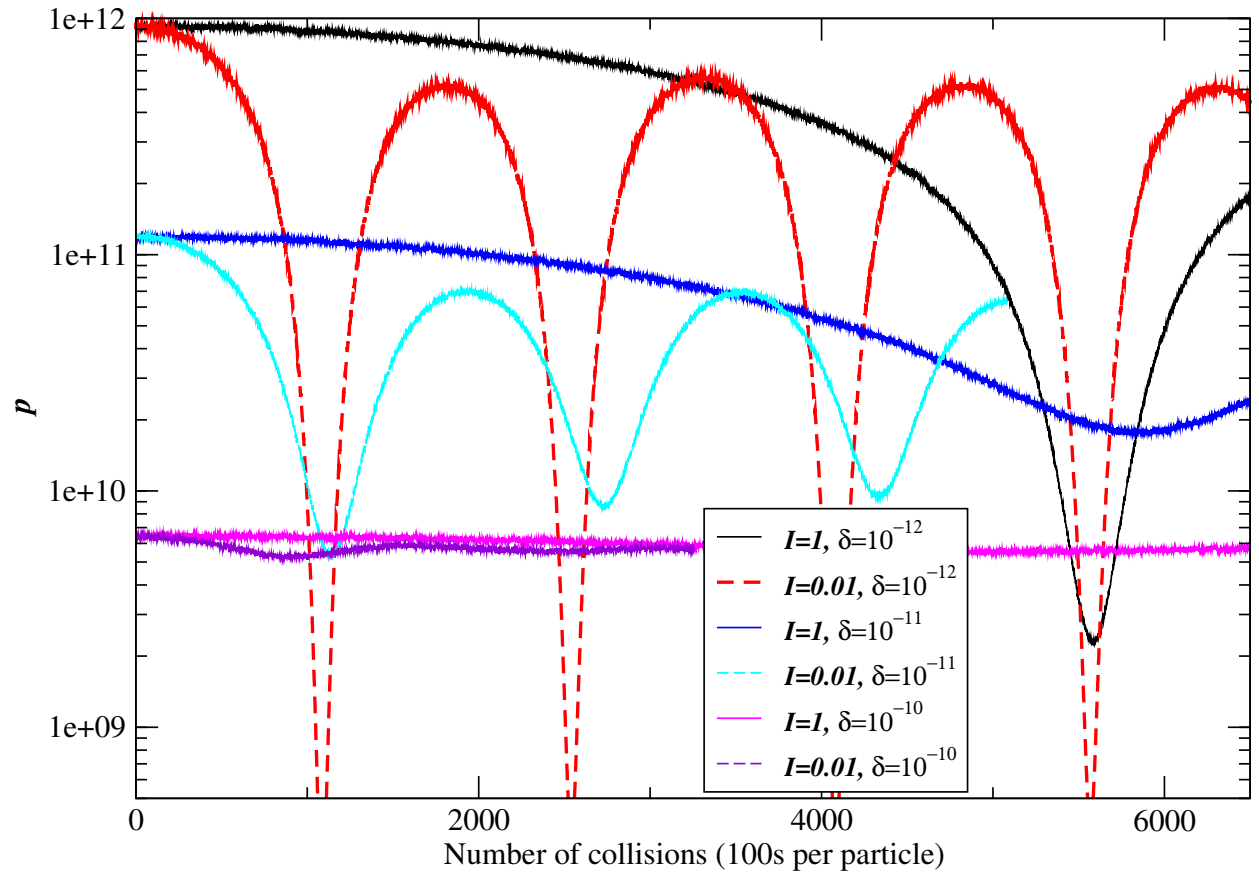

Figure 9: (Color online) The "instantaneous" reduced pressure for a jammed hypoconstrained packing of three-dimensional ellipsoids with semiaxes ratio $1.025^{-1} : 1 : 1.025$, at different (estimated) distances from the jamming point $\delta$. Molecular dynamics runs using a natural moment of inertia of the particles as well as ones using a much smaller moment of inertia are shown. The pressure oscillations are sustained for very long periods of time, however, it is not clear whether they eventually dissipate.

as

$$p \approx \frac{d_f s}{1-\phi/\phi_J}.$$

In Fig. 10 we show $C = p(1-\phi/\phi_J)$ as a function of

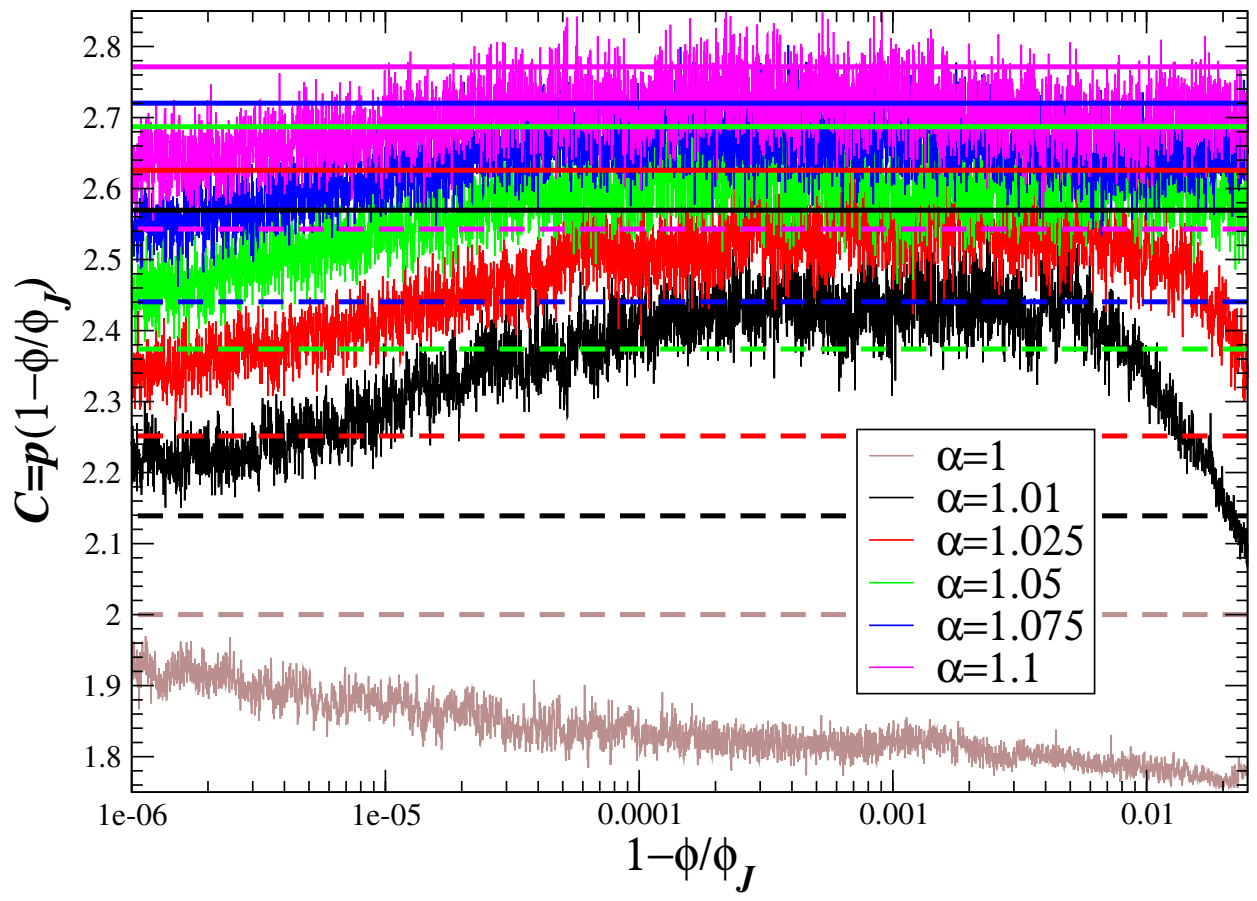

Figure 10: (Color online) The pressure scaling coefficient $C = p(1-\phi/\phi_J)$ as systems of hard ellipses are compressed from a dense liquid to the jamming point. The value of $C$ is not constant however it seems to remain between the bounds $C_L$ (shown with a dashed line in the same color as $C$) and $C_U$ (shown with a solid line).

the jamming gap for compressions of systems of ellipses of different aspect ratios close to unity. The compression started with a dense liquid and the particles were grown slowly at an expansion rate $\gamma = 10^{-5}$ to a high pressure (jamming) $p = 10^9$. The figure shows for each aspect ratio the lower bound $C_L = d_f s = 3s$ and the upper bound $C_U = d_f(1+s)/2 = 1.5(1+s)$, where $s$ was calculated by counting the almost perfect contacts at the highest pressure [2]. As expected from the arguments above, we see that very close to the jamming point $C \approx C_L$, however, further away from jamming $C \approx C_U$. For packings that are not hypoconstrained $C_L = C_U = d_f$, and for disks $C_U = C_L = 2$.

## VIII. ENERGY MINIMA IN SYSTEMS OF DEFORMABLE PARTICLES

In this section we consider the connections between jamming in hard particle packings and stable (local) energy minima (inherent structures [28]) for systems of deformable (soft) particles. This has a two-fold purpose. Firstly, in physical systems particles are always deformable, and therefore it is important to establish that the hard-particle conditions for jamming we established in Section V are relevant to systems of deformable particles. We expect that if the particles are sufficiently stiff, to be made more quantitative shortly, the behavior of the soft-particle system will approach that of the corresponding hard-particle packing. Secondly, considering the conditions for the existence of a stable energy minimum will enable us to derive in a simpler fashion and better understand the jamming conditions from the previous section.

We consider systems with short-ranged continuous interparticle potentials that are a monotonically decreasing function $E$ of the overlap between particles,

$$U_{ij} = E\left[\zeta\left(\mathbf{q}_i, \mathbf{q}_j\right)\right]. \tag{31}$$

That is, we assume that the elastic behavior of the particles is such that the interaction energy only depends on the distance between the particles as measured by the overlap potential $\zeta$. An example of such an elastic potential is an inverse power-law

$$E(\zeta) = (1+\zeta)^{-\nu}, \tag{32}$$

which in the limit $\nu \to \infty$ approaches a hard-particle interaction

$$E_H(\zeta) = \begin{cases} 0 \text{ if } \zeta > 0 \\ \infty \text{ if } \zeta < 0 \end{cases}.$$

For sufficiently large power exponents $\nu$ the interaction is localized around particles in contact and the overall energy

$$U = \sum_{ij} U_{ij} \to \max_{ij} U_{ij} = \left(1 + \min_{ij} \zeta_{ij}\right)^{-\nu} = \left(1 + \tilde{\delta}^2\right)^{-\nu}$$

is dominated by the most overlapping pair of particles [see Eq. (28) for the definition of $\tilde{\delta}$]. Additionally, as $\nu$ grows the interparticle potential becomes stiff in the sense that small changes in the distance between the particles cause large changes of the interparticle force

$$f = -\frac{dE}{d\zeta} \geq 0,$$

and the stiffness coefficient

$$k = \frac{d^2E}{d\zeta^2} \geq 0$$

becomes very large and positive. This indicates a physical interpretation of the hard-particle interaction potential: It is the limit of taking an infinite stiffness coefficient while the force between particles is kept at some non-negative value, which can be tuned as desired by infinitesimal changes in the distance between the particles (but note that the forces in different contacts are correlated since the motion of particles affects all of them simultaneously).

### A. Stable Energy Minima Correspond to Jammed Packings

Assume that we have a packing of hard particles and that we can find a set of interparticle interaction potentials $U_{ij}$ for the geometric contacts such that the configuration is a stable energy minimum. This means that any motion of the particles leads to increasing the energy $U$, i.e., to overlap of some pair of particles. Therefore, the packing of hard particles is jammed. This gives a simple way to prove that a given packing is jammed: Find a set of interparticle potentials that makes the configuration a stable energy minimum [12, 13]. We examine the conditions for a stable energy minimum when the interaction potentials are twice differentiable next.

The converse is also true, in the sense that arbitrarily near a jammed packing there is an energy minimum for a sufficiently "hard" interaction potential (in some cases the potential energy $U$ may have to be discontinuous at the origin [12]). We demonstrate this on the examples from Figs. 7 and 8 for a power-law interaction potential with increasing exponent $\nu$ in Figs. 11 and 12, respectively. It is clear that in the limit $p \rightarrow \infty$, the contours of the interaction potential become those of $\tilde{\delta}(\Delta\mathbf{Q})$ and are thus closed near the origin, i.e., the energy has a minimum. The higher the exponent $p$ is, however, the more anharmonic the interaction potential becomes and the contours are no longer ellipsoidal near the energy minimum.

It should be emphasized that the energy minima in soft-particle systems have a variable degree of overlap between neighboring particles and therefore do not correspond to hard-particle packings. In particular, at large pressures or applied forces the deformability of the particles becomes important and the energy minima no longer have the geometric structure of packings. However, in the limit of no externally-applied forces, i.e., $\mathbf{f} \rightarrow 0$, the only interacting particles are those that barely overlap, i.e., that are nearly touching. Therefore energy minima for purely-repulsive interaction potentials and a finite cutoff correspond to jammed packings of hard particles in the limit of zero external pressure (alternatively, one can keep the applied forces constant and make the grains infinitely stiff [6]). Therefore, the packings of soft particles studied in Ref. [3] very slightly above the "jamming threshold" $\phi_c$ are closely related to collectively jammed ideal packings of spheres of diameter $D = \sigma$ (polydispersity is trivial to incorporate) [40].

### B. Hessian Eigenvalues and Jamming

It is well-known that for smooth interactions a given configuration is a stable energy minimum if the gradient of the energy is zero and the Hessian is positive-definite, and the converse is also true if positive-definite is replaced with positive-semidefinite. This has been used as a criterion for jamming in systems of deformable particles [3, 40].

The gradient of $U = \sum_{ij} U_{ij}$ is

$$\begin{aligned} \nabla_{\mathbf{Q}} U &= \sum_{ij} \frac{dE}{d\zeta_{ij}} \left(\nabla_{\mathbf{Q}} \zeta_{ij}\right) = \left(\nabla_{\mathbf{Q}} \boldsymbol{\zeta}\right) \left(\nabla_{\boldsymbol{\zeta}} E\right) \\ &= \mathbf{A} \left(\nabla_{\boldsymbol{\zeta}} E\right) = -\mathbf{A}\mathbf{f}. \end{aligned}$$

The first-order *necessary* condition for a stable energy minimum is therefore exactly the force/torque balance condition

$$\mathbf{A}\mathbf{f} = 0 \text{ and } \mathbf{f} \geq 0,$$

as we derived using linear programming and duality theory for hard-particle packings. The Hessian is

$$\begin{aligned} \nabla^2_{\mathbf{Q}} U &= \left[\nabla_{\mathbf{Q}} \left(\nabla_{\boldsymbol{\zeta}} E\right)\right] \mathbf{A}^T + \left(\nabla^2_{\mathbf{Q}} \boldsymbol{\zeta}\right) \left(\nabla_{\boldsymbol{\zeta}} E\right) \\ &= \left[\mathbf{A} \left(\nabla^2_{\boldsymbol{\zeta}} E\right)\right] \mathbf{A}^T + \left(\nabla_{\mathbf{Q}} \mathbf{A}\right) \left(\nabla_{\boldsymbol{\zeta}} E\right) \\ &= \mathbf{A}\mathbf{K}\mathbf{A}^T - \mathcal{H}\mathbf{f} = \mathbf{A}\mathbf{K}\mathbf{A}^T - \mathbf{H}, \end{aligned}$$

where $\mathbf{K} = \nabla^2_{\boldsymbol{\zeta}} \epsilon = \text{Diag}\{k_{ij}\}$ is an $[M \times M]$ diagonal matrix with the stiffness coefficients along the diagonal, and $\mathcal{H} = \nabla_{\mathbf{Q}} \mathbf{A} = \nabla^2_{\mathbf{Q}} \boldsymbol{\zeta}$ is the Hessian of the overlap constraints. Note that more careful notation with derivatives of vectors and matrices can be developed and should in principle be employed in calculations to avoid confusions about the order of matrix multiplications and transpositions [41].

The Hessian

$$\mathbf{H}_U = \nabla^2_{\mathbf{Q}} U = \mathbf{A}\mathbf{K}\mathbf{A}^T - \mathbf{H}$$

consists of two terms, the *stiffness matrix* $\mathbf{H}_K = \mathbf{A}\mathbf{K}\mathbf{A}^T$, and the *stress matrix* $\mathbf{H}$ that we already encountered in the second-order expansion of the impenetrability constraints. The importance of not neglecting the stress matrix is also noted independently in Ref. [21], where also expressions are given for this matrix for certain types of contact geometry.

The second-order *sufficient* condition for a strict energy minimum is

$$\mathbf{H}_U \succ \mathbf{0}.$$

Since $\mathbf{K} > \mathbf{0}$, the stiffness matrix $\mathbf{H}_C$ is positive-semidefinite: For any vector $\Delta\mathbf{Q}$ that is not a floppy mode, $\Delta\mathbf{Q}^T \mathbf{H}_K \Delta\mathbf{Q} > 0$, while $\Delta\mathbf{Q}^T \mathbf{H}_K \Delta\mathbf{Q} = 0$ if $\Delta\mathbf{Q}$ is a floppy mode (i.e., $\mathbf{A}^T \Delta\mathbf{Q} = \mathbf{0}$). Therefore, for any direction of particle motion that is not a floppy mode, one can make the stiffness coefficients large enough to make $\Delta\mathbf{Q}^T \mathbf{H}_K \Delta\mathbf{Q} > 0$, regardless of the value of $\Delta\mathbf{Q}^T \mathbf{H} \Delta\mathbf{Q}$. Floppy modes, however, correspond to negative curvature directions of the Hessian $\mathbf{H}_U$ if they are positive-curvature directions of the stress matrix, $\Delta\mathbf{Q}^T \mathbf{H} \Delta\mathbf{Q} > 0$. Therefore, *the energy minimum is strict if and only if the stress matrix is negative-definite on the space of floppy modes*. This is exactly the same result as the second-order condition for jamming we derived in Section V using duality theory.

For deformable particles, the stiffness coefficients are finite. Therefore, for sufficiently large interparticle forces, the stress matrix may affect the eigenspectrum of the Hessian $\mathbf{H}_U$ and therefore the stability of potential energy minima. For spheres, as we derived earlier, $\mathbf{H} \succ \mathbf{0}$ and therefore interparticle forces may only destabilize packings: This is the well known result that increasing the interparticle forces leads to *buckling modes* in sphere packings [10]. Jamming in systems of soft spheres is therefore considered in the limit of $\mathbf{f} \to \mathbf{0}$, i.e., the point when particles first start interacting [3, 34]. For ellipsoids however, the forces can, and in practice they do, provide stability against negative or zero-frequency vibrational modes. The magnitude of the forces becomes important, and will determine the shape of the density of states (DOS) spectrum [34] for small vibrational frequencies. To quote from Ref. [10], "*The basic claim...is that one cannot understand the mechanical properties of amorphous materials if one does not explicitly take into account the direct effect of stresses.*"

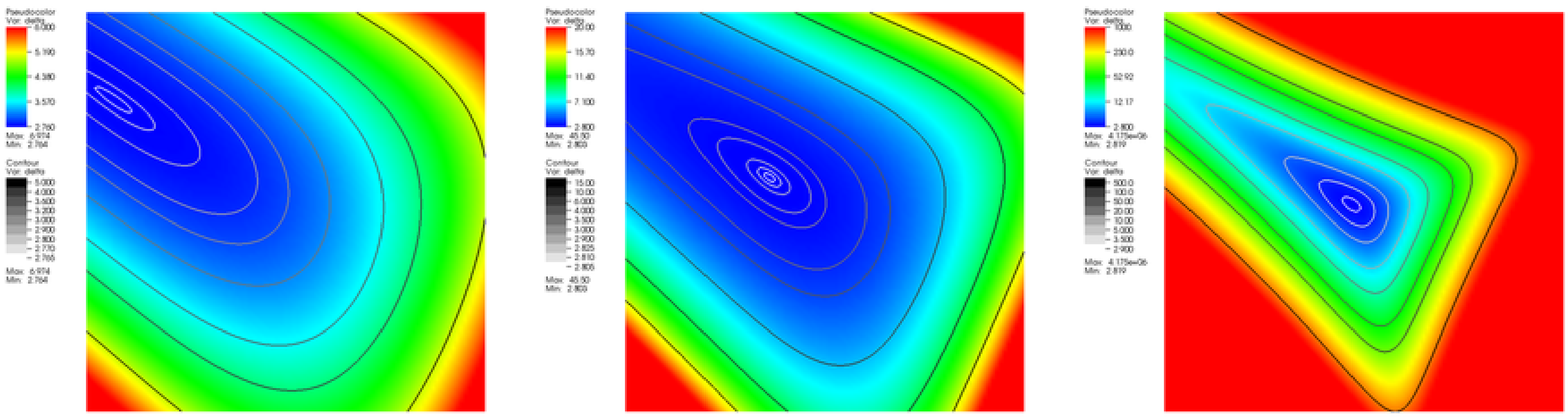

Figure 11: (Color online) The total interaction energy $U(\Delta\mathbf{Q})$ for the example in Fig. 7 when the disks are deformable and interact via a power-law potential. We show $U$ as a color plot with overlayed contours for power exponents $\nu = 12, 25,$ and 100 (going from left to right). Compare the $\nu = 100$ case to the contours of $\tilde{\delta}(\Delta\mathbf{Q})$ in Fig. 7.

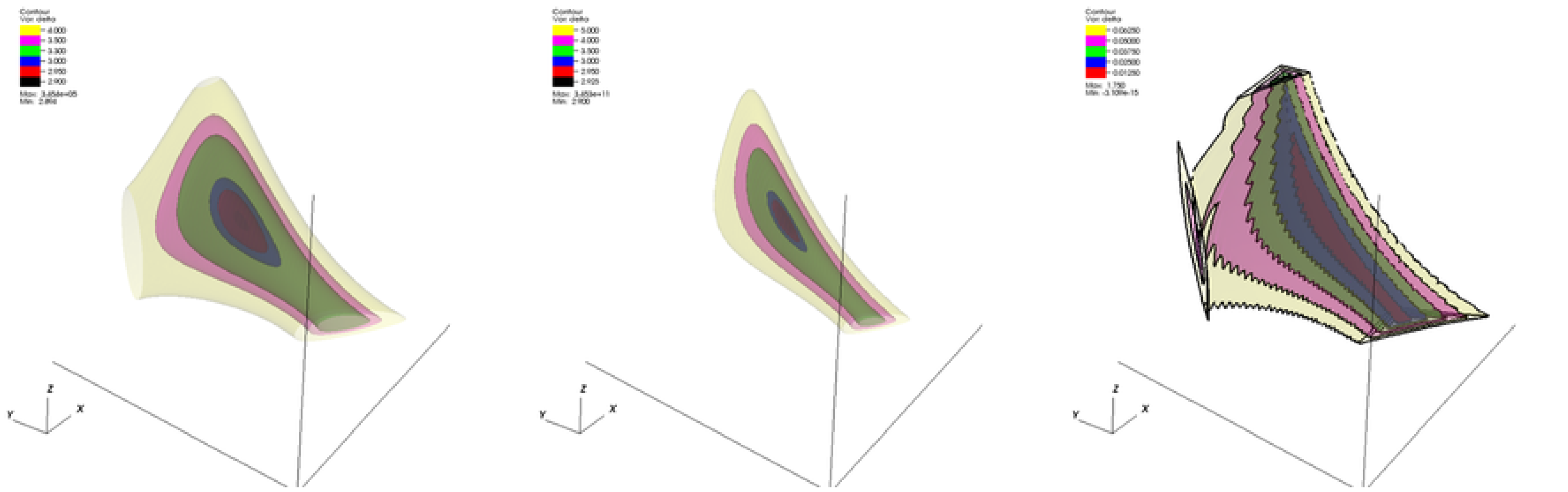

Figure 12: (Color online) The contours (iso-surfaces) of the total interaction energy $U(\Delta\mathbf{Q})$ for the example in Fig. 8 when the ellipses are deformable and interact via a power-law potential. Going from left to right, we show $\nu = 12$ and 25, as well as the hard ellipsoid $\tilde{\delta}(\Delta\mathbf{Q})$, corresponding to the limit $\nu \to \infty$.

The density of states (vibrational modes) in packings of soft spheres has been the subject of recent interest [34, 42, 43]. In particular, a Boson peak of low-frequency modes has been identified and attributed to the marginal rigidity (isostaticity) of the packings [42]. The effect of pre-stresses (pressure) on the density of vibrational modes has also been studied [43]. Such studies should be carried out also for packings of soft ellipsoids. In this case additional low-frequency modes will appear due to the floppy modes, especially at low pressures and for nearly spherical ellipsoids. These floppy modes will affect the mechanical response of the system, and there will be a subtle interplay between the low-frequency modes due to the marginal rigidity and those that appear because of the floppy modes inherent to hypoconstrained systems.

### C. An Example of Pre-Stress Stability

Figure 13 shows a very simple example in which pre-stressing, i.e., pre-existing forces, stabilize a structure. Although the example is not a packing, it illustrates well some of the essential features. First, the geometry of the system is degenerate, since the two springs are exactly parallel. This degeneracy insures that a self-stress exists, since one can stretch/compress both springs by an identical amount and still maintain force balance.

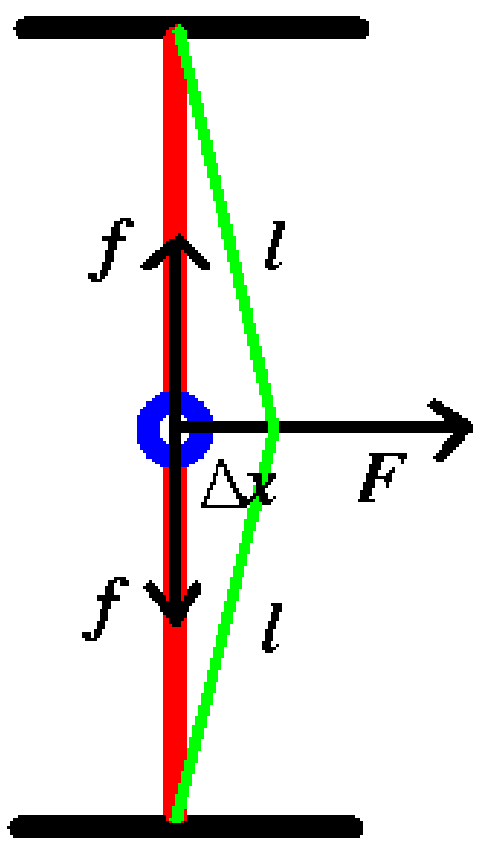

Figure 13: (Color online) An example of a pre-stress stable system. Two elastic springs of stiffness $k$ and length $l$ are connected via a joint that can move in the horizontal direction under the influence of an external force $F$.

Observe that geometrically the change in the position of the joint $\Delta x$ causes a *quadratic* change in the length of each spring $\Delta l \approx \Delta x^2$. To balance an applied force $F$, the force inside each spring $f$ needs to be $f\Delta x = F$. If the system is not pre-stressed, then the potential energy is *quartic* around the origin, $\Delta U = \frac{1}{2}k\Delta l^2 \approx \frac{1}{2}k\Delta x^4$, and the applied force causes a very large deformation of the structure $\Delta x = (F/k)^{1/3}$. The structure is stable (i.e., corresponds to a jammed packing), however, its response to perturbations is not harmonic. If however there is an initial force $f$ in the springs, then the potential energy is *quadratic* around the origin $\Delta U \approx f\Delta l = f\Delta x^2$ and the deformation is linear in the applied force $\Delta x = F/f$. If $f < 0$, then the system is unstable and will buckle, and if $f > 0$ the system is stable and its response to perturbations is harmonic. This is exactly the form of stability that hypoconstrained ellipsoid packings have.

It is instructive to compare the simple example in Fig. 13 with the example given in Fig. 5. In the latter there is also a single floppy mode. Let the small displacement of the central mobile particle along this floppy mode, due to an applied torque $\tau$, be $\Delta q$. This involves both a small rotation and a small displacement of the centroid, and causes a quadratic change in the contact distances $\Delta\mathbf{l} \sim \Delta q^2$. If the packing is pre-stressed by a slight compression (or expansion of the central ellipse), so that the contact forces are a positive multiple of the self-stress, $\mathbf{f} = \lambda\mathbf{f}_{\text{self}}$, $\lambda > 0$, then the potential energy is quadratic, $\Delta U = \mathbf{f}^T\Delta\mathbf{l} \sim \lambda\Delta q^2$. The deformation needed to resist the applied torque is determined from $\tau = \mathbf{f}^T\Delta\mathbf{N} = \lambda\Delta q\left(\mathbf{f}_{\text{self}}^T\boldsymbol{\nabla}_q\mathbf{N}\right)$, i.e., $\Delta q \sim \tau/\lambda$. Here $\boldsymbol{\nabla}_q\mathbf{N}$ denotes the sensitivity of the normal vectors $\mathbf{N}$ (represented in a suitable matrix form) at the points of contact with respect to the position of the mobile ellipse. The response of the system is therefore strongly-dependent upon the magnitude of the pre-stress $\lambda$, just as the response in the example in Fig. 13 is dependent upon $f$.

## IX. PACKINGS OF NEARLY SPHERICAL ELLIPSOIDS

In this section we will consider nearly spherical ellipsoids, that is, ellipsoids with aspect ratio $\alpha$ close to unity. In particular, we try to understand why these packings are hypoconstrained and to quantitatively explain the sharp rise in the density and contact numbers of disordered packings as asphericity is introduced. We propose that the packings of nearly spherical ellipsoids should be looked at as continuous perturbations of jammed disordered sphere packings, and establish the leading order terms in the expansion around the sphere point.

### A. Rotational and Translational Degrees of Freedom Are Not Equal

One might at first sight expect a discontinuous change in the contact number, and therefore the structure, as asphericity is introduced. After all, the number of degrees of freedom jumps suddenly from $d_f = d$ to (for non-spheroids) $d_f = d(d+1)/2 > d$. However, such an expectation is not reasonable. Firstly, the number of degrees of freedom is $d_f = d(d+1)/2$ even for spheres, since spheres can rotate too. This rotation does not affect the non-overlap conditions and therefore is not coupled to translational degrees of freedom. If the ellipsoids are nearly spherical, particle rotation is only mildly coupled to particle translations and rotation only affects the non-overlap conditions very close to the jamming point. This is seen, for example, through a violation of the equipartition theorem in non-equilibrium MD simulations of hard ellipsoids, depending on the moment of inertia of the particles and the time-scale of the system evolution. We therefore expect that thermodynamically and kinetically, at least at the level of translations, systems of nearly spherical ellipsoids will behave identically to systems of spheres until the interparticle gaps become comparable to the difference between the semiaxes. It is therefore not really surprising that the properties of the jammed packings such as $\phi_J$ or $\bar{Z}$ change continuously with $\alpha$.

What is somewhat surprising however is that $\phi_J$ and $\bar{Z}$ are not differentiable functions of particle shape. In particular, starting with a unit sphere and changing a given semiaxes by $+\epsilon \ll 1$ increases the density linearly in $\epsilon$, and changing it by $-\epsilon$ also increases the density by the same amount, $\Delta\phi_J \sim |\epsilon|$. As we will show through our calculations, this non-differentiability is a consequence of the breaking of rotational symmetry at the sphere point. The particle orientations themselves are not differentiable functions of particle shape and change discontinuously as the sphere point is crossed.

Finally, there is little reason to expect packings of nearly spherical particles to be rotationally jammed. After all, sphere packings are never rotationally jammed, since the spheres can rotate in place arbitrarily. Similarly, near the jamming point, it is expected that particles can rotate significantly even though they will be translationally trapped and rattle inside small cages, until of course the actual jamming point is reached, at which point rotational jamming will also come into play. It is therefore not surprising that near the sphere point, the parameters inside the packing-generation protocol, such as the moment of inertia of the particles and the expansion rate of the particles, can significantly affect the final results. In particular, using fast particle expansion or too large of a moment of inertia leads to packings that are clearly not rotationally jammed, since the torques are not balanced, however, they are translationally jammed and have balanced centroid forces. We do not have a full understanding of the dynamics of our packing-generation algorithm, even near the jamming point.

In this paper we will focus on packings that are also rotationally jammed. In general one may need to distinguish between *translational* and *rotational jamming.* For example, the ellipsoid packing produced by simply stretching the crystal packing of spheres along a certain axis by a scaling factor of $\alpha$ is translationally but not rotationally (strictly [11]) jammed. This is because by changing the axis along which the stretch is performed one gets a whole family of ellipsoid packings with exactly the same density. Therefore, it is possible to shear the packing by changing the lattice vectors used in the periodic boundary conditions, without changing the density, as illustrated in Fig. 14 in two dimensions.

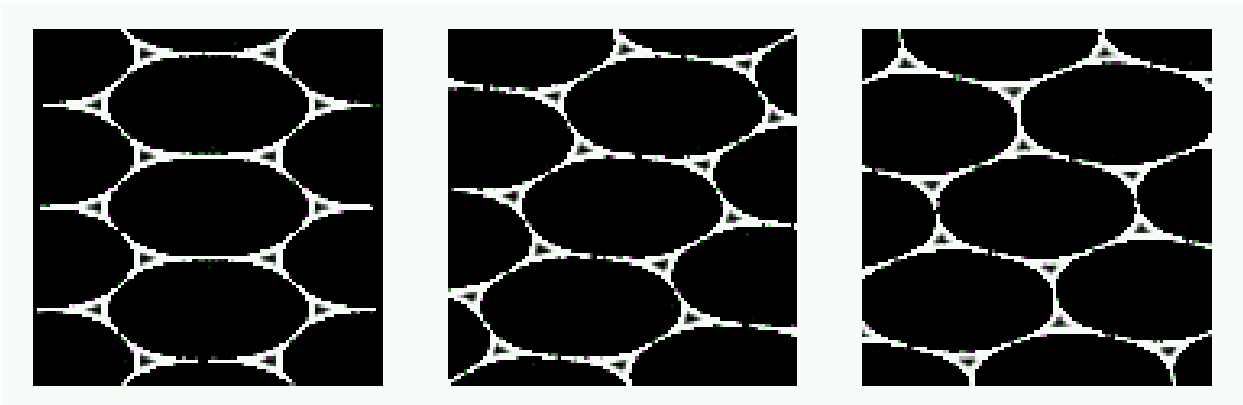

Figure 14: The triangular packing of ellipses is not rotationally jammed since one can shear the packing continuously, without introducing overlap or changing the density. The figure shows a sequence of snapshots as this shearing motion proceeds. The packing is however (strictly [11]) translationally jammed.

#### 1. *Isostatic Packings are Translationally Ordered*

As we already demonstrated, in order for a hypoconstrained packing of ellipsoids to be jammed, the packing geometry must be degenerate. The existence of a self-stress $\mathbf{f}$ requires that the orientations of particles be chosen so that the torques are balanced in addition to the forces on the centroids. This leads to a loss of "randomness" in a certain sense, since the number of jammed configurations is reduced greatly by the fact that geometrically "special" (not generic) configurations are needed to balance the torques.

However, it is also important to point out that disordered isoconstrained packings of nearly spherical ellipsoids are hard to construct. In particular, achieving isocounting near the sphere point requires translational ordering. In two dimensions, the average number of contacts per particle needed is $\bar{Z} = 6$, however, the maximal kissing number near the sphere point is also $Z_{\max} = 6$. Therefore the only possibility is that every particle have exactly $Z = 6$ contacts. This inevitably leads to translational ordering on a triangular lattice. In other words, the only isoconstrained packing of ellipses in the limit $\alpha \to 1$ is the hard disk triangular crystal. Similarly, in three dimensions, $\bar{Z} = Z_{\max} = 12$ for non spheroids, and therefore every particle must have exactly $Z = 12$ neighbors. While it not rigorously known what are the sphere packings with all particles having twelve neighbors, it is likely that only stacking variants of the FCC/HCP lattice achieve that property. For spheroids, the isoconstrained number of contacts is $\bar{Z} = 10$ and the results in Fig. 1 indicate that this value is nearly reached for sufficiently large aspect ratios. For nonspheroids, however, we only observe a maximum of 11.4 contacts per particle, consistent with the fact that achieving the isoconstrained value requires more translational ordering.

### B. Two Near-Spheres (Nearly) Touching

In what follows we will need first-order approximations of the impenetrability constraints between two nearly spherical ellipsoids. Assume there are two spheres $A$ and $B$ of radius $O_{A/B}$ touching. Transform the spheres into ellipsoids with semiaxes $O\mathbf{I} + \Delta\mathbf{O}$, and orientation described by the rotation matrix $\mathbf{Q}$, and denote $\boldsymbol{\epsilon}_O = O^{-1}\Delta\mathbf{O}$. Finally, define the matrix

$$\mathbf{T} = \mathbf{Q}^T \boldsymbol{\epsilon}_O \mathbf{Q},$$

which in the case of turning a disk into an ellipse with semiaxes $O$ and $O(1-\epsilon)$, i.e., aspect ratio $\alpha = 1 + \epsilon$, $\epsilon \ll 1$, becomes

$$\mathbf{T} = -\epsilon \begin{bmatrix} \sin^2\phi & -\sin\phi\cos\phi \\ -\sin\phi\cos\phi & \cos^2\phi \end{bmatrix} = -\epsilon \mathbf{T}_\phi,$$

where $\theta$ is the angle of orientation of the ellipse. It can be shown that to first order in $\epsilon$ the new distance between the ellipsoids is

$$\Delta\zeta = 2\mathbf{u}_{AB}^T \mathbf{S} \mathbf{u}_{AB},$$

where

$$\mathbf{S} = \frac{O_A}{O_A + O_B}\mathbf{T}_A + \frac{O_B}{O_A + O_B}\mathbf{T}_B.$$

The torque exerted by the contact force $\mathbf{f} = f\mathbf{n}$ on a given particle, to first order in asphericity $\epsilon$, comes about because the normal vector no longer passes through the centroid of the particle (as it does for spheres). One can ignore the small changes in the magnitude of the normal force or the change in the contact point $\mathbf{r}_C$, and only consider the change in the normal vector

$$\mathbf{n} \approx \mathbf{X}\mathbf{u} \approx (\mathbf{I} - 2\mathbf{T})\,\mathbf{u} = \mathbf{u} - 2\mathbf{T}\mathbf{u},$$

giving a torque

$$\boldsymbol{\tau} = \mathbf{r}_C \times \mathbf{f} \approx 2Of\,(\mathbf{T}\mathbf{u}) \times \mathbf{u}.$$

### C. Maintaining Jamming Near the Sphere Point

Assume now that we have a collectively jammed *isoconstrained* sphere packing with density $\phi_J^S$ and that we want to make the disks slightly ellipsoidal by shrinking them along a given set of axes, while still *preserving* jamming. Keeping orientations fixed, one can expand each near-sphere by a scaling factor $\Delta\mu$ *and* displace each centroid by $\Delta\mathbf{r}$, so that all particles that were initially in contact are still in contact. Note that because the matrix $\mathbf{S}$ is proportional to $\epsilon$, so will $\Delta\mu$ and $\Delta\mathbf{R}$. In other words, the change in the density will be linear in asphericity. However, the value of the *slope* depends on the choice of orientations of the ellipsoids. Referring back to Section V D 2 we see that to first order in $\epsilon$, $\Delta\mu$ is

$$\begin{aligned}\Delta\mu &= \frac{1}{M}\mathbf{f}^T\Delta\boldsymbol{\zeta} = \frac{1}{M}\sum_{\{i,j\}} f_{ij}\mathbf{u}_{ij}^T\mathbf{S}_{ij}\mathbf{u}_{ij} \\ &= \frac{1}{2M}\sum_i \sum_{j\in\mathcal{N}(i)} f_{ij}\mathbf{u}_{ij}^T\mathbf{T}_i\mathbf{u}_{ij},\end{aligned}$$

giving a new jamming density

$$\phi_J/\phi_J^S = (1+\Delta\mu)^d \prod_{k=1}^d (1+\epsilon_i^O) \approx 1 + d\Delta\mu + \mathbf{e}^T\boldsymbol{\epsilon}_O.$$

Keeping all ellipsoids aligned produces an affine deformation of the sphere packing that has the same jamming density, but is not (first-order) jammed. Therefore, the true jamming density must be higher, $\phi_J \geq \phi_J^S$. This explains why the jamming density increases with aspect ratio near the sphere point. The added rotational degrees of freedom allow one to increase the density beyond that of the aligned (nematic) packing, which for ellipsoids has exactly the same density as the sphere point.

Can we find a set of orientations for the ellipsoids so that the resulting packing is jammed? The first condition for jamming is that there exist a self-stress that balances both forces and torques on each particle. Just from the force-balance condition, one can already determine the interparticle forces $\mathbf{f}$. These will change little as one makes the particles slightly aspherical, because the normal vectors barely change. Therefore, the self-stress is already known a priori, without regard to the choice of particle orientations. The orientations must be chosen so that the *torques* are also balanced. As shown above, to first order in asphericity $\epsilon$, the torque balance condition for particle $i$ is

$$\sum_{j\in\mathcal{N}(i)} f_{ij}\,(\mathbf{T}_i\mathbf{u}_{ij}) \times \mathbf{u}_{ij} = \sum_j f_{ij}\mathbf{U}_{ij}\mathbf{T}_i\mathbf{u}_{ij} = 0. \qquad (33)$$

This gives for each particle a set of possible orientations, given the contact network of the isoconstrained sphere packing. The torque balance condition (33) is in fact the first-order optimality condition for maximizing the jamming density, as expected. It is worth pointing out that for a random assignment of orientations to ellipses the expected change in density is identically zero; in order to get an increase in the density one must use orientations correlated with the translational degrees of freedom.

#### 1. Ellipses

In two dimensions, for a particular contact with $\mathbf{u} = \langle\cos\theta, \sin\theta\rangle$ we have the simple expressions

$$\begin{aligned}\mathbf{u}\mathbf{T}_\phi\mathbf{u} &= \sin^2(\phi-\theta) \\ \mathbf{u}\times(\mathbf{T}_\phi\mathbf{u}) &= \frac{1}{2}\sin\left[2(\phi-\theta)\right].\end{aligned}$$

Considering $2\phi$ as the variable, one easily finds the solution to Eq. (33)

$$2\phi = \arctan(\pm\sum_i f_i\sin 2\theta_i, \pm\sum_i f_i\cos 2\theta_i). \qquad (34)$$

If we calculate the second derivative for the density increase we find that

$$\frac{d^2}{d\phi^2}\left[\sum_i f_i\sin^2(\phi-\theta_i)\right] = \pm 1,$$

and therefore in order to maximize the jamming density we need to choose the minus signs in Eq. (34). Once we find the unique orientation of each ellipse that ensures torque balance, we can calculate the jamming density

$$\phi_J/\phi_J^S \approx 1 + s_\phi\epsilon, \qquad (35)$$

where

$$s_\phi = 2\frac{\sum_i\sum_{j\in\mathcal{N}(i)} f_{ij}\left(\mathbf{u}_{ij}^T\mathbf{T}_i^\phi\mathbf{u}_{ij}\right)}{\sum_i\sum_{j\in\mathcal{N}(i)} f_{ij}} - 1.$$

We have calculated the slope $s_\phi$ for disordered binary disk packings (with $\phi_J^S \approx 0.84$) numerically, and find a value $s_\phi \approx 0.454$. We compare this theoretical value with numerical calculations in Fig. 15. The first comparison is directly to the packing fractions obtained using the Lubachevsky-Stillinger algorithm, which do not have anything to do with perturbing a sphere packing. Although the simulation jamming densities are not linear over a wide range of aspect ratios, near $\alpha = 1$ they are and the slope is close to the theoretically-predicted $s_\phi$. We also compare to results obtained by perturbing a jammed disk packing using MD. Specifically, we start with a jammed disk packing at a relatively high pressure ($p = 1000$) and assign an orientation according to Eq. (34) to every disk, and then we start growing the large semiaxes slowly while performing a form of constant-pressure MD. The density changes automatically to keep the pressure constant, and from the instantaneous density we estimate the jamming density using Eq. (30). In Fig. 15 we show how the (estimated) jamming density changes with aspect ratio. If we freeze the orientations (i.e., use an infinite moment of inertia), we obtain results that follow the theoretical slope prediction closely. Very good agreement with the results from the LS algorithm is obtained over a wide range of $\alpha$ if we start with the correct orientations and then allow the ellipse orientations to change dynamically. For comparison, in the inset we show that the packing density actually decreases if we use the LS algorithm and freeze orientations at their initial (random) values, demonstrating that balancing the torques and (maximally) increasing the density requires a particular value for the particle orientations.

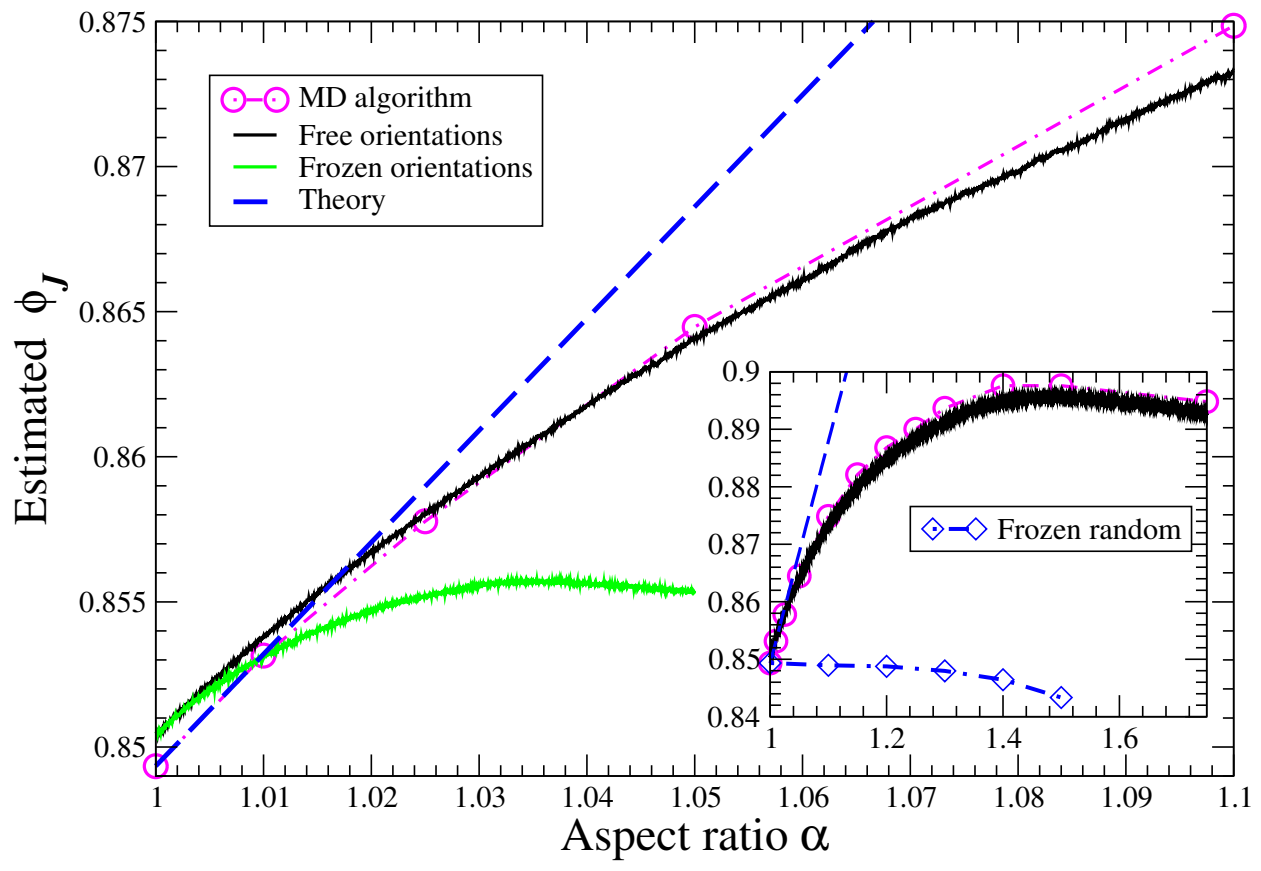

Figure 15: (Color online) The estimated jamming density near the disk point for binary packings of hard ellipses, as obtained from the LS packing algorithm, from perturbing the disk packing using constant-pressure MD, and from the first-order perturbation theory. The inset shows some of the data over a larger range of aspect ratio and also shows the packing densities obtained when the ellipses have infinite moment of inertia in the LS algorithm.

For ellipses, there are unique orientations that guarantee the existence of self-stresses near a given isoconstrained jammed disk packing. Do these orientations actually lead to jammed packings, that is, are the second-order conditions for jamming also satisfied? If one starts with a jammed disk packing and transforms the disks into ellipses of aspect ratio sufficiently close to unity, the packing will remain translationally jammed [13]. Subsequent increase in the size of the particles must eventually lead to a packing of maximal density. It is not however *a priori* obvious whether this packing is rotationally and translationally jammed or has some kind of peculiar unjamming motions that preserve the density, such as the ones shown in Fig. 14. For small disk packings, we have found the perturbed ellipse packings to be second-order jammed sufficiently close to the sphere point. For larger systems, even for very small asphericities, it is difficult numerically to perturb a given disk packing into an ellipse packing without leading to new contacts or breaking of old ones, as discussed shortly. An analytical investigation may be able to prove that the perturbed packings are actually second-order jammed, and therefore prove that there exist (large) jammed ellipse packings with $\bar{Z} = 4$, the absolute minimum contact number possible for a jammed packing.

Finally, we note that in three dimensions the torque balance equations (33) involve quaternions and are quartic, and it does not seem an analytical solution is possible as it is in two dimensions. We however expect that the calculations performed here in $d = 2$ can be generalized to higher dimensions as well. One interesting question to answer theoretically in $d = 3$ is whether the middle axes ($\beta$) affects the slope of the density $s_\phi$ or whether only the ratio of the largest to the smallest semiaxes ($\alpha$), matters. In Ref. [4] we proposed that the rapid increase in packing fraction could be attributed to the need to increase the contact numbers, since forming more contacts requires a denser packing of the particles. This is supported by the observation that the maximal packing density is achieved for the most aspherical shape ($\beta = 1/2$). However, numerical results very close to the sphere point are consistent with a slope $s_\phi$ independent of $\beta$. The arguments of this section indicate that the density rise is independent of the rise of the coordination number, at least near the sphere point.

### D. Contact Number Near the Sphere Point

In our perturbation approach to ellipsoid packings near the sphere point, we assumed that the contact network remains that of the disk packing even as the aspect ratio moves away from unity. However, as the aspect ratio increases and the packing structure is perturbed more and more, some new contacts between nearby particles will inevitably close, and some of the old contacts may break. In Fig. 16 we show a system that the linear perturbation prediction produces at $\alpha = 1.025$. While the original contacts in the jammed disk packing are maintained relatively well, we see that many new overlaps form that were not contacts in the disk packing. This means that the contact number will increase from $\bar{Z} = 4$ as asphericity is introduced.

These observations suggest a way to calculate the leading order term of $\bar{Z}(\alpha)-2d$: We simply count the overlaps introduced by orienting and displacing the centroids of the ellipsoids according to the linear perturbation theory. It is well-known that jammed disordered sphere packings have an unusual multitude of nearly-touching particles, as manifested by a power-law divergence in the pair correlation function near contact of the form $g_2(r) \sim (r - D)^{-0.4}$ (once rattlers are removed) [2]. For binary disks in two dimensions the exact exponent has not been calculated, but it appears close to a half [52]. These near contacts will close to form true contacts and cause the rapid increase in $\bar{Z}(\alpha)$, and we expect that the growth will be of the form

$$\bar{Z}(\alpha) - 2d \approx Z_\alpha \sqrt{\alpha - 1}. \tag{36}$$

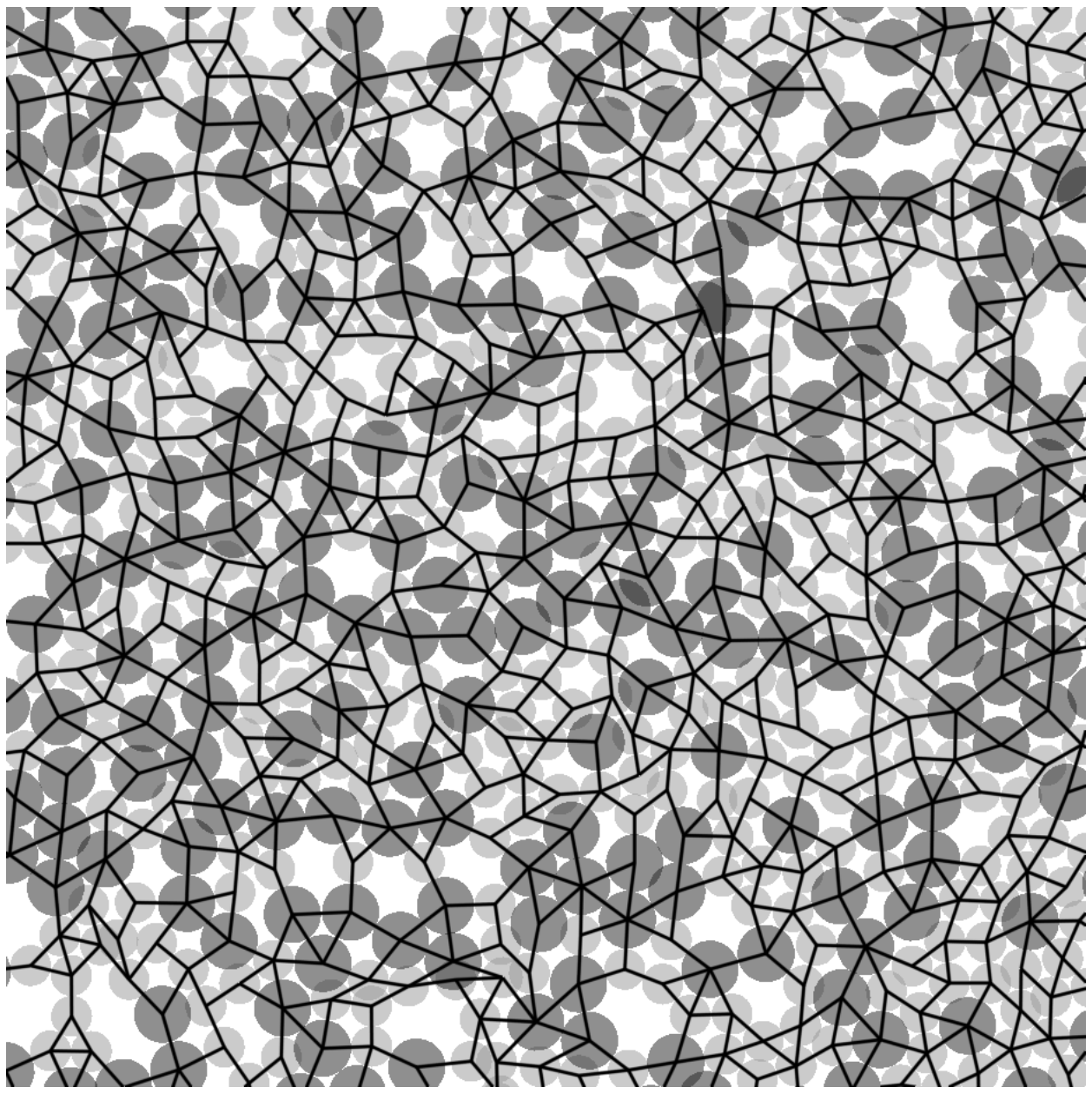

Figure 16: Overlaps introduced at $\alpha = 1.025$ by the naive linear perturbation theory which only takes into account the original contact network of the disk packing (black lines). We see many overlaps forming between particles that were nearly touching when $\alpha = 1$.

A more rigorous analysis is difficult since we do not really have an understanding of the geometry of the near contacts. We have numerically estimated the coefficient $Z_\alpha$ and plotted the prediction of Eq. (36) in Fig. 2. It is seen that the prediction matches the actual simulation results well sufficiently close to the sphere point.

## X. CONCLUSIONS

In this paper we presented in detail the mathematical theory of jamming for packings of nonspherical particles and tried to understand the properties of jammed packings of nonspherical particles of aspect ratio close to unity, focusing on hard ellipses and ellipsoids. In this section we summarize our findings and also point to directions for future investigation.

Mathematically, understanding jamming in hard-particle packings is equivalent to understanding the behavior of large systems of nonlinear inequalities as given by the impenetrability conditions. These inequalities can be written explicitly by introducing a continuously differentiable overlap potential whose sign determines whether two particles overlap. In Section III we generalized the overlap potential proposed by Perram and Wertheim for hard ellipsoids to arbitrary smooth strictly convex particle shapes and determined its first order derivatives.

In Section IV, we discussed the conjecture that large disordered jammed packings of hard particles are isoconstrained, i.e., that they have an equal number of constraints and degrees of freedom, $\bar{Z} = 2d_f$. It is not possible to make this conjecture into a theorem since the term “disordered” is highly nontrivial to define [16]. However, arguments have been made in the literature in support of isocounting. We showed that this conjecture can be supported with reasonable arguments only for spheres, where particle rotations are not considered. In particular, while it is expected that $\bar{Z} \leq 2d_f$ for “random” packings, the converse inequality $\bar{Z} \geq 2d_f$ only applies to spheres. Packings of nonspherical particles can be jammed and have less than $2d_f$ contacts per particle, i.e., be hypoconstrained. A minimally rigid ellipsoid packing, i.e., a packing that has the minimal number of contacts needed for jamming, satisfied only the inequality $\bar{Z} \geq 2d$, since at least $2d$ contacts per particle are needed to block particle translations. Particle rotations, however, and combined rotation/translation motions, can be blocked by the curvature of the particle surfaces at the point of contact. In essence, if the radii of curvatures at the point of contact are sufficiently large, i.e., the particle contact is sufficiently “flat”, rotation of the particles is blocked. This can be visualized by considering the limit of infinite radii of curvatures, when have a contact between two flat surfaces. Such contacts, in a certain sense, count as several “contact points” and block several degrees of freedom.

In Section V, we generalized the mathematics of first and second-order rigidity for tensegrity frameworks developed in Ref. [12] to packings of nonspherical particles. We proved that in order for a packing to be jammed there must exist a set of (nonzero) non-negative interparticle forces that are in equilibrium, i.e., the packing must have a self-stress. Furthermore, we considered second-order terms for hypoconstrained packings that do have a self-stress but also have floppy modes, that is, particle motions that preserve interparticle distances to first order. The second-order analysis showed that jammed packings of strictly convex particles cannot have less than $2d$ contacts per particle. We found that floppy modes involving particle rotations can be blocked (rigidified) by the stress-matrix, which includes second-order information about the particle surfaces at the point of contact. We proposed that this is exactly the type of jamming found in disordered ellipsoid packings near the sphere point, and in Section VI we presented a numerical algorithm for testing hypoconstrained ellipsoid packings for jamming and applied it to some computer-generated samples. We demonstrated that the packings are indeed jammed even very close to the sphere point, where they have close to $2d$ contacts per particle.

In Section VII we considered the thermodynamics of packings that are close to, but not exactly at, the jamming point, so that particles have some room to rattle (free volume). We found that for hypoconstrained packings the jamming basin $\mathcal{J}_{\Delta\mathbf{Q}}$, which is localized around the jamming point in configuration space, is very elongated along the space of floppy modes. For iso- or hyperstatic packings, as jammed sphere packings always are, the jamming basin approaches a polytope in the jamming limit, whereas for hypoconstrained packings it approaches a (hyper) banana. The latter leads to very large oscillations of the instantaneous pressure near the jamming point and a violation of the asymptotic free-volume equation of state (pressure scaling).

Real packings are always made from deformable (albeit very stiff) particles, i.e., particles that interact via some elastic interaction potential. The analog of a jammed hard-particle packing for deformable particles are strict energy minima (inherent structures), i.e., structures where any motion of the particles costs energy (quadratic in the displacements). In Section VIII we analyzed the first- and second-order conditions for a strict energy minimum for twice-differentiable interaction potentials. We found that the first-order condition is exactly the requirement for the existence of a self-stress, and that the second-order condition is exactly the condition that the stress-matrix blocks the floppy modes. This deep analogy between jamming in hard-particle packings and energy minima in soft-particle packings is not unexpected since a "soft" potential can approximate the singular hard-particle potential arbitrarily closely. As the potential becomes stiffer, the energy minimum will become highly anharmonic and its shape will closely resemble that of the jamming basin $\mathcal{J}_{\Delta\mathbf{Q}}$ (even at very small temperatures).

Finally, in Section IX we developed a first-order perturbation theory for packings of nearly spherical ellipsoids, expanding around the sphere point. The theory is based on the idea that packings of ellipsoids with aspect ratio $\alpha = 1 + \epsilon$ near unity have the same contact network as a nearby isostatic packing of hard spheres. In order for the ellipsoid packing to also be jammed, the orientations of the ellipsoids must be chosen so as to balance the torques on each particle. These orientations also maximize the jamming density, increasing it beyond that of the disk packing, and we analytically calculated the linear slope of the density increase with $\epsilon$ for binary ellipse packings. The calculated coefficient is in good agreement with numerical results. The perturbation of the sphere packing also leads to a rapid increase in the average particle coordination $\bar{Z}$, which we attributed to the closing of the multitude of near contacts present in disordered disk packings. The predicted $\bar{Z} \sim \sqrt{\epsilon}$ is also in good agreement with numerical observations.

The observed peculiar behavior of packings of non-spherical particles near the sphere point is a consequence of the breaking of rotational symmetry. Near the sphere point the coupling between particle positions and orientations is weak and translations dominate the behavior of the system. In this sense sphere packings are a good model system, and particle shapes close to spherical can be treated as a continuous perturbation of sphere packings. However, even for aspect ratios relatively close to unity, the perturbation changes the properties of the system such as density and contact number in a sharp fashion, making sphere packings a quantitatively unreliable reference point for packings of more realistic particle shapes. Furthermore, even qualitative understanding of jamming and mechanical rigidity for packings of non-spherical particles requires consideration of phenomena that simply do not have a sphere equivalent.

Future work should consider the mathematics of jamming for packings of hard particles that are convex, but not necessarily smooth or strictly convex. In particular, particles with sharp corners and/or flat edges are of interest, such as, for example, cylinders and cubes. We also believe that understanding jamming in frictional hard-particle packings, even for the case of spheres, requires a more thorough mathematical foundation. It is also important to consider packings of soft ellipsoids and in particular develop algorithms to generate them computationally and to study their mechanical properties and vibrational spectra. Investigations of the thermodynamics of very dense ellipsoid systems also demand further attention.

## Acknowledgments

A.D. and S.T. were supported in part by the National Science Foundation under Grant No. DMS-0312067. R.C. was supported in part by the National Science Foundation under Grant No. DMS-0510625. We thank Paul Chaikin for many inspiring discussions of ellipsoid packings.

## Appendix A: THE RECTANGULAR LATTICE OF ELLIPSES

In this Appendix we consider a simple example of a jammed hypoconstrained packing of ellipses having $\bar{Z} = 4$, the minimum necessary for jamming even for disks. Namely, the rectangular lattice of ellipses, i.e., the stretched version of the square lattice of disks, is collectively jammed, and in particular, it is second-order jammed. More specifically, freezing all but a finite subset of the particles, the remaining packing is second-order jammed. An illustration is provided in Fig. 17. At first glance, it appears that one can rotate any of the ellipses arbitrarily without introducing overlap. However, this is only true up to first order, and at the second-order level the "flat" contacts between the ellipses, that is, the contacts whose normals are along the small ellipse semiaxes, block this rotation through the curvature of the particles at the point of contact.

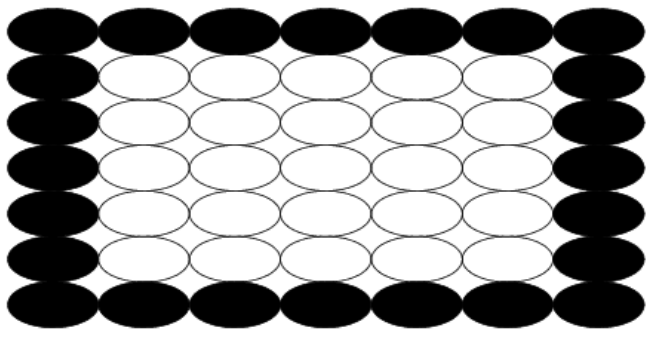

Figure 17: The rectangular lattice of ellipses (i.e., affinely stretched square lattice of hard disks) with "hard-wall" boundary conditions created by freezing the ellipses on the boundary. This packing is jammed since the curvature of the flat contacts blocks the rotations (including collective ones) of the ellipses.

The set of first-order flexes, i.e., particle motions which preserve contact distances to first order, can easily be constructed in this example due to the simple geometry. Namely, a basis vector for this set is a single ellipse rotating around its centroid, giving the total number of first-order flexes $N_f = N$ [19]. The basis formed by these first-order flexes is not orthogonal. However, its advantage is that it is easier to calculate the stress matrix, or more specifically, the matrix $\mathbf{H}_V$; we only need to consider ellipsoid rotations without considering translations. The same observation applies whenever one takes a jammed sphere packing and makes the particles nonspherical but does not change the normal vectors at the point of contact. This can be done, for example, by simply taking a jammed sphere packing and swelling the particles to be nonspherical, without changing the geometry or connectivity of the contact network. If the particles swell enough to make all of the contacts sufficiently flat, the new packing will be jammed, since all of the first-order flexes consist of particle rotations only and are blocked by the flat curvature of the contacts.

The fact that "flat" (the contacts among vertical neighbors in Fig. 17) contacts block rotations can easily be seen analytically by considering the case of one ellipse jammed among four fixed ellipses (two horizontally, two vertically). Specifically, any self-stress for which the contact force in the "flat" contacts is larger than the force in the "curved" contacts, $f_{\text{flat}} > f_{\text{curv}}$, makes the mobile ellipse jammed, more specifically, pre-stress rigid [19]. The same result can be shown to apply to the square lattice of ellipses for an arbitrary number of ellipses. If the ellipses are not hard but rather deformable, the packing would not support a compression along the curved contacts, but it would along the flat contacts. This is a very intuitive result: If one takes a smooth ellipsoid and presses it against a table with its most curved tip, it will buckle and the only stable configuration is one where the flat tip presses against the table. Note however that the hard-ellipse equivalent is jammed and can resist any finite external forces, including a compression along the curved contacts. The anharmonicity of the hard-sphere potential becomes essential in this example, since the packing can choose the correct internal (self) stresses (forces) needed to provide mechanical rigidity. In (realistic) systems of deformable particles, the internal stresses are fixed and determined by the state of compression.